\documentclass{aastex62}

\newcommand{\Eexc}{$E_{\rm exc}$}
\newcommand{\Teff}{$T_{\rm eff}$}  % Command for Teff, text mode
\newcommand{\kms}{km\,s$^{-1}$}
\def\ione{\,{\sc i}}
\def\ii{\,{\sc ii}}
\def\iii{\,{\sc iii}}

\accepted{ by ApJ for publication September 11, 2018 }
\shorttitle{NLTE line formation for Mg\ione\ and Mg\ii\ in atmospheres of B-A-F-G-K stars }
\shortauthors{Alexeeva et al.}

\begin{document}

\title{NLTE line formation for Mg\ione\ and Mg\ii\ in atmospheres of B-A-F-G-K stars}

\author{Sofya Alexeeva}
\affiliation{Shandong Provincial Key Laboratory of Optical Astronomy and Solar-Terrestrial Environment, Institute of Space Sciences, Shandong University, Weihai 264209, China \\}
\email{alexeeva@sdu.edu.cn}
\nocollaboration

\author{Tatiana Ryabchikova}
\affiliation{Institute of Astronomy of the RAS, Moscow, 119017, Russia \\}
\nocollaboration

\author{Lyudmila Mashonkina}
\affiliation{Institute of Astronomy of the RAS, Moscow, 119017, Russia \\}
\nocollaboration

\author{Shaoming Hu} 
\affiliation{Shandong Provincial Key Laboratory of Optical Astronomy and Solar-Terrestrial Environment, Institute of Space Sciences, Shandong University, Weihai 264209, China \\}
\email{husm@sdu.edu.cn}
\nocollaboration

\begin{abstract}

Non-local thermodynamical equilibrium (NLTE) line formation for Mg\ione\ and Mg\ii\ lines is considered in classical 1D-LTE model atmospheres of the Sun and 17 stars with reliable atmospheric parameters and in a broad range of spectral types: 3900 $\le$ \Teff\ $\le$ 17500~K, 1.1 $\le$ log$g$ $\le$ 4.7, and $-$2.6 $\le$ [Fe/H] $\le$ +0.4.

We find that, for each star, NLTE leads to smaller line-to-line scatter. For 10 stars, NLTE leads to consistent abundances from Mg\ione\ and Mg\ii, while the difference in the LTE abundance varies between $-0.21$ and +0.23~dex. We obtain an abundance discrepancy betweeen Mg\ione\ and Mg\ii\ in the two very metal-poor stars, HD~140283 and HD~84937. An origin of these abundance differences remains unclear.

Our standard NLTE modelling predicts Mg\ione\ emission lines at 7.736, 11.789, 12.224, and 12.321~$\mu$m in the atmospheres with \Teff\ $\le$ 7000~K. We reproduce well the Mg\ione\ 12.2 and 12.3~$\mu$m emission lines in Procyon. However, for the Sun and 3 K-giants, the predicted
Mg\ione\ emission lines are too weak compared with the observations.

For stars with 7000~K $\le$ \Teff\ $\leq$ 17500~K, we recommend the Mg\ii\ 3848, 3850, 4384, 4390, 4427, and 4433~\AA\ lines for Mg abundance determinations even at the LTE assumption due to their small NLTE effects. The Mg\ione\ 4167, 4571, 4702, 5528, 5167, 5172, and 5183~\AA\ lines can be safely used in the LTE analysis of stars with 7000~K $<$ \Teff\ $\leq$ 8000~K. For the hotter stars, with \Teff\ from 8000 to 9500~K, the NLTE effects are minor only for Mg\ione\ 4167, 4702, and 4528~\AA.

\end{abstract}


\section{Introduction} \label{sec:intro}

 Magnesium is one of the best-observable elements in A to K-type stars.  
 The Mg\ione b 5167, 5172, 5183~\AA\ lines can be measured over a wide range of effective temperatures up to 13\,000~K \citep{2009AA...503..945F} and over a wide range of metallicity, even in the most metal-poor stars ever discovered in our Galaxy, with [Fe/H] $\simeq -7$ \citep{2014Natur.506..463K} and $-5.5$ \citep{Aoki_he1327}, and the Galaxy satellites, with [Fe/H] $\sim -4$ \citep[for example, Scl-50 in][]{2010A&A...524A..58T}.
 The Mg\ii\ 4481~\AA\ line is strong and unblended in BA-type stars and commonly used as abundance and rotational velocity indicator.  
 This line can be detected in luminous objects outside the Local Group which are accessible to low-resolution spectroscopy. 

The knowledge of accurate magnesium abundances is important for studies of the history of $\alpha$-process nucleosynthesis in the Universe, formation and evolution of a large number of galaxies, and understanding stellar physics and planetary systems. Most of magnesium originates from a single astrophysical site, core-collapse supernovae \citep{1995ApJS..101..181W}, where it is produced by the $\alpha$-process. Therefore, magnesium is suitable as a reference element in the galactic chemical evolution studies \citep{Fuhrmann1998}, and the [Mg/Fe] abundance ratio is commonly used to estimate an initial mass function and star-formation timescale in a system \citep[e.g.,][]{1990ApJ...365..539M}.

Lines of magnesium combined with the other spectroscopic indicators are widely applied to determine stellar atmosphere parameters, that is, effective temperature \Teff\ and surface gravity log~$g$. The Mg\ione b lines with pressure-broadened wings are used as a surface gravity diagnostic for cool stars \citep[e.g.,][]{1960MNRAS.121...52D, 1988A&A...190..148E, 1997A&A...323..909F}.
The simultaneous presence of lines of the two ionization stages, Mg\ione\ and Mg\ii, in A and F-type stars
provides an opportunity to determine their effective temperatures 
 \citep[e.g.,][]{1995ApJS...99..659V,2001A&A...369.1009P,2009AA...503..945F}.
 
 In stellar atmospheres with \Teff\ $>$ 4500~K, neutral magnesium is a minority species, and its statistical equilibrium (SE) can easily deviate from local thermodynamic equilibrium (LTE) owing to deviations
of the mean intensity of ionizing radiation from the Planck function. A clear signature of the departures from LTE for Mg\ione\ is the emission lines at 7.3~$\mu$m (5g -- 6h), 12.3 $\mu$m (7i -- 6h), and 12.2 $\mu$m (7h -- 6g) observed in the Sun \citep{1981ApJ...247L..97M, 1983ApJ...275L..11C, 1983ApJ...269L..61B}, Arcturus \citep{1996ASPC..109..723U}, Procyon \citep{2004ApJ...611L..41R}, and Pollux \citep{2008A&A...486..985S}. These lines arise in the transitions between the high-excitation levels of Mg\ione\ and should form in deep atmospheric layers, which are not affected by the chromospheric temperature rise. The LTE calculations with classical hydrostatic model atmosphere result always in the absoption line profile. 

There have been a large number of studies based on the non-local thermodynamic equilibrium (NLTE) line formation for Mg\ione\ in the Sun and FGK-type stars. The original model atoms were from 
\citet{1969BAAS....1..272A, 1988ApJ...330.1008M, 1992A&A...253..567C, 1996ARep...40..187M,
  1998A&A...333..219Z, Gratton1999, 2000ApJ...541..207I, 2004A&A...418..551M, 2011MNRAS.418..863M, 2015AA...579A..53O, 2017A&A...597A...6N}. 
 It was understood that the main NLTE mechanism for Mg\ione\ is ultra-violet (UV) overionization of the low-excitation levels. This results in an underpopulation of neutral magnesium that tends to make the Mg\ione\ lines weaker compared with their LTE strengths. Using classical hydrostatic and plane-parallel (1D) model atmosphere without chromosphere, \citet{1992A&A...253..567C} explained the formation of two Mg\ione\ 12-micron emission features in the solar spectrum as a result of population depletion by line photon losses followed by population replenishment from the ionic reservoir in the highly excited levels. \citet{2011MNRAS.418..863M} 
 evaluated the NLTE effects for the IR lines of Mg\ione, which are important for the Gaia project, in the grid of model atmospheres of cool giants.
Using a comprehensive model atom based on accurate collisional data, \citet{2016A&A...586A.120O} calculated departure coefficients for the Mg\ione\ levels and LTE and NLTE equivalent widths for the Mg\ione\ lines in a grid of the MARCS atmospheric models \citep{2008A&A...486..951G}. The magnesium NLTE abundances of extended samples of cool stars were determined by \citet{2010A&A...509A..88A} and \citet{2017ApJ...847...16B}. Taking advantage of the NLTE line formation for not only  Mg\ione\ but also Fe\ione-Fe\ii, \citet{2016ApJ...833..225Z} and \citet{2017A&A...608A..89M} showed that the Galactic dwarf and giant stars at [Fe/H] $\le -0.8$ and the very metal-poor ([Fe/H] $< -2$) giants in the classical dwarf spheroidal galaxies Sculptor, Ursa Minor, Sextans, and Fornax reveal a similar plateau at [Mg/Fe] $\simeq$ 0.3. A step forward in modelling the Mg\ione\ lines in cool stars was self-consistent NLTE calculations with a three-dimensional, time-dependent, and hydrodynamical (3D) model atmosphere. They were performed for the most iron-deficient star,
SMSS0313-6708 \citep{2017A&A...597A...6N}, and two models representing the atmospheres of a metal-poor giant and a metal-rich dwarf \citep{2017ApJ...847...15B}. 

 As for Mg\ii\ in late-type stars,  NLTE modelling was performed by \citet{2004MNRAS.350.1127A} and \citet{2015AA...579A..53O}, with detailed NLTE abundance results from Mg\ii\ 7877, 7896~\AA\ reported in the first paper.
% the results of the NLTE modelling and abundance determinations were reported only in the papers by \citet{2004MNRAS.350.1127A} and \citet{2015AA...579A..53O}.

  For A and the earlier type stars, NLTE studies of the magnesium lines are few in number. 
 \citet{1972ApJ...177..115M} and \citet{1975A&A....41..245S}  investigated the
Mg\ii\ lines in O-B type stars and found  very large deviations from LTE in the hot low-gravity models.
 \citet{1988A&A...192..264G}  determined the NLTE abundances from the Mg\ione\ and Mg\ii\ lines in Vega (\Teff = 9500 K, log$g$ = 3.90) and obtained a satisfactory agreement between abundances inferred from the two ionization stages. The calculated NLTE abundance corrections were small for both Mg\ione\ and Mg\ii\ lines, with the largest one of $-$0.1~dex for Mg\ii\ 4481~\AA.
   The NLTE calculations for Mg\ione-Mg\ii\ in three A0-type stars  were performed by \citet{2001A&A...369.1009P}. They showed that accounting for the departures from LTE is
essential for an accurate effective temperature determination from the Mg\ione-Mg\ii\ ionization equilibrium.

  This study aims to construct a comprehensive model atom for Mg\ione-Mg\ii\ based on the most up-to-date atomic data available so far, which can be applicable to analysis of the magnesium lines in a broad range of spectral types, from K to late B.
 As a first application of the treated model atom, we obtain the magnesium abundances of the reference stars, Vega and Sun, with well-determined atmospheric parameters, using an extensive list of Mg\ione\ and Mg\ii\ lines.
Based on high-resolution observed spectra, we determine the magnesium NLTE abundances also for eight representatives of A and late B stars
and eight representatives of F-G-K stars.

The article is organized as follows. Section\,\ref{Sect:atom} describes the model atom of Mg\ione -Mg\ii, adopted atomic data, and the departures from LTE for Mg\ione --Mg\ii\ depending
 on the stellar parameters. Analysis of the solar Mg\ione --Mg\ii\ lines is presented in Section\,\ref{subsec:sun}. 
 In Section\,\ref{Sect:Stars}, we determine the Mg abundances of the reference K--B stars. Comparisons with the previous studies are presented in Section\,\ref{sec:comparison}.
We summarize our conclusions in Section\,\ref{Sect:Conclusions}.

 \begin{figure*}
 \begin{minipage}{190mm}
 \includegraphics[scale=0.65]{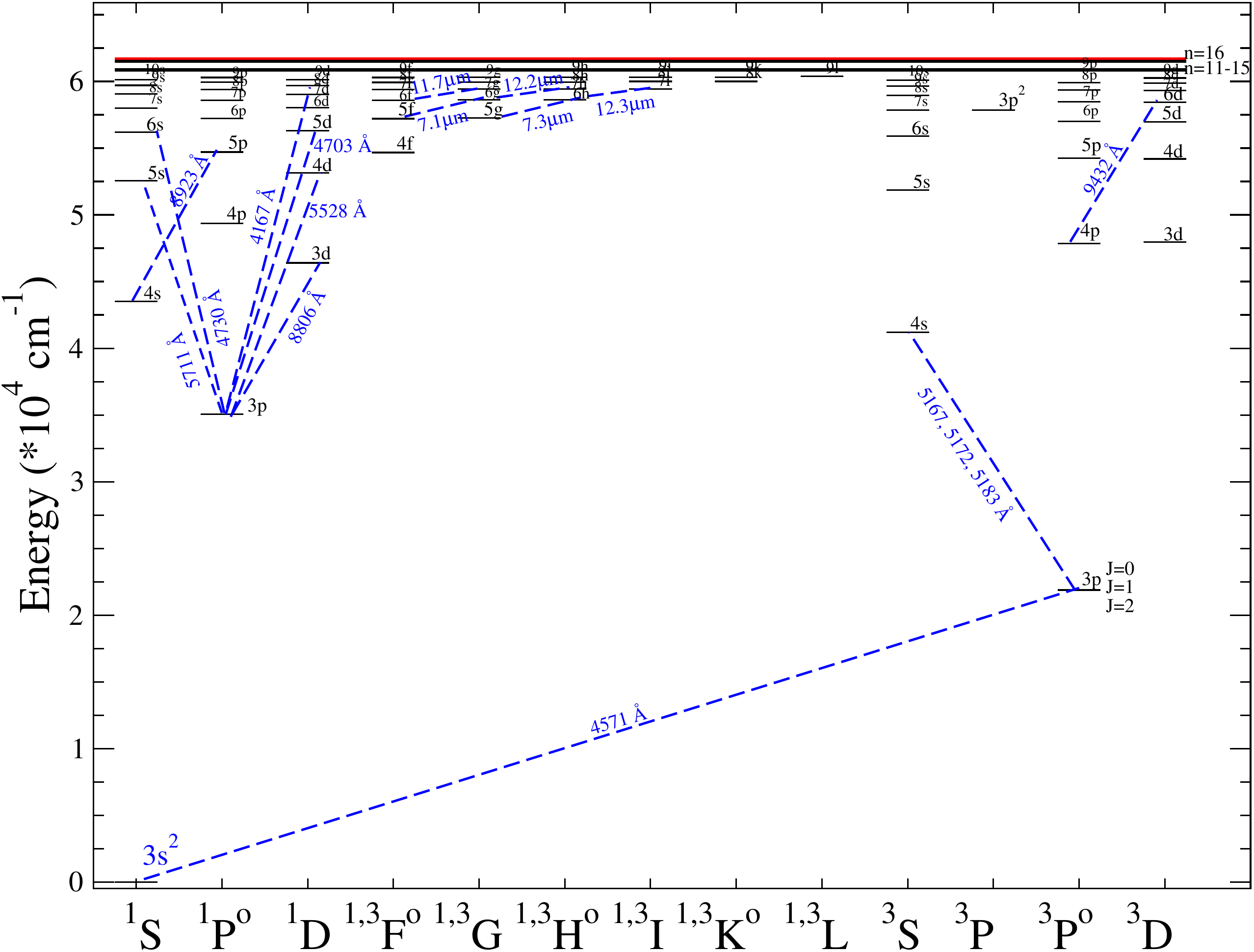}
 \caption{Term diagram for Mg\ione. The dashed lines indicate the transitions, where the investigated spectral lines arise. }
 \label{Grot_Mg1}
 \end{minipage} 
 \end{figure*}

  \begin{figure}
\begin{center}
 \includegraphics[scale=0.54]{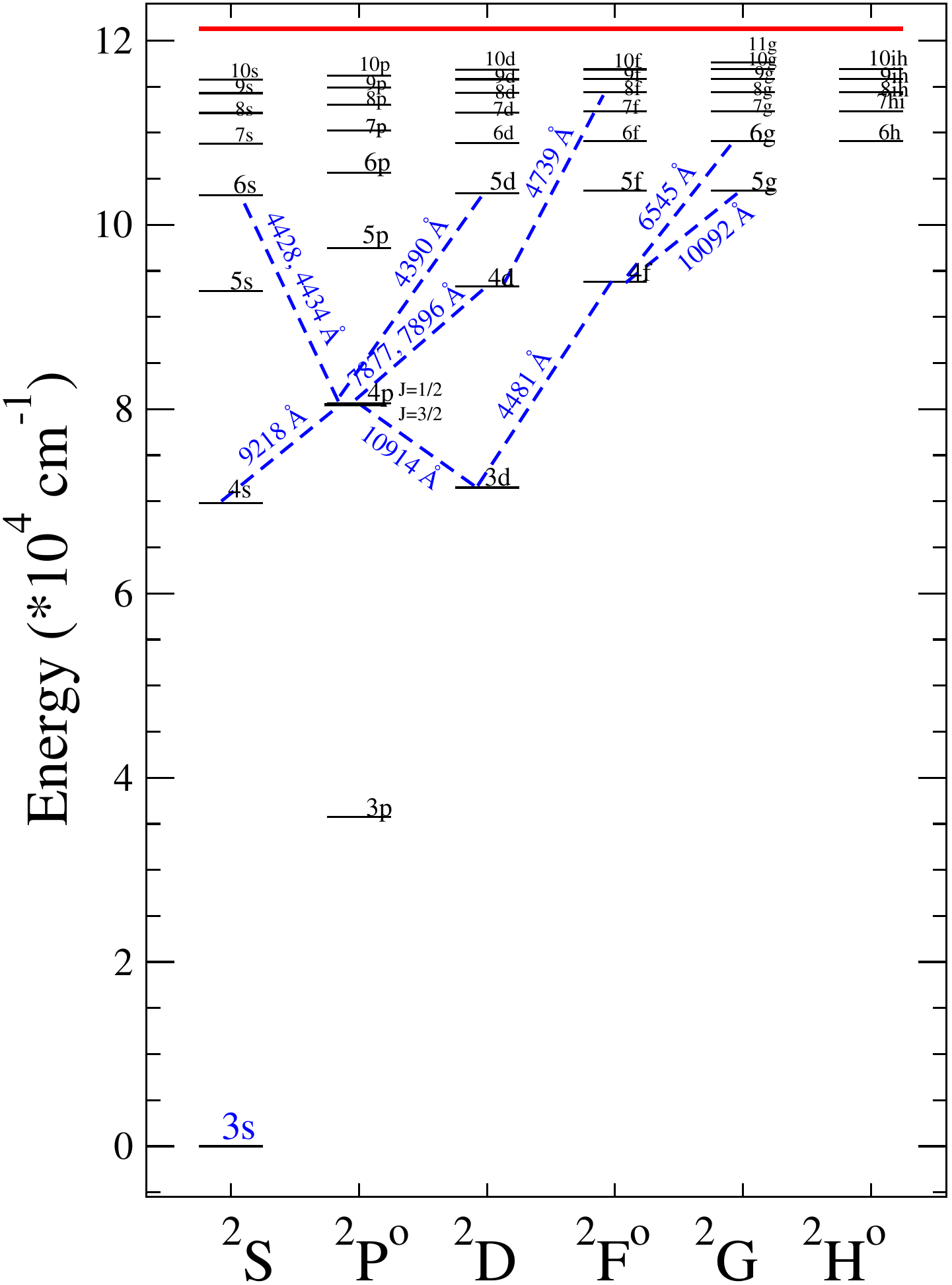}
 \caption{Term diagram for Mg\ii. The dashed lines indicate the transitions, where the investigated spectral lines arise. }
 \label{Grot_Mg2}
 \end{center}
 \end{figure}

   \begin{figure*}
   \begin{minipage}{179mm}
 \begin{center}
 \parbox{0.258\linewidth}{\includegraphics[scale=0.26]{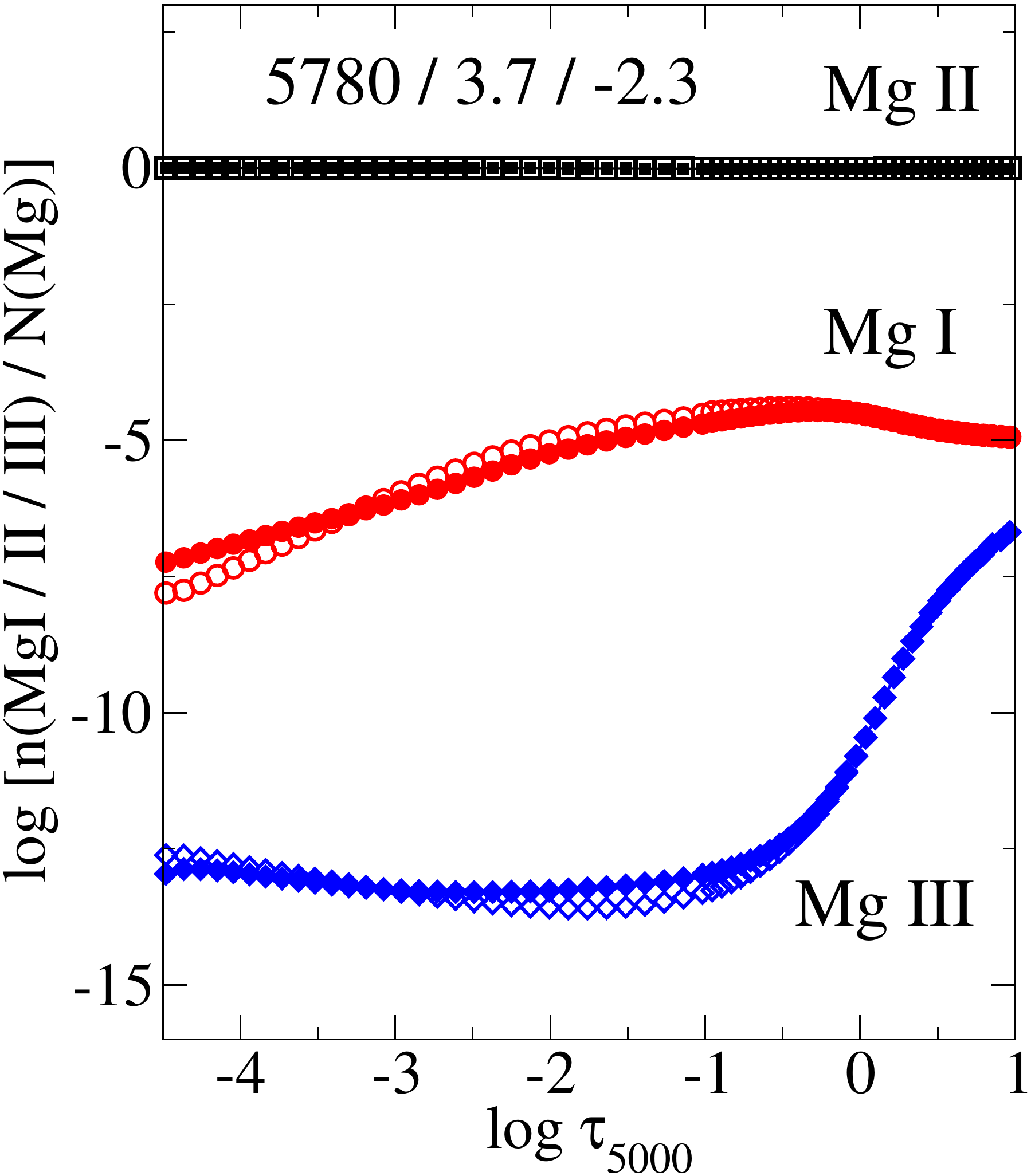}\\
 \centering}
 \parbox{0.216\linewidth}{\includegraphics[scale=0.26]{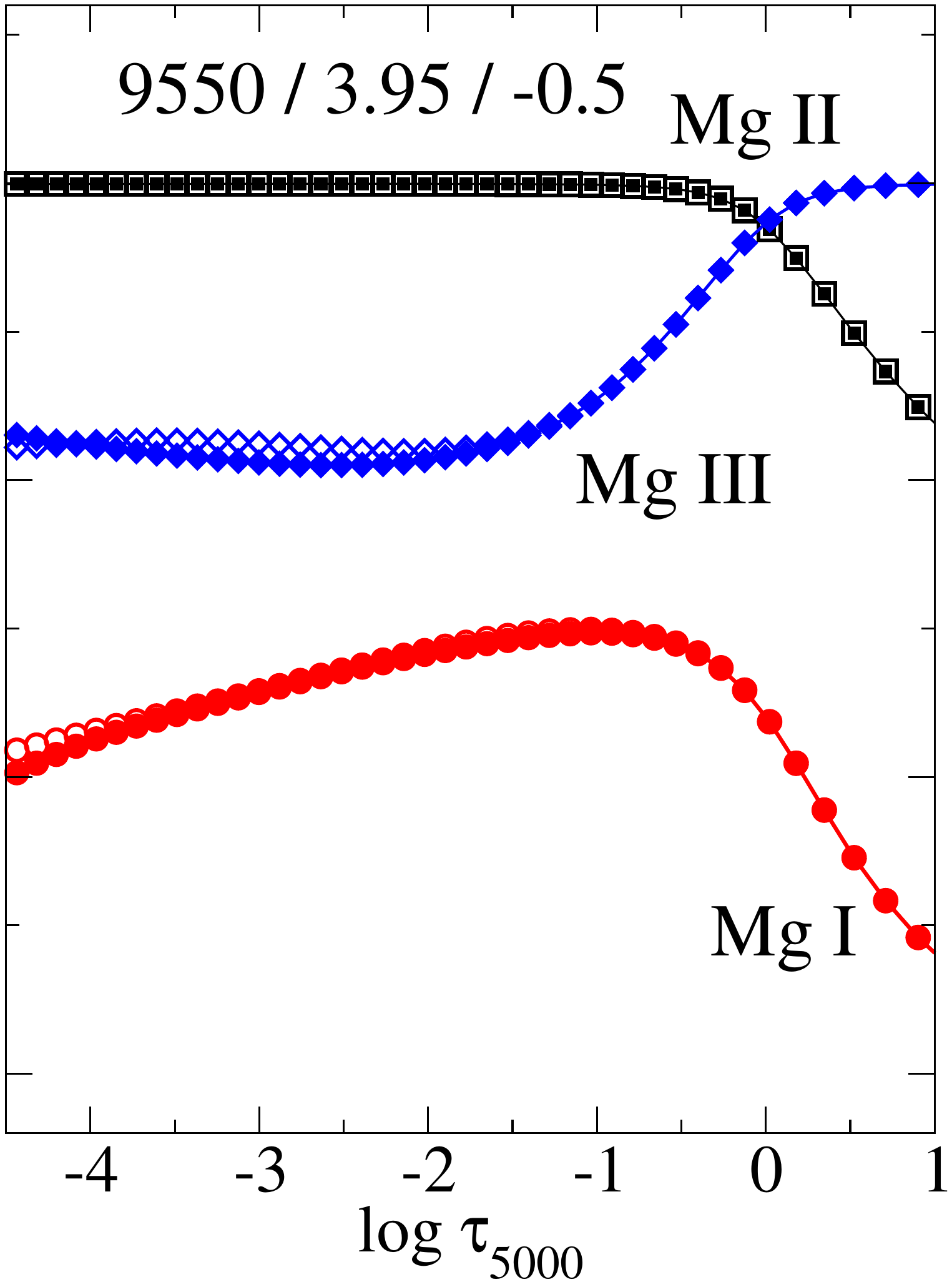}\\
 \centering}
  \parbox{0.22\linewidth}{\includegraphics[scale=0.26]{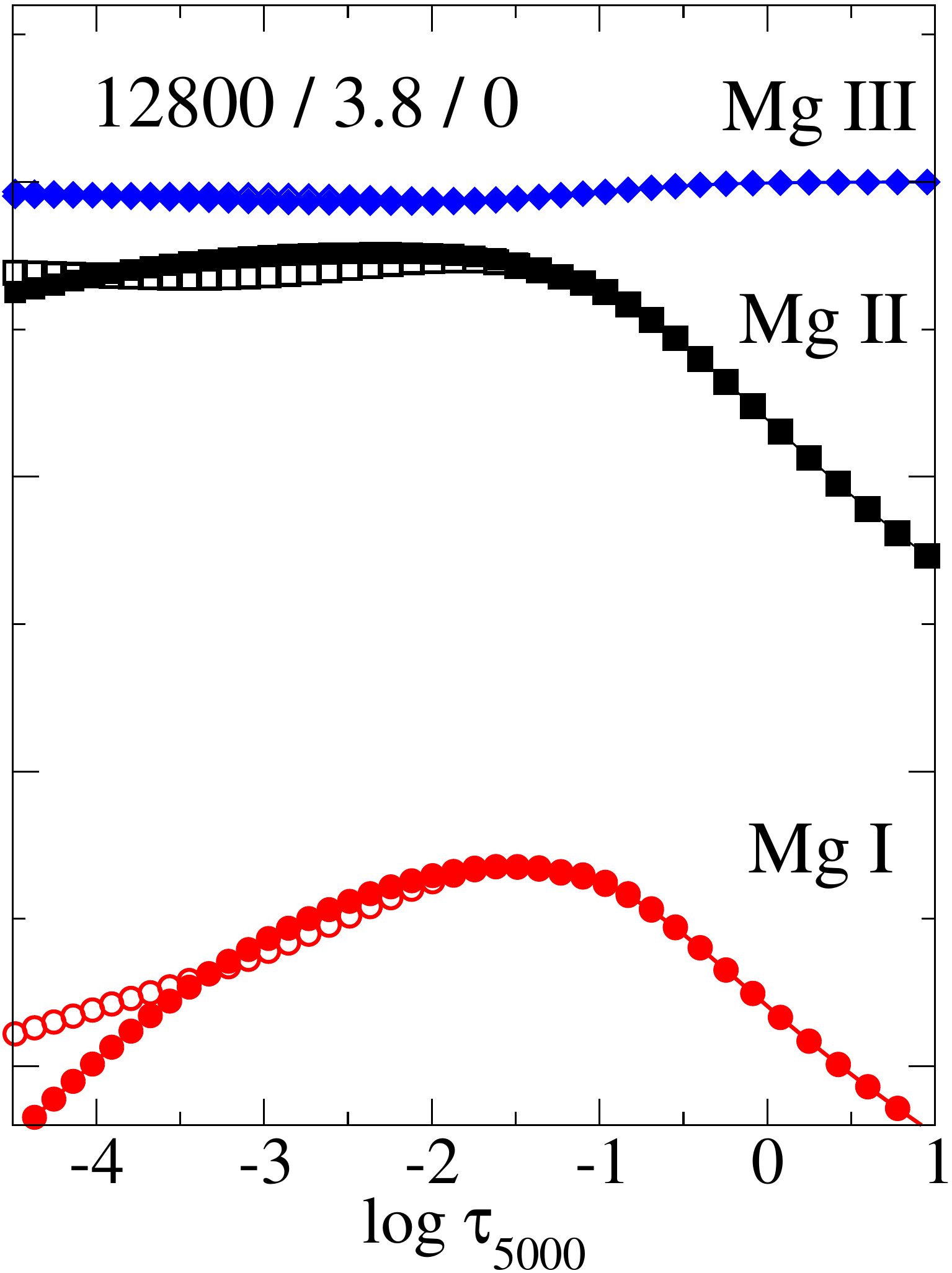}\\ 
 \centering}
 \parbox{0.18\linewidth}{\includegraphics[scale=0.26]{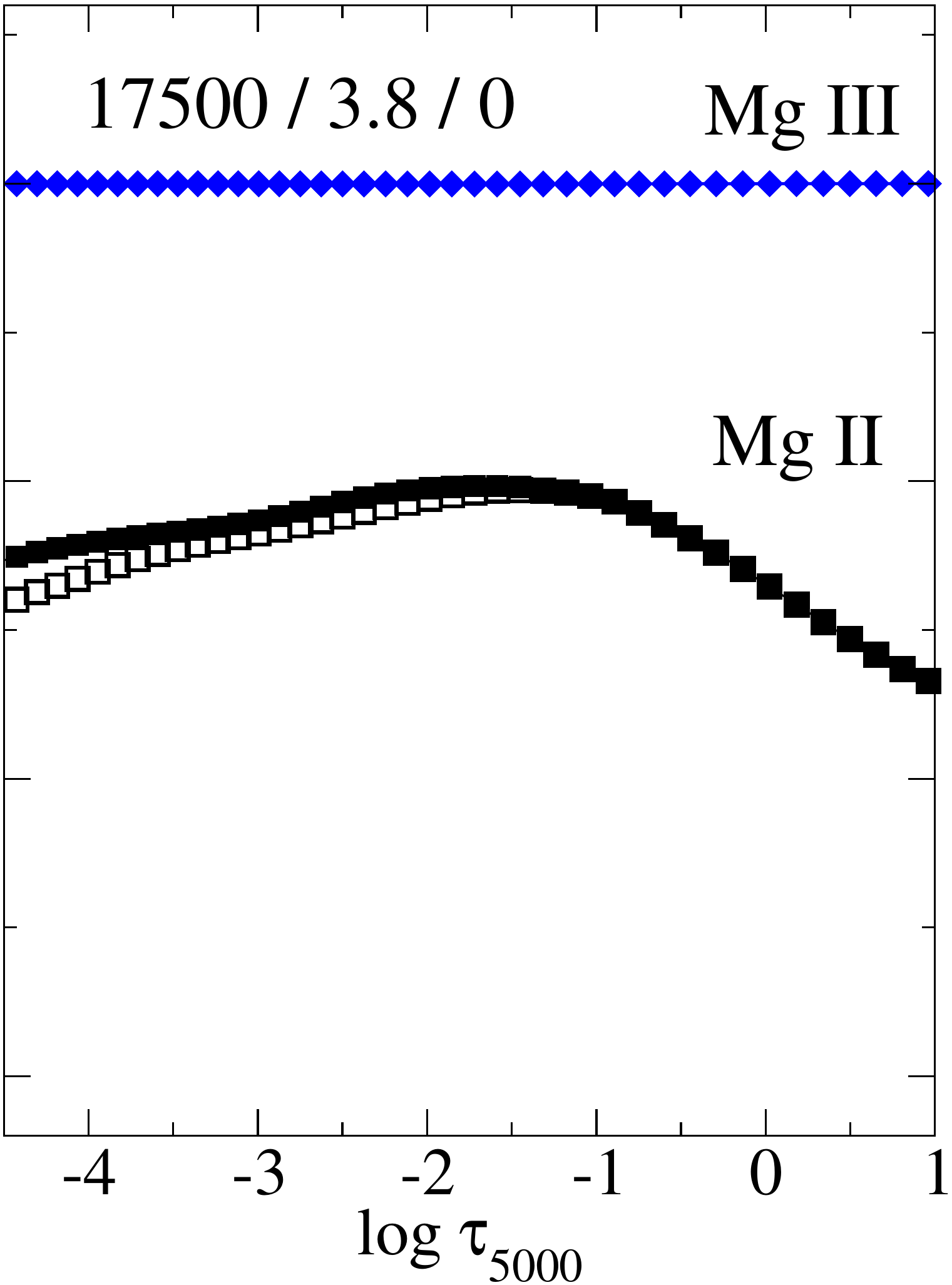}\\
 \centering}
 \hspace{1\linewidth}
 \hfill
 \\[0ex]
 \caption{NLTE and LTE fractions of Mg\ione, Mg\ii, and Mg\iii\ in the model atmospheres of different effective temperatures. LTE -- open symbols, NLTE -- filled symbols.}
 \label{balance}
 \end{center}
 \end{minipage}
 \end{figure*}

\section{NLTE line formation for Mg\ione\ -- Mg\ii}\label{Sect:atom}

 \subsection{Model atom and atomic data}
 
 {\bf Energy levels.} 
Model atom includes the Mg\ione\ levels belonging to singlet and triplet terms of the 3s$nl$ ($n \le$ 16, $l \le$ 8) and 3p$^2$ electronic configurations, the Mg\ii\ levels belonging to doublet terms of the 2p$^6$$nl$ ($n \le$ 10, $l \le$ 6) electronic configurations, and the ground state of Mg\iii.
 Most energy levels were taken from the Version 5 of NIST database \footnote{\url{https://www.nist.gov/pml/atomic-spectra-database}} \citep{NIST_ASD}.
 The Mg\ione\ levels with $n$ = 11--16 have small energy differences and their populations must be in LTE relative to one another. These levels were used to make  
two superlevels.
 Triplet fine structure was neglected except for the 3s3p$^3$P$^\circ$ splitting and doublet fine structure  was neglected except for the 4p$^2$P$^\circ$ splitting.
 Energy levels up to 0.02/0.45 eV below the ionization threshold are explicitly included in our Mg\ione/Mg\ii\ model atom. 
 The term diagrams for Mg\ione\ and Mg\ii\ in our model atom are shown in Fig.\,\ref{Grot_Mg1} and Fig.\,\ref{Grot_Mg2}. 
 
 \noindent {\bf Radiative data.} 
 Our model atom includes 673 allowed bound-bound ($b-b$) transitions. Their transition probabilities were taken from the Opacity Project (OP) database 
 TOPbase \footnote{\url{http://cdsweb.u-strasbg.fr/topbase/topbase.html}} \citep{1993BICDS..42...39C, 1989JPhB...22..389L, 1993AAS...99..179H} except for the
 transitions betweeen levels with $n$ = 7$-$15 and $l$ = 4$-$8 in Mg\ione, for which oscillator strengths were calculated following \citet{1957ApJS....3...37G}.
 Photo-ionization cross-sections for levels of Mg\ione\ with $n \le 9$, $l \le 4$ and Mg\ii\ with $n \le 10$, $l \le 4$ were taken from TOPbase, and we adopted the hydrogen-like cross-sections for the higher excitation levels. 
 
 \noindent {\bf Collisional data.} 
For all the transitions between the lowest 17 energy levels of Mg\ione\ (up to 5p $^1$P$^\circ$) and  transitions from these levels to the more excited levels up to 3p$^2$ $^1$S, in total 369 transitions, we adopted accurate electron-impact excitation data from \citet{2015A&A...577A.113M}. \citet{1995JPhB...28.4879S} provided accurate electron collision strengths for transitions between the lowest 10 levels of Mg\ii\ (up to 5d $^2$D).
%For electron-impact excitation of all the transitions between the lowest 37 energy levels of Mg\ione\ (up to 3p$^2$ $^1$S) and the lowest 10 levels of Mg\ii\ (up to 5d $^2$D) we adopted accurate effective collision strengths from \citet{2015A&A...577A.113M} and \citet{1995JPhB...28.4879S}, respectively.  
 For the remaining transitions, we use the impact parameter method \citep[IPM,][]{1962PPS....79.1105S} in case of the allowed transitions and assume that the effective collision strength $\Omega_{ij}$ = 1 in case of the forbidden transitions.
  
 The electron-impact ionization was considered using the formula of \citet{Seaton1962} and the threshold photoionization cross-sections.
 As discussed in the literature, the statistical equilibrium of Mg\ione\ in cool atmospheres is affected by inelastic collisions with neutral hydrogen atoms \citep[e.g][]{2013AA...550A..28M}.
  We used detailed quantum mechanical calculations of \citet{2012A&A...541A..80B} for H\ione\ impact excitations and charge transfer processes.
%We used detailed quantum mechanical calculations of \citet{2012A&A...541A..80B} for Mg\ione +H\ione\ collisions.
 
 \subsection{Method of calculations}
 
  To solve the radiative transfer and statistical equilibrium equations, we used the code \textsc{DETAIL} \citep{detail} based on the accelerated $\Lambda$-iteration method \citep{rh91}. 
  The \textsc{DETAIL} opacity package was updated by \citet{2011JPhCS.328a2015P} by including bound-free opacities of neutral and ionized species.
  The obtained departure coefficients, $b_{\rm{i}}$ = $n_{\rm{NLTE}}$ / $n_{\rm{LTE}}$, were then used by the code \textsc{synthV\_NLTE} \citep{2016MNRAS.456.1221R} to calculate the synthetic NLTE line profiles. Here, $n_{\rm{NLTE}}$ and $n_{\rm{LTE}}$ are the statistical equilibrium and thermal (Saha-Boltzmann) number densities, respectively. Comparison with the observed spectrum and spectral line fitting were performed with the \textsc{binmag} code\footnote{\url{http://www.astro.uu.se/~oleg/binmag.html}} \citep{binmag3,2018ascl.soft05015K}.

 % Calculations were performed using plane-parallel (1D), chemically homogeneous model atmospheres from the Kurucz's grid\footnote{\url{http://www.oact.inaf.it/castelli/castelli/grids.html}} \citep{2004astro.ph..5087C}, the MARCS grid \citep{2008A&A...486..951G}, and also computed with the \textsc{LLmodels} code \citep{2004AA...428..993S}.
 % For Sirius, we took the Kurucz's model atmosphere\footnote{\url{http://kurucz.harvard.edu/stars/SIRIUS/ap04t9850g43k0he05y.dat}} computed with the parameters close to those derived by \citet{1993AA...276..142H}. 
  
   Calculations were performed using plane-parallel (1D) and chemically homogeneous model atmospheres. 
  %For late-type stars (3900 K $<$ \Teff $<$ 7000 K), we applied classical plane-parallel model atmospheres from the MARCS model grid \citep{2008A&A...486..951G}, which were interpolated for given \Teff, log$g$, and [Fe/H] using a   FORTRAN-based routine written by Thomas Masseron\footnote{See http://marcs.astro.uu.se/software.php}. For A-B-type stars, the model atmospheres were calculated under the LTE assumption with the \textsc{LLmodels} code \citep{2004AA...428..993S}.  
  For consistency with our NLTE studies of C\ione -C\ii\ \citep{2015MNRAS.453.1619A,2016MNRAS.462.1123A}, Ti\ione -Ti\ii\ \citep{2016MNRAS.461.1000S}, and Ca\ione -Ca\ii\ \citep{2018MNRAS.477.3343S}, here we use exactly the same model atmosphere for each star, as in the earlier papers. For late-type stars (3900~K $<$ \Teff\ $<$ 7000~K), these are the model atmospheres, which were interpolated in the MARCS model grid \citep{2008A&A...486..951G} for given \Teff, log$g$, and [Fe/H] using a FORTRAN-based routine written by Thomas Masseron\footnote{See http://marcs.astro.uu.se/software.php}. For A-B-type stars, the model atmospheres were calculated  with the code \textsc{LLmodels} \citep{2004AA...428..993S}. An exception is Sirius, for which the model atmosphere was computed by R.~Kurucz\footnote{http://kurucz.harvard.edu/stars/SIRIUS/ap04t9850g43k0he05y.dat}.

  Table~\ref{tab1} lists lines of Mg\ione\ and Mg\ii\ used in our abundance analyses together with the adopted line data.
  For lines of Mg\ione\ in the visible spectral range, oscillator strengths were taken from \citet{2017AA...598A.102P} and, for the infrared (IR) lines, were calculated following \citet{1957ApJS....3...37G}. The NIST data \citep{NIST_ASD} were mostly employed for lines of Mg\ii.  It is worth noting that the Mg\ione\ line arising from the ground state, 4571\,\AA, was not used for late type stars of close-to-solar metallicity due to possible affecting by stellar chromospheres \citep[see, for example,][]{1975SoPh...42..289A,2017A&A...604A..50S}.
%Oscillator strengths were taken from the NIST database \citep{NIST_ASD}, and from \citet{2007AA...461..767A} for Mg\ione\ at 5167.32, 5172.68 and 5183.60~\AA\ lines.
%  For the Mg\ione\ infrared lines, oscillator strengths were calculated following \citet{1957ApJS....3...37G}.

  For most lines of Mg\ione\ and Mg\ii\ Stark collisional data are taken from \citet{1996AAS..117..127D} and \citet{1995BABel.151..101D}, respectively. They are given separately for
  collisions with electrons and protons in logarithmic form for T=10\,000~K. \textsc{SynthV\_NLTE} code was changed to account for
  both colliders. In the atmospheres of cool stars both $\Gamma_{4e}$ and $\Gamma_{4p}$ may be coadded and
  multiplied by number of electrons only, while in hotter stars collisions should be treated separately and coadded after multiplying by the number of colliders.

  In cool atmospheres, profiles of the Mg\ione\ lines are sensitive to the van der Waals broadening.
  For most part of neutral magnesium lines collisional broadening due to collisions with neutral H is described
  via cross-sections and velocity parameters from the ABO theory \citep{1995MNRAS.276..859A, 1997MNRAS.290..102B}. 
  The van der Waals broadening constants, $\Gamma_6$ /$N_H$, were taken from \citet{BPM} and are presented in
  Table~\ref{tab1} in logarithmic form for T = 10\,000~K. For infrared lines of Mg\ione\ van der Waals broadening parameters were taken from \citet{2015AA...579A..53O}.
   For the remaining lines of Mg\ione\ and for all Mg\ii\ lines the corresponding broadening data were extracted from Kurucz' site\footnote{\url{http://kurucz.harvard.edu/atoms/1200/gfemq1200/}}$^,$\footnote{\url{http://kurucz.harvard.edu/atoms/1201/gfemq1201/}}.

 \subsection{Departures from LTE for Mg\ione/Mg\ii\ in B- to K-type star}\label{Sect:departure}

 \begin{figure*}
  \begin{minipage}{175mm}
 \parbox{0.35\linewidth}{\includegraphics[scale=0.6]{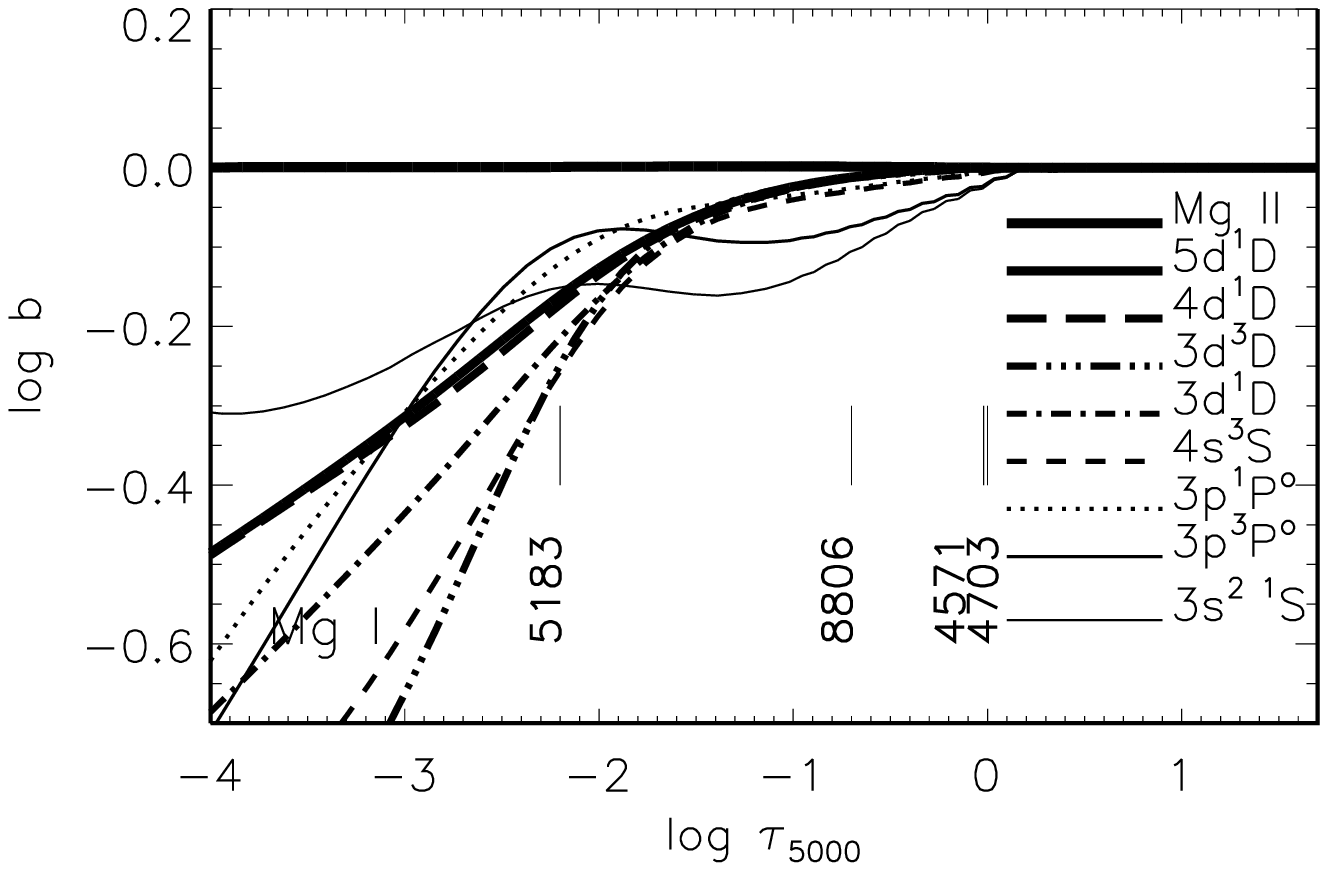}\\
 \centering}
 \hspace{0.1\linewidth}
 \parbox{0.35\linewidth}{\includegraphics[scale=0.6]{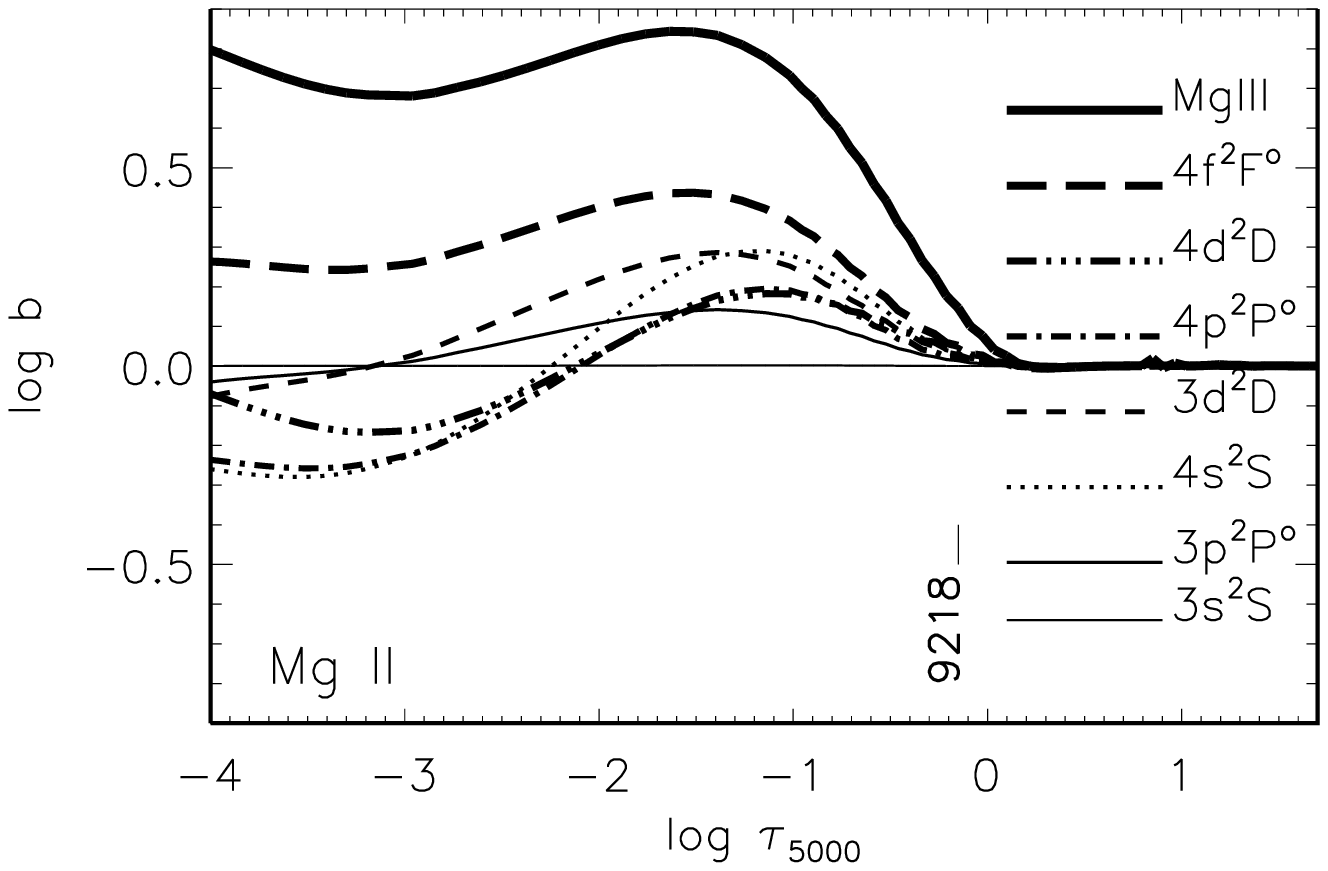}\\
 \centering}
 \hspace{0.00\linewidth}
 \hfill
 \\[0ex]
 \parbox{0.35\linewidth}{\includegraphics[scale=0.6]{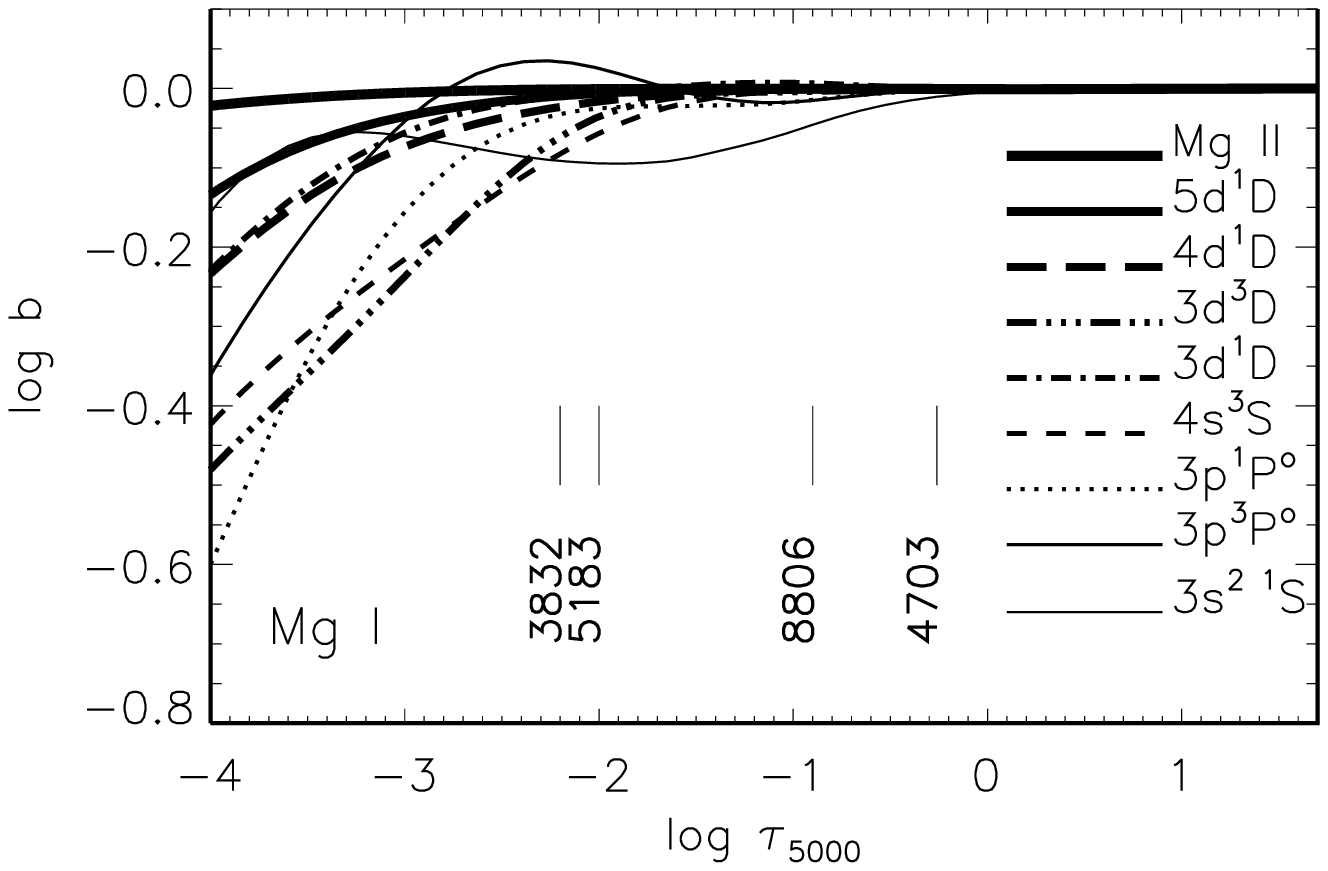}\\
 \centering}
 \hspace{0.1\linewidth}
 \parbox{0.35\linewidth}{\includegraphics[scale=0.6]{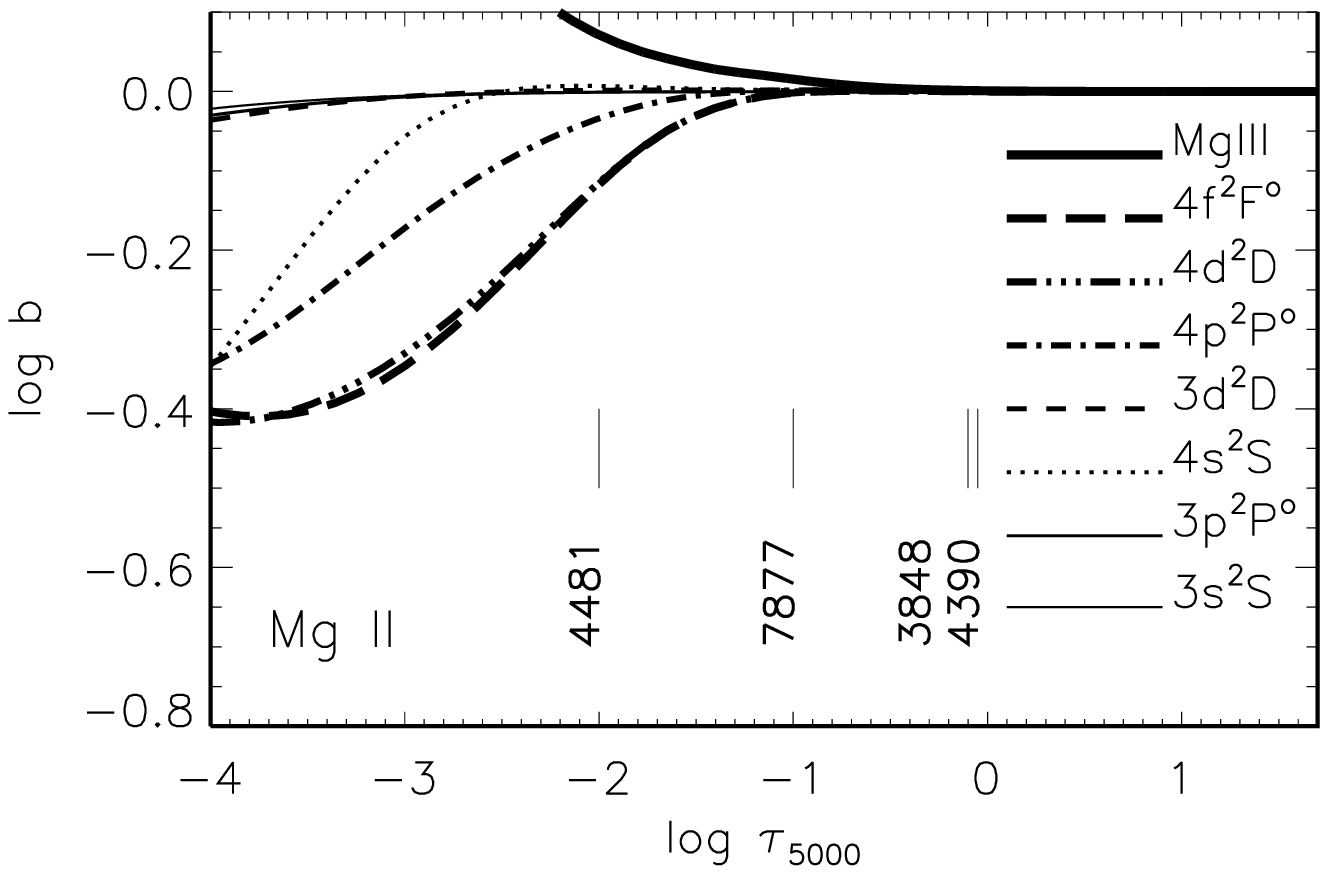}\\
 \centering}
 \hfill
 \\[0ex]
 \parbox{0.35\linewidth}{\includegraphics[scale=0.6]{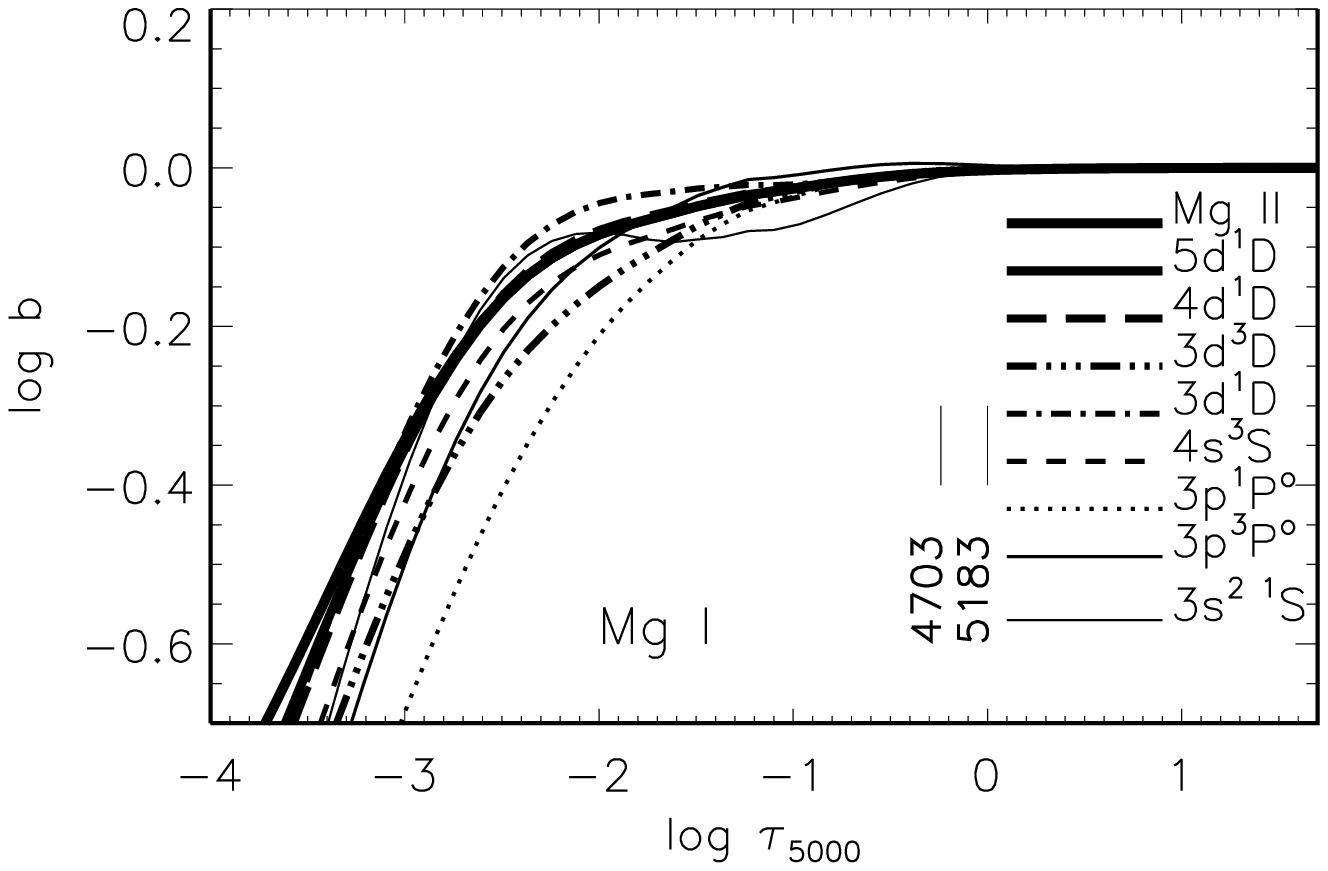}\\
 \centering}
 \hspace{0.1\linewidth}
 \parbox{0.35\linewidth}{\includegraphics[scale=0.6]{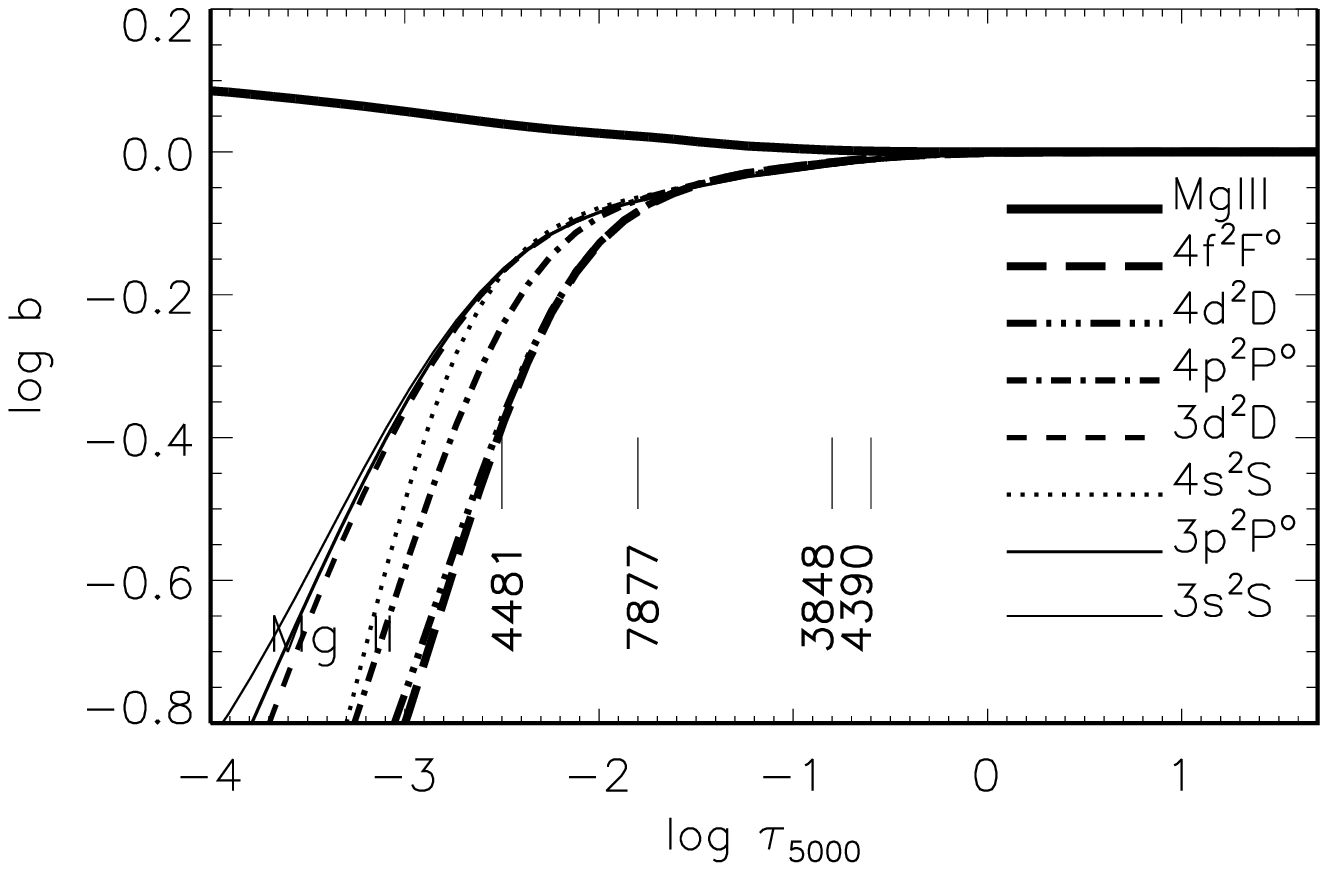}\\
 \centering}
 \hfill
 \caption{Departure coefficients for the Mg\ione\ levels (left column) and the Mg\ii\ levels and the ground state Mg\iii\ (right column) as a function of $\log \tau_{5000}$ in the  model atmospheres with 5780 / 3.8 / $-$2.3 (top row), 9550 / 3.95 / $-$0.5 (middle row) and 12800 / 3.8 / 0.0 (bottom row).  }
 \label{DC}
 \end{minipage}
 \end{figure*}

 \begin{deluxetable*}{lcccccccccccc}
\tablecaption{Lines of Mg\ione\ and Mg\ii\ used in abundance analysis and solar magnesium abundances, log~$\epsilon$ (cols. 10 -- 12). \label{tab1}}
\tabletypesize{\scriptsize}
\tablehead{
\colhead{$\lambda$} & \colhead{Transition} & \colhead{\Eexc} & \colhead{log~$gf$} & \colhead{Ref.} & \colhead{log $\Gamma_4$/$N_e$}  & \colhead{log $\Gamma_4$/$N_p$}  &
\colhead{Ref.} & \colhead{log $\Gamma_6$ /$N_H$} & \colhead{Ref.} & \colhead{LTE} & \multicolumn2c{NLTE}  \\
\colhead{\AA\,} & \colhead{} & \colhead{eV} & \colhead{}& \colhead{}&\multicolumn2c{rad s$^{-1}$cm$^3$ } & \colhead{} & \colhead{rad s$^{-1}$cm$^3$} & \colhead{} & \colhead{} & \colhead{M15} & \colhead{IPM}
}
\colnumbers
\startdata
 Mg\ione\ &                                                    &      &           &   &           &           &       &          &        &  &  & \\
 3829.36 & 3p $^{3}$P$_{0}^{\circ}$ -- 3d $^{3}$D$_{1}$        &  2.71&  $-$0.214 & 1 &  $-$4.67  &  $-$5.06  & 5     &  $-$7.29 &~7      &  &  &  \\
 3832.29 & 3p $^{3}$P$_{1}^{\circ}$ -- 3d $^{3}$D$_{1}$        &  2.71&  $-$0.339 & 1 &  $-$4.67  &  $-$5.06  & 5     &  $-$7.29 &~7      &  &  &  \\
 3832.30 & 3p $^{3}$P$_{1}^{\circ}$ -- 3d $^{3}$D$_{2}$        &  2.71&    ~0.138 & 1 &  $-$4.67  &  $-$5.06  & 5     &  $-$7.29 &~7      &  &  &  \\
 3838.29 & 3p $^{3}$P$_{2}^{\circ}$ -- 3d $^{3}$D$_{1}$        &  2.71&  $-$1.515 & 1 &  $-$4.67  &  $-$5.06  & 5     &  $-$7.29 &~7      &  &  &  \\
 3838.29 & 3p $^{3}$P$_{2}^{\circ}$ -- 3d $^{3}$D$_{3}$        &  2.71&    ~0.409 & 1 &  $-$4.67  &  $-$5.06  & 5     &  $-$7.29 &~7      &  &  &  \\
 3838.29 & 3p $^{3}$P$_{2}^{\circ}$ -- 3d $^{3}$D$_{2}$        &  2.71&  $-$0.339 & 1 &  $-$4.67  &  $-$5.06  & 5     &  $-$7.29 &~7      &  &  &  \\
 4167.27 & 3p $^{1}$P$_{1}^{\circ}$ -- 7d $^{1}$D$_{2}$        &  4.35&  $-$0.746 & 1 &  $-$3.51  &  $-$5.00  & 5     &  $-$6.89 &~8      & 7.47 & 7.48  & 7.50 \\    
 4571.09 & 3s$^{2}$ $^{1}$S$_{0}$   -- 3p $^{3}$P$_{1}^{\circ}$&  0.00&  $-$5.397 & 1 &  $-$6.15  &           & 8     &  $-$7.77 &~9      & 7.45 & 7.47  & 7.49 \\    
 4702.99 & 3p $^{1}$P$_{1}^{\circ}$ -- 5d $^{1}$D$_{2}$        &  4.35&  $-$0.456 & 1 &  $-$4.17  &  $-$4.75  & 5     &  $-$6.69 &~7      & 7.35 & 7.35  & 7.39 \\    
 4730.03 & 3p $^{1}$P$_{1}^{\circ}$ -- 6s $^{1}$S$_{0}$        &  4.35&  $-$2.379 & 1 &  $-$4.23  &  $-$4.91  & 5     &  $-$7.12 &~8      & 7.77 & 7.77  & 7.72  \\   
 5167.32 & 3p $^{3}$P$_{0}^{\circ}$ -- 4s $^{3}$S$_{1}$        &  2.71&  $-$0.865 & 1 &  $-$5.43  &  $-$6.06  & 5     &  $-$7.27 &~7      & 7.51 & 7.57  & 7.54  \\   
 5172.68 & 3p $^{3}$P$_{1}^{\circ}$ -- 4s $^{3}$S$_{1}$        &  2.71&  $-$0.387 & 1 &  $-$5.43  &  $-$6.06  & 5     &  $-$7.27 &~7      & 7.48 & 7.53  & 7.51  \\   
 5183.60 & 3p $^{3}$P$_{2}^{\circ}$ -- 4s $^{3}$S$_{1}$        &  2.72&  $-$0.166 & 1 &  $-$5.43  &  $-$6.06  & 5     &  $-$7.27 &~7      & 7.49 & 7.52  & 7.53  \\   
 5528.41 & 3p $^{1}$P$_{1}^{\circ}$ -- 4d $^{1}$D$_{2}$        &  4.35&  $-$0.513 & 1 &  $-$4.63  &  $-$5.19  & 5     &  $-$6.98 &~7      & 7.44 & 7.45  & 7.45  \\   
 5711.08 & 3p $^{1}$P$_{1}^{\circ}$ -- 5s $^{1}$S$_{0}$        &  4.35&  $-$1.742 & 1 &  $-$4.71  &  $-$5.38  & 5     &  $-$7.29 &~8      & 7.58 & 7.59  & 7.64  \\   
 8806.76 & 3p $^{1}$P$_{1}^{\circ}$ -- 3d $^{1}$D$_{2}$        &  4.35&  $-$0.128 & 1 &  $-$5.46  &  $-$5.90  & 5     &  $-$7.41 &~7      & 7.53 & 7.53  & 7.58  \\   
 Mean    &                                                     &      &               &   &           &           &       &          &        &  7.52 & 7.54  &  7.54  \\   
$\sigma$ &                                                     &      &               &   &           &           &       &          &        &  0.11 & 0.11  &  0.09  \\    
% 8923.57 & 4s $^{1}$S$_{0}$ -- 5p $^{3}$P$_{1}^{\circ}$        &  5.39&  $-$1.679 & 1 &  $-$4.23  &  $-$4.23  & 3     &  $-$7.23 & 3      &  &  &  \\
 73736.40& 5g $^{1,3}$G -- 6h $^{1,3}$H$^{\circ}$              &  7.09&   ~1.780  & 4 &  $-$3.06  &  $-$3.08  & 5     &  $-$6.56 &~9      &$e$ &  7.92 &  7.66 \\
122240.00& 6g $^{1,3}$G -- 7h $^{1,3}$H$^{\circ}$              &  7.27&   ~1.670  & 4 &  $-$2.39  &  $-$2.41  & 5     &  $-$6.54 &~9      &$e$ &  7.74 &  7.63 \\
123217.00& 6h $^{1,3}$H$^{\circ}$ -- 7i $^{1,3}$I              &  7.27&   ~1.947  & 4 &  $-$2.55  &  $-$2.55  & 5     &  $-$6.58 &~9      &$e$ &  7.74 &  7.63 \\
   Mg\ii\ &                                                    &      &           &   &           &           &       &          &        &  &  &  \\                                         
 3848.21  & 3d $^{2}$D$_{5/2}$ -- 5p $^{2}$P$_{3/2}^{\circ}$   &  8.86&$-$1.580   & 2 & $-$4.80   & $-$5.48   & 6     &  $-$7.42 &~8      &  &  &  \\
 3848.34  & 3d $^{2}$D$_{3/2}$ -- 5p $^{2}$P$_{3/2}^{\circ}$   &  8.86&$-$2.534   & 2 & $-$4.80   & $-$5.48   & 6     &  $-$7.42 &~8      &  &  &  \\
 3850.38  & 3d $^{2}$D$_{3/2}$ -- 5p $^{2}$P$_{1/2}^{\circ}$   &  8.86&$-$1.842   & 2 & $-$4.80   & $-$5.48   & 6     &  $-$7.42 &~8      &  &  &  \\
 4384.64  & 4p $^{2}$P$_{1/2}^{\circ}$ -- 5d $^{2}$D$_{3/2}$   &  9.99& $-$0.790  & 2 & $-$4.20   & $-$4.86   & 6     &  $-$7.34 &~8      &  &  &  \\
 4390.51  & 4p $^{2}$P$_{3/2}^{\circ}$ -- 5d $^{2}$D$_{3/2}$   &  9.99& $-$1.478  & 2 & $-$4.20   & $-$4.86   & 6     &  $-$7.34 &~8      &  &  &  \\
 4390.57  & 4p $^{2}$P$_{3/2}^{\circ}$ -- 5d $^{2}$D$_{5/2}$   &  9.99& $-$0.530  & 2 & $-$4.20   & $-$4.86   & 6     &  $-$7.34 &~8      &  &  &  \\
 4427.99  & 4p $^{2}$P$_{1/2}^{\circ}$ -- 6s $^{2}$S$_{1/2}$   &  9.99& $-$1.208  & 2 & $-$4.46   & $-$5.38   & 6     &  $-$7.32 &~8      &  &  &  \\
 4433.98  & 4p $^{2}$P$_{3/2}^{\circ}$ -- 6s $^{2}$S$_{1/2}$   &  9.99& $-$0.910  & 2 & $-$4.46   & $-$5.38   & 6     &  $-$7.32 &~8      &  &  &  \\
 4481.13  & 3d $^{2}$D$_{5/2}$ -- 4f $^{2}$F$_{7/2}^{\circ}$   &  8.86& ~0.749    & 2 & $-$4.70   & $-$5.58   & 6     &  $-$7.56 &~8      &  &  &  \\
 4481.15  & 3d $^{2}$D$_{5/2}$ -- 4f $^{2}$F$_{5/2}^{\circ}$   &  8.86& $-$0.560  & 2 & $-$4.70   & $-$5.58   & 6     &  $-$7.56 &~8      &  &  &  \\
 4481.33  & 3d $^{2}$D$_{3/2}$ -- 4f $^{2}$F$_{5/2}^{\circ}$   &  8.86& ~0.590    & 2 & $-$4.70   & $-$5.58   & 6     &  $-$7.56 &~8      &  &  &  \\
 4739.59  & 4d $^{2}$D$_{5/2}$ -- 8f $^{2}$F$_{7/2}^{\circ}$   & 11.57& $-$0.660  & 2 & $-$2.53   &           & 8     &  $-$7.00 &~8      &  &  &  \\
 4739.71  & 4d $^{2}$D$_{3/2}$ -- 8f $^{2}$F$_{5/2}^{\circ}$   & 11.57& $-$0.816  & 2 & $-$2.53   &           & 8     &  $-$7.00 &~8      &  &  &  \\
 6545.94  & 4f $^{2}$F$_{5/2}^{\circ}$ -- 6g $^{2}$G$_{7/2}$   & 11.63&  ~0.040   & 3 & $-$2.98   &           & 8     &  $-$7.25 &~8      &  &  &  \\
 6545.99  & 4f $^{2}$F$_{7/2}^{\circ}$ -- 6g $^{2}$G$_{7/2}$   & 11.63&  ~0.150   & 3 & $-$2.98   &           & 8     &  $-$7.25 &~8      &  &  &  \\
% 6545.99  & 4f $^{2}$F$_{7/2}^{\circ}$ -- 6g $^{2}$G$_{9/2}$   & 11.63& $-$1.390  & 4 &           &           &       &          &        &  &  &  \\ 
 7877.05  & 4p $^{2}$P$_{1/2}^{\circ}$ -- 4d $^{2}$D$_{3/2}$   &  9.99& ~0.391    & 2 & $-$4.61   & $-$4.61   & 6     &  $-$7.52 &~8      & 7.63 & 7.59 & 7.59 \\
 7896.04  & 4p $^{2}$P$_{3/2}^{\circ}$ -- 4d $^{2}$D$_{3/2}$   &  9.99& $-$0.308  & 2 & $-$4.61   & $-$4.61   & 6     &  $-$7.52 &~8      &      &      &      \\
 7896.37  & 4p $^{2}$P$_{3/2}^{\circ}$ -- 4d $^{2}$D$_{5/2}$   &  9.99& ~0.643    & 2 & $-$4.61   & $-$4.61   & 6     &  $-$7.52 &~8      & 7.66 & 7.59 & 7.59 \\
 9218.25  & 4s $^{2}$S$_{1/2}$ -- 4p $^{2}$P$_{3/2}^{\circ}$   &  8.65& ~0.268    & 2 & $-$5.06   & $-$5.06   & 6     &  $-$7.61 &~8      & 7.90 & 7.68 & 7.68 \\
 9244.26  & 4s $^{2}$S$_{1/2}$ -- 4p $^{2}$P$_{1/2}^{\circ}$   &  8.65& $-$0.034  & 2 & $-$5.06   & $-$5.06   & 6     &  $-$7.61 &~8      &      &      &      \\
 10914.24 & 3d $^{2}$D$_{5/2}$ -- 4p $^{2}$P$_{3/2}^{\circ}$   &  8.86& ~0.038    & 2 & $-$5.11   & $-$5.11   & 6     &  $-$7.61 &~8      & 7.67 & 7.52 & 7.52 \\
 10951.78 & 3d $^{2}$D$_{3/2}$ -- 4p $^{2}$P$_{1/2}^{\circ}$   &  8.86& $-$0.219  & 2 & $-$5.11   & $-$5.11   & 6     &  $-$7.61 &~8      & 7.68 & 7.56 & 7.56 \\
 Mean     &                                                    &      &           &   &           &           &       &          &        & 7.72 & 7.59 & 7.59 \\
$\sigma$  &                                                    &      &           &   &           &           &       &          &        & 0.11 & 0.05 & 0.05 \\ \hline
\multicolumn2l{Mean of Mg\ione\ and Mg\ii\ }                   &      &           &   &           &           &       &          &        & 7.60 & 7.56  & 7.56 \\
$\sigma$     &                                                 &      &           &   &           &           &       &          &        &  0.14 & 0.10  & 0.09  \\ \hline
\enddata
\tablecomments{ {\bf References:}  \bf{(1) \citet{2017AA...598A.102P} };
   (2) \citet{NIST_ASD};  (3) \citet{KP}; (4) \citet{1957ApJS....3...37G};(5) \citet{1996AAS..117..127D}; 
	(6) \citet{1995BABel.151..101D}; (7) \citet{BPM};  (8) Kurucz\footnote{\url{http://kurucz.harvard.edu/atoms.html}}; 
(9) \citet{2015AA...579A..53O}. 
%(12) \citet{1955U}.  
$\Gamma_4$/$N_{e,p}$ and $\Gamma_6$/$N_H$ are given for T=10000~K.
Emission lines are marked by symbol $e$. $\sigma$ means standard deviation.}
\end{deluxetable*}

%\subsection{The NLTE effects}
 
 { Figure}\,\ref{balance} displays fractions of Mg\ione, Mg\ii, and Mg\iii\ in the model atmospheres of different effective temperatures
 from the LTE and NLTE calculations. In the atmospheres with \Teff\ $\lesssim$ 12\,000 K, Mg\ii\ is the dominant ionization stage in the line-formation region, with small admixtures of Mg\ione\ (typically a few parts in a thousand) and Mg\iii, while  Mg\ii\ becomes the minority species in the hotter atmospheres. 
  NLTE leads to 
depleted population of Mg\ione\ in the cool atmosphere (\Teff\/ log~$g$ / [Fe/H] = 5780 / 3.7 / $-$2.3), while
only small deviations from LTE for Mg\ione\ are obtained in the 9550 / 3.95 / $-$0.5 model. In the 17500 / 3.8 / 0 model, NLTE leads to enhanced number density of Mg\ii.

 Figure\,\ref{DC} shows the departure coefficients for the selected levels 
 in the models 5780 / 3.8 / $-$2.3, 9550 / 3.95 / $-$0.5, and 12800 / 3.8 / 0.0, which represent the atmospheres of HD~140283, Vega and $\pi$~Cet, respectively.  As expected, the departure coefficients are equal to unity deep in the atmosphere, where the gas density is large and collisional processes dominate, enforcing the LTE. 
 
 In the 5780 / 3.8 / $-$2.3 and 9550 / 3.95 / $-$0.5 models, the Mg\ione\ 3p $^{3}$P$^{\circ}$ level ($\lambda_{thr}$ = 2510\,\AA) and the ground state ($\lambda_{th}$ = 1622\,\AA) are subject to the UV overionization, and the population depletion is extended to the remaining Mg\ione\ levels due to their coupling to the low-excitation levels. Since singly-ionized magnesium dominates the element abundance over all atmospheric depths, the Mg\ii\ ground-state keeps its LTE populations.

The Mg\ii\ excited levels are overpopulated relative to their LTE populations in the 5780 / 3.8 / $-$2.3 model, outward log~$\tau = 0$, due to radiative pumping the ultraviolet transitions from the ground state, 
but the levels 4p$^2$P$^\circ$, 4d$^2$D, and 4f$^2$F$^\circ$ have depleted populations in the hotter 9550 / 3.95 / 0.5 model, outward log~$\tau = -1$, due to photon losses in the near-IR and visible lines of Mg II arising in the transitions 4s$^2$S-4p$^2$P$^\circ$ (9218~\AA), 4p$^2$P$^\circ$-4d$^2$D (7877~\AA), and 3d$^2$D-4f$^2$2F$^\circ$ (4481~\AA).

 In the 12800 / 3.8 / 0.0 model, outwards log~$\tau =-1.0$, NLTE leads to depleted populations of the Mg\ii\ levels because of the UV overionization.
The formation depths of the Mg\ii\ lines shift to the outer layers, where the deviations from LTE are strong. 
 
 The NLTE effects for a given spectral line can be understood from analysis of the departure coefficients of the lower (b$_l$) and upper (b$_u$) levels at the line-formation depths. 
 A NLTE strengthening of lines occurs, if b$_l>$ 1 and (or) the line source function is smaller than the Planck function, that is, b$_l>$ b$_u$. The Mg\ione\ and Mg\ii\ lines used in this study are listed in Table~\ref{tab1}. 
 First, we consider lines of Mg\ione\ at 8806 and 4571~\AA\ in the 5780 / 3.8 / $-$2.3 model. 
The Mg\ione\ 8806~\AA\ line is strong, and its core forms around $\log \tau_{5000} = -0.7$, where the departure coefficient of the upper level 3d $^{1}$D$_{2}$ drops rapidly (Figure~\ref{DC}) due to photon escape from the line itself, resulting in dropping the line source function below the Planck function and enhanced absorption in the line core. In contrast, in the line wings, absorption is weaker compared with the LTE case due to overall overionization in deep atmospheric layers. 
As a result, the NLTE effect on the total energy absorbed in this line tends to be minor. 
%the abundance correction, $\Delta_{\rm NLTE}$ = log$\epsilon_{\rm NLTE}$ - log$\epsilon_{\rm LTE}$, for this line tends to be zero.
  The Mg\ione\ 4571~\AA\ line is weak and its core forms at atmospheric depths ($\log \tau_{5000} \simeq -0.05$) dominated by overionization of Mg\ione\ resulting in the weakened line
  and positive abundance correction of $\Delta_{\rm NLTE}$ = log$\epsilon_{\rm NLTE}$ - log$\epsilon_{\rm LTE}$ = +0.17~dex.
 
  In the 12800 / 3.8 / 0.0 model, we consider Mg\ii\ 4481, 7877, and 4390~\AA. 
The  Mg\ii\ 4390~\AA\ line is weak and forms in the layers around $\log \tau_{5000} = -0.7$, where b$_l<$ 1 and b$_l<$ b$_u$ resulting in weakening the line and positive NLTE abundance correction of 0.04~dex.
  In contrast, the Mg\ii\ 4481 and 7877~\AA\ lines are strong and their cores form around $\log \tau_{5000} = -2.5$ and $\log \tau_{5000} = -1.7$, where b$_l>$ b$_u$ and the line source function drops below the Planck function resulting in strengthened lines and negative NLTE abundance corrections of $\Delta_{\rm NLTE}$ = $-$0.22~dex and $-$0.21~dex, respectively.

\subsection{Mg\ione\ IR emission lines}
 
\begin{figure*}
   \begin{minipage}{190mm}
 \begin{center}
 \parbox{0.38\linewidth}{\includegraphics[scale=0.25]{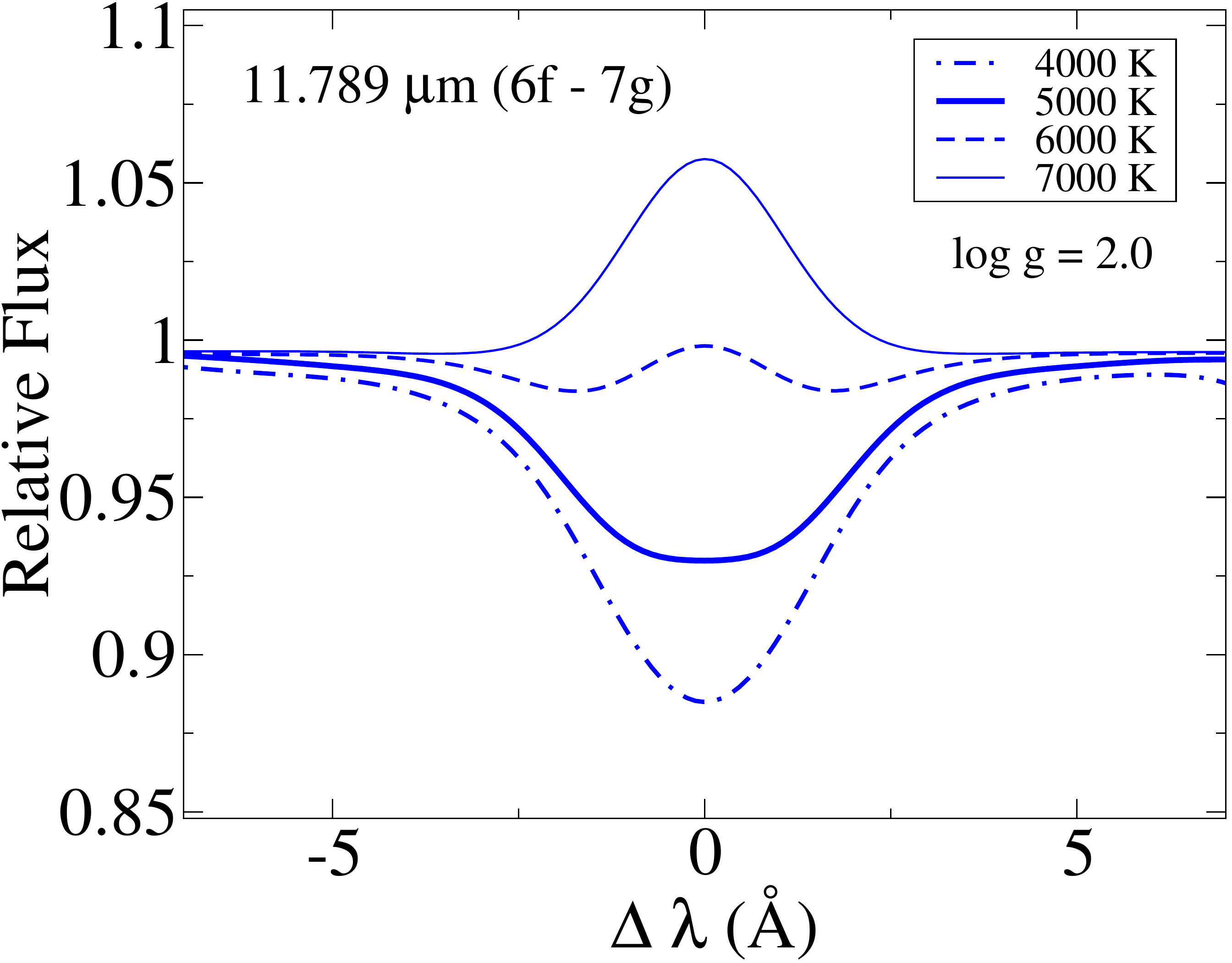}\\
 \centering}
 \parbox{0.38\linewidth}{\includegraphics[scale=0.25]{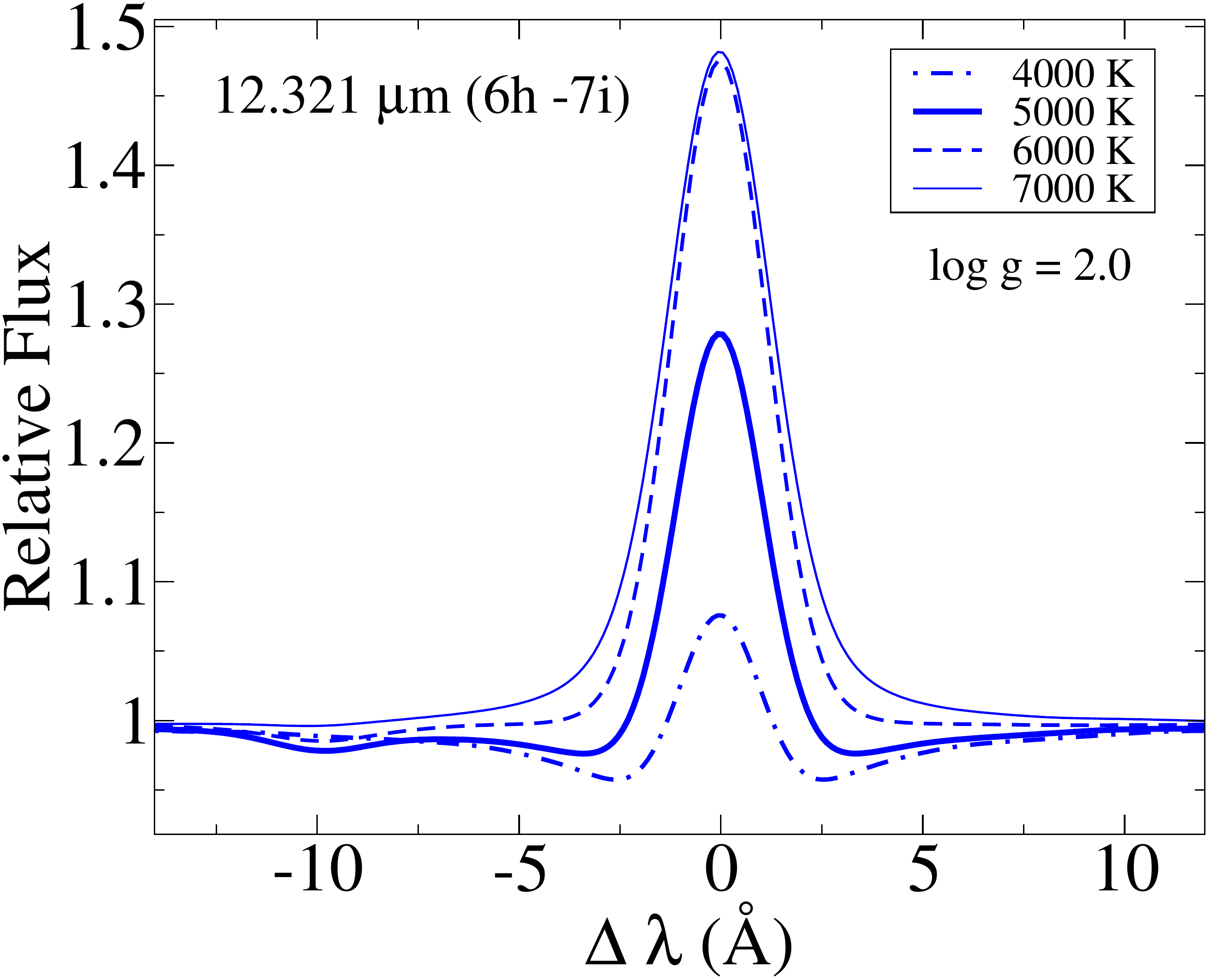}\\
 \centering}
 \hspace{1\linewidth}
 \hfill
 \\[0ex]
 \parbox{0.38\linewidth}{\includegraphics[scale=0.25]{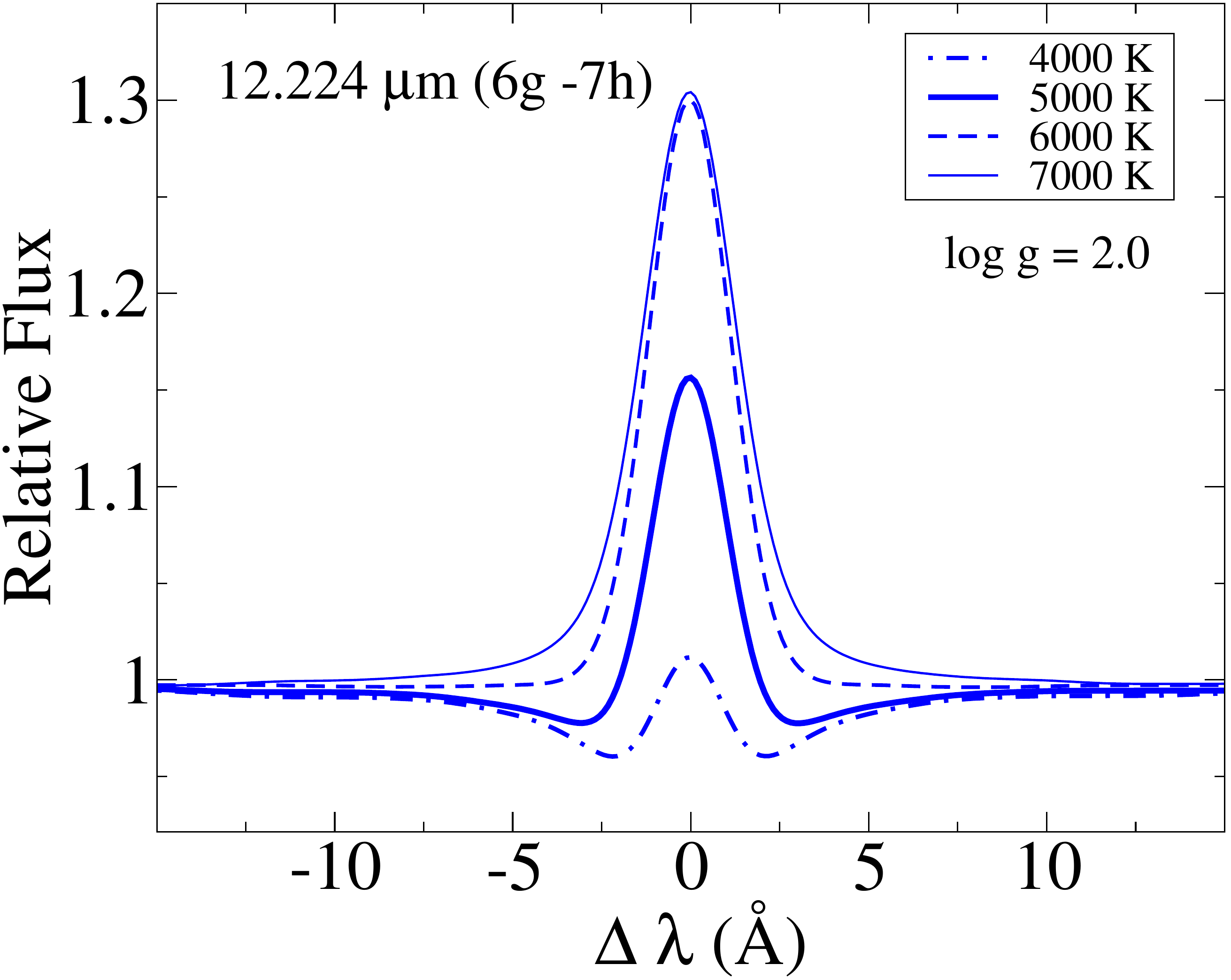}\\
 \centering}
 \parbox{0.38\linewidth}{\includegraphics[scale=0.25]{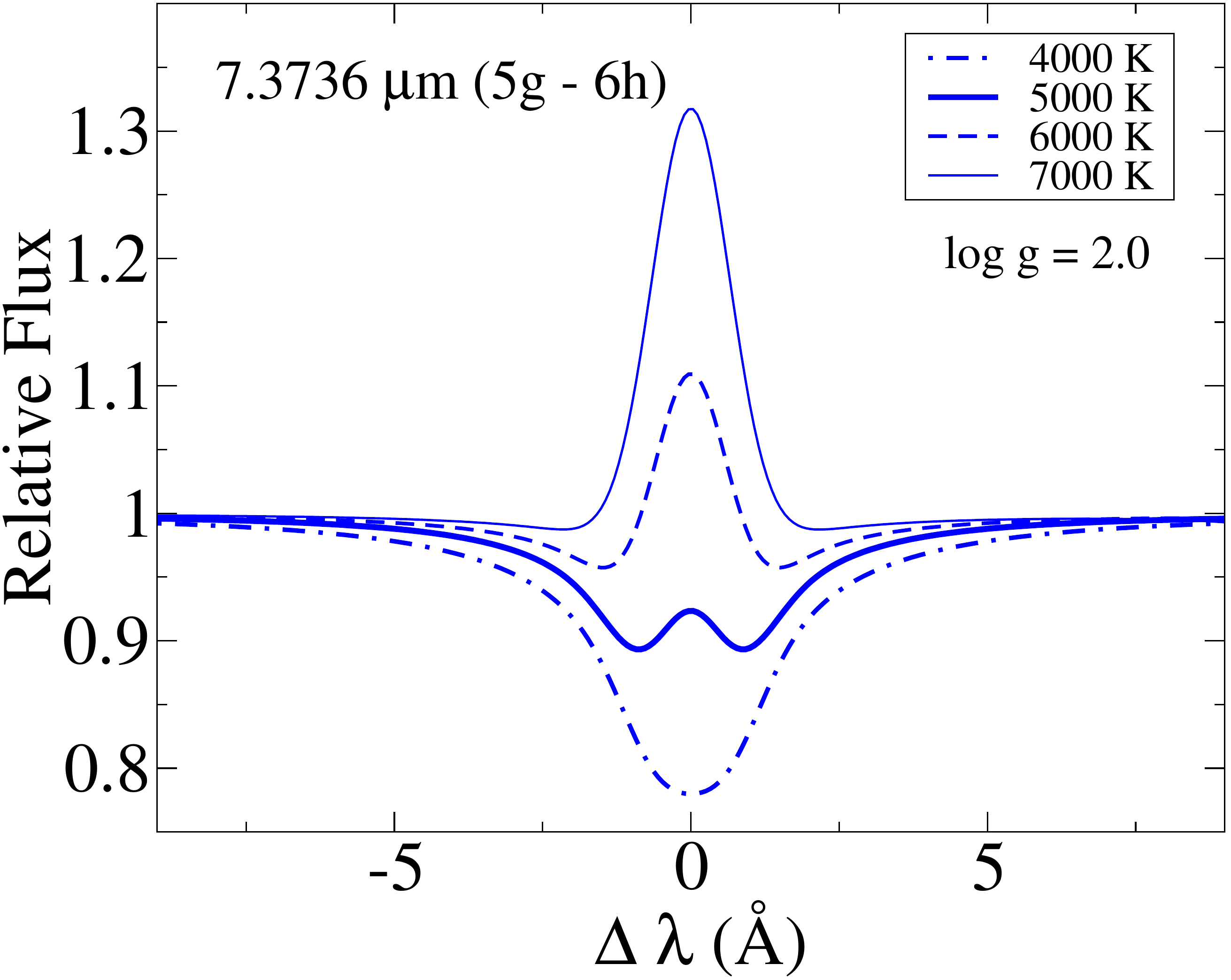}\\
 \centering}
 \hspace{1\linewidth}
 \hfill
 \\[0ex]
 \caption{Evolution of the Mg\ione\ 11.789 $\mu$m, 12.321 $\mu$m, 12.224 $\mu$m, and 7.736 $\mu$m line profiles with effective temperature.
 Everywhere, [Fe/H] = 0, [Mg/Fe] = 0, and $\xi_t$ = 2~\kms. The theoretical spectra are convolved with an instrumental profile of R~=~65\,000.} 
 \label{param34}
 \end{center}
 \end{minipage}
 \end{figure*}

 Our NLTE calculations in the grid of model atmospheres with 4000~K $\le$ \Teff\ $\le$ 7000~K,  4.0 $\le$ log$g$ $\le$ 2.0, and 0 $\le$ [Fe/H] $\le -$3 predict several Mg\ione\ lines, which arise between the high-excitation levels, to appear in emission. This concerns with  
 the Mg\ione\ infrared lines at 7.736~$\mu$m (5g-6h), 11.789~$\mu$m (6f-7g), 12.224~$\mu$m (6g-7h), and 12.321~$\mu$m (6h-7i). 
 Figure~\ref{param34} shows how a variation in \Teff\ affects theoretical profiles of the Mg\ione\ infrared lines and at which atmospheric parameters the line absorption changes by the emission.
 For different lines, an emission feature appears at different temperatures: \Teff\ $\ge$~6000~K for 11.789~$\mu$m, \Teff\ $\ge$~5000~K for 7.736~$\mu$m. The 12.321 and 12.224~$\mu$m lines are quite strong and they reveal emission profiles over the 4000 -- 7000~K temperature range.
 
 \citet{1992A&A...253..567C} were the first who computed the 12~$\mu$m emission peaks employing standard plane-parallel NLTE modelling for Mg\ione\ with a theoretical model atmosphere without chromosphere. They showed that competitive processes of the overionization of the low-excitation levels of Mg\ione\ and close collisional coupling of the Rydberg levels to the Mg\ii\ ground state, which contains most of the element, result in depleted level populations of Mg\ione\ compared with the LTE case and a systematic departure divergence pattern with the higher levels closer to LTE ($b_{\rm u} > b_{\rm l}$ for any pair l $<$ u). In the infrared, even tiny population divergences have large effects on the line source function due to the increasing importance of stimulated emission. The 12~$\mu$m line source function increases outwards in the atmosphere, resulting in an emission profile. 

The NLTE calculations of \citet{1996ASPC..109..723U}, \citet{2004ApJ...611L..41R}, \citet{2008A&A...486..985S} and this study show that similar physical processes produce emission not only in 12~$\mu$m, but also in the other Mg\ione\ IR lines and not only in the solar atmosphere, but in extended range of atmospheric parameters. 
  
 It is worth noting that, in the hotter atmospheres, the emission phenomena are observed in the near IR lines of C\ione, at 8335, 9405, 9061-9111, and 9603-9658~\AA, and Ca\ii, at 8912-27 and 9890~\AA, and they were well reproduced using standard plane-parallel NLTE modelling of C\ione -C\ii\ \citep{2016MNRAS.462.1123A} and Ca\ione -Ca\ii\ \citep{2018MNRAS.477.3343S}.
  
%  The mechanism of the solar Mg\ione\ 12.224~$\mu$m and 12.321~$\mu$m emission lines was explained fo the first time by \citet{1992A&A...253..567C}.   The NLTE calculations of \citet{1996ASPC..109..723U}, \citet{2004ApJ...611L..41R}, \citet{2008A&A...486..985S} and our calculations show that very similar physical processes produce emission in   these and also the other Mg\ione\ IR lines in extended atmospheric parameter range. We describe the  mechanism briefly.
  
%  The high-excitation (\Eexc\ $> $ 5.5~eV) levels of Mg\ione\ have strong collisional coupling to the Mg\ii\ ground state.  Mg\ii\ is the dominant ionization stage in the investigated stellar parameter range.   Therefore, they are depopulated to a lesser extent than the lower levels, which are subject to the UV overionization.   The overionization-recombination mechanism resulting in $b_{\rm u} > b_{\rm l}$ for each transition favors an emission phenomenon in radiative cascade transitions from the high-excitation levels. 

\section{Solar lines of Mg\ione\ and Mg\ii} \label{subsec:sun}

  As a first application of the developed model atom, we analyze
  lines of Mg\ione\ and Mg\ii\ in the Sun. 
  We used the MARCS model atmosphere 5777 / 4.44 / 0 and a depth-independent microturbulence of 0.9~\kms. The element abundance was determined from line-profile fitting.
  
\begin{figure*}
  \begin{minipage}{170mm}
  %\begin{center}
  \parbox{0.3\linewidth}{\includegraphics[scale=0.19]{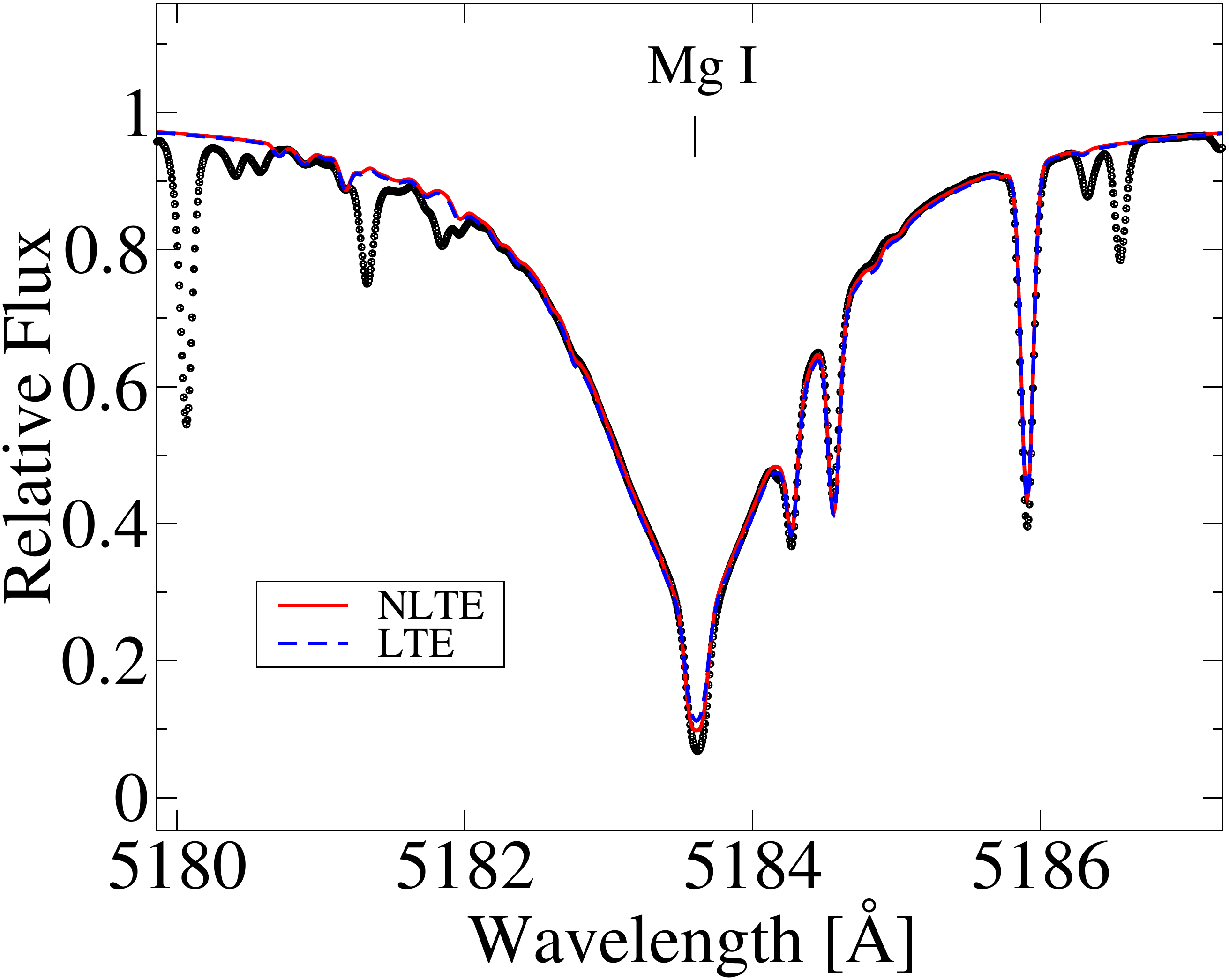}\\
  \centering}
  %\hspace{0.33\linewidth}
  \parbox{0.3\linewidth}{\includegraphics[scale=0.19]{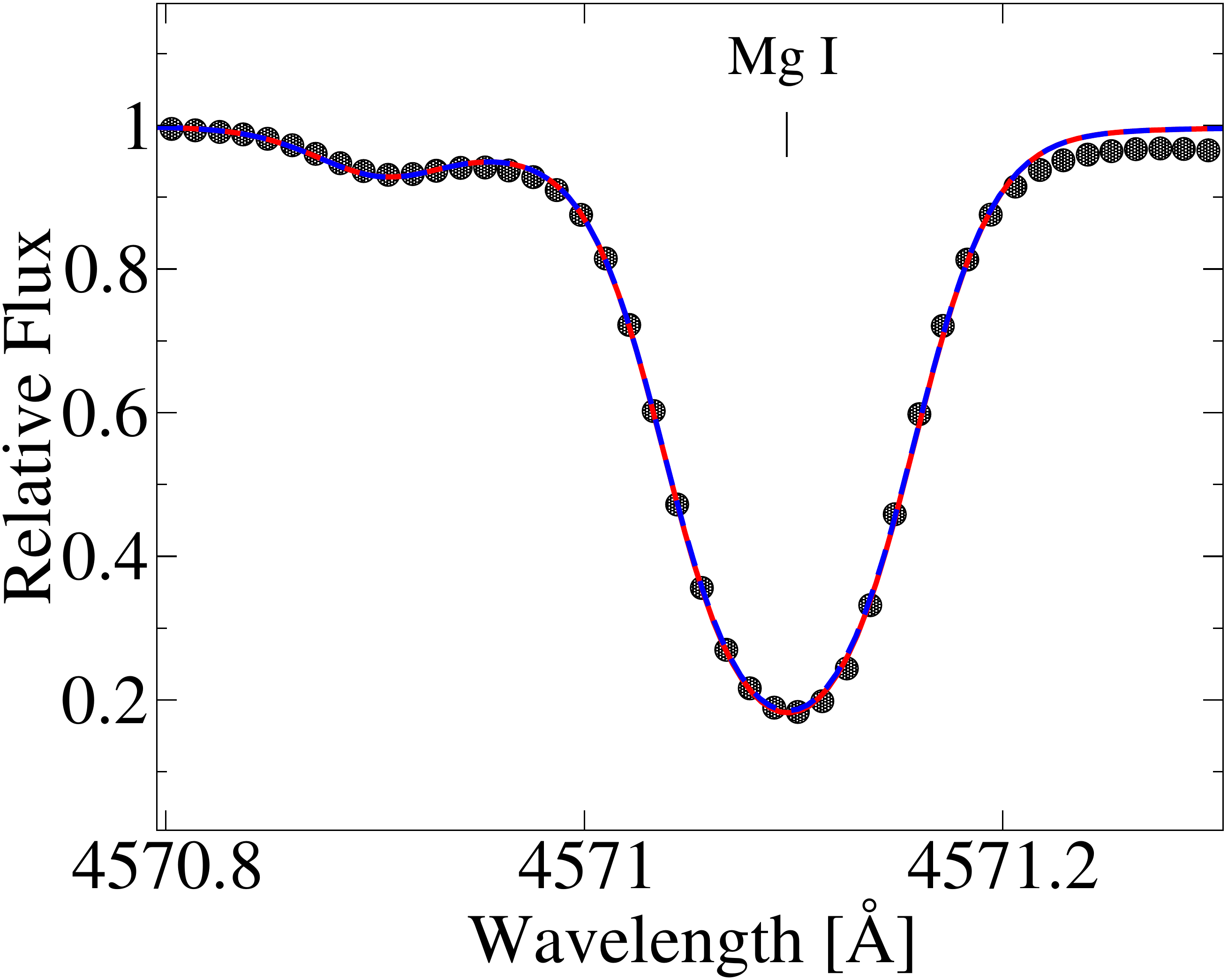}\\
  \centering}
  \parbox{0.3\linewidth}{\includegraphics[scale=0.19]{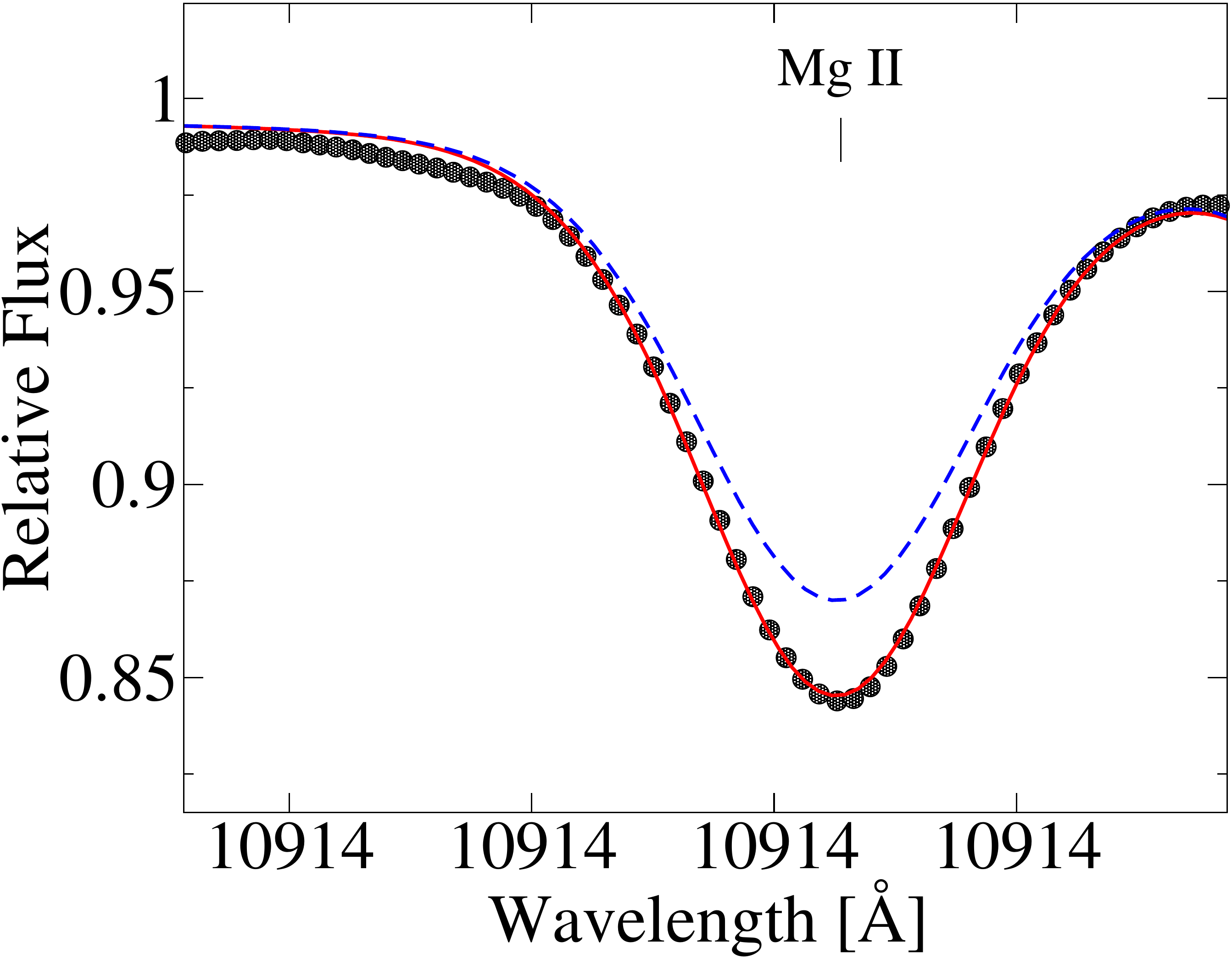}\\
  \centering}
  \hspace{1\linewidth}
  \hfill
  \\[0ex]
  %\hspace{0.33\linewidth}
  \parbox{0.3\linewidth}{\includegraphics[scale=0.19]{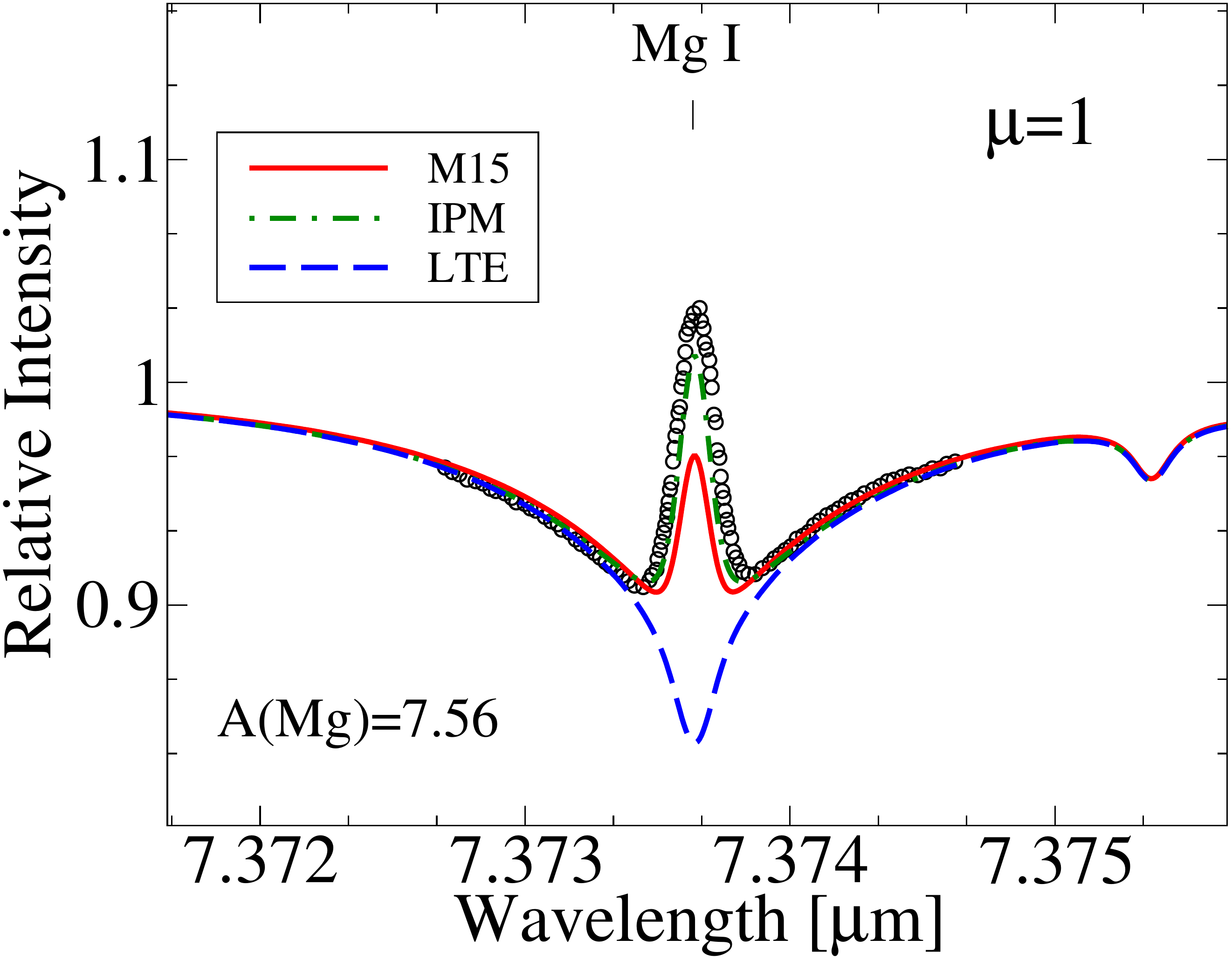}\\
  \centering}
  \parbox{0.3\linewidth}{\includegraphics[scale=0.19]{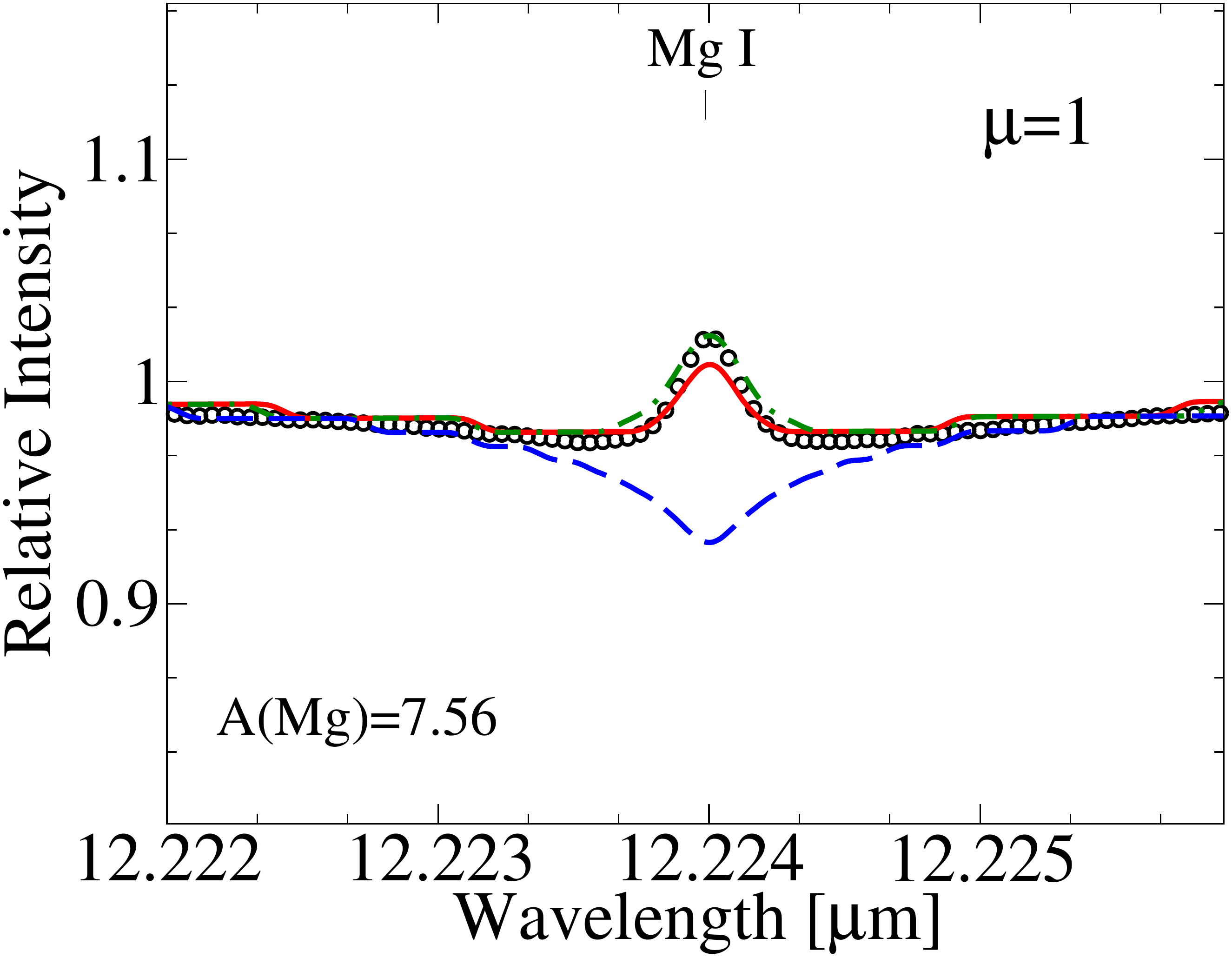}\\
  \centering}
  \parbox{0.3\linewidth}{\includegraphics[scale=0.19]{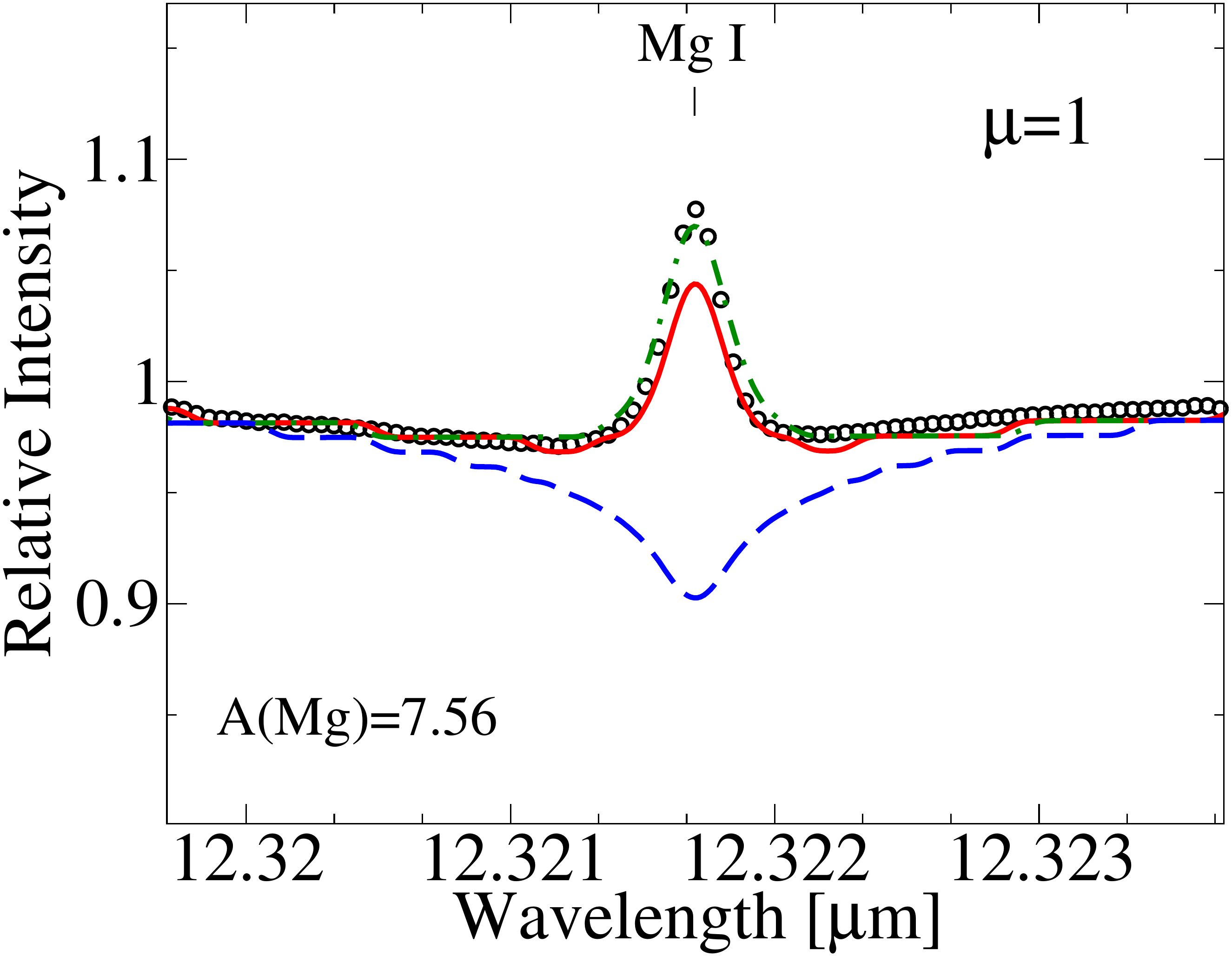}\\
  \centering}
  \hspace{1\linewidth}
  \hfill
  \\[0ex]
  \caption{Top row: Best NLTE fits (continuous curve) to the solar flux line profiles (black circles) of Mg\ione\ 5183, 4571~\AA\ and Mg\ii\ 10914~\AA. For each line, the LTE profile (dashed curve) was
computed with the magnesium abundance obtained from the NLTE analysis. Bottom row: solar center disk intensity line profiles of Mg\ione\ 7.3, 12.2, and 12.3~$\mu$m (open circles) compared with the  synthetic spectra computed with log~$\epsilon_{\rm Mg}$ = 7.56 at the LTE assumption (dashed curve) and in NLTE with the M15 (continuous curve) and IPM (dash-dotted curve) model atom. 
   }
  \label{pics}
  \end{minipage}
  \end{figure*}

{\it Lines in the visible and near IR spectral range.}
  The solar flux observations were taken from the Kitt Peak Solar Atlas \citep{1984sfat.book.....K}.  
Our synthetic flux profiles were convolved with a profile that combines a rotational broadening of 1.8~\kms\ and broadening by macroturbulence with a radial-tangential profile. For different lines of Mg\ione\ and Mg\ii, 
  the most probable macroturbulence velocity, V$_{mac}$, was varied between 2 and 4~\kms.
  As a rule, the uncertainty in fitting the observed profile is less than 0.02~dex for weak lines and 0.03~dex for strong lines. 
  The best NLTE fits of the selected absorption lines of Mg\ione\ and Mg\ii\ are shown in Fig.~\ref{pics}.  
  
  Individual line abundances are presented in Table~\ref{tab1}. 
  Hereafter, the element abundances are given in the scale, where for hydrogen log~$\epsilon_{\rm H}$ = 12. 
  We obtain that the mean NLTE abundance from the Mg\ione\ lines,  log~$\epsilon_{\rm Mg}$ = 7.54$\pm$0.11, 
  agrees well with that from the Mg\ii\ lines, log~$\epsilon_{\rm Mg}$ = 7.59$\pm$0.05, while in LTE the abundance difference between Mg\ione\ and Mg\ii, $\Delta$log~$\epsilon_{\rm Mg}$(Mg\ione\ -- Mg\ii), amounts to $-$0.20~dex. 

  Our derived solar mean abundance from the Mg\ione\ and Mg\ii\ lines, log~$\epsilon_{\rm Mg}$ = 7.56$\pm$0.10, is consistent within the error bars with the meteoritic value, log~$\epsilon_{\rm Mg}$ = 7.55$\pm$0.01,  
  \citep{2009LanB...4B...44L} and also with the solar photospheric abundances from the 1D NLTE calculations of \citet{2015AA...579A..53O} [7.57$\pm$0.08] and \citet{2017ApJ...847...15B} [7.50$\pm$0.05]. 

{\it Mg\ione\ IR emission lines.}
  The solar center disk intensity observations of Mg\ione\ 7.3, 12.2, 12.3 $\mu$m were taken from \citet{1991ApJ...383..450C} and \citet{1983ApJ...269L..61B}. 
  The synthetic intensity profiles were computed using the NLTE abundance derived from the Mg\ione\ lines, log~$\epsilon_{\rm Mg}$ = 7.56, and were broadened by macroturbulence with a radial-tangential profile of V$_{mac}$ = 2~\kms. 
 
   Using our standard model atom as described in Section~\ref{Sect:atom} (hereafter, referred to as M15), we obtain emission peaks in the Mg\ione\ 7.3, 12.2, 12.3~$\mu$m lines (Fig.~\ref{pics}), however, the fits to the solar center disk intensity profiles are not perfect, in particular, for Mg\ione\ 7.3~$\mu$m. We note that, for Mg\ione\ 12.2 and 12.3~$\mu$m, their wings can be better fitted with a change in abundance of order 0.2~dex. 

 \citet{2015AA...579A..53O} showed that a magnitude of the emission and a shape of the Mg\ione\ 7.3, 12.2, 12.3~$\mu$m line profiles depend on a treatment of inelastic collisions in the NLTE calculations, although, even with their most advanced collisional recipe, they could not reproduce perfectly the observed line profiles. We performed test NLTE calculations, which are fully based on the 
impact parameter method for all the allowed transitions and using $\Omega = 1$ for all the forbidden ones. We refer to this recipe to as IPM. In this case, the emission lines of Mg\ione\ appear to be stronger and fit better to the observed line profiles compared with the case of M15 (Fig.~\ref{pics}). Our calculations support the conclusion of earlier studies that collisional data for transitions between the Rydberg levels in Mg\ione\ need to be accurately calculated.

\section{Magnesium abundances of the selected stars}\label{Sect:Stars}
 
 \begin{deluxetable*}{lccccccc}
%\tablenum{2}
\tablecaption{Atmospheric parameters of the selected stars and sources of the data. \label{tab_param}}
\tablewidth{0pt}
\tablehead{
\colhead{Star} & \colhead{Name} & \colhead{Sp. T.} & \colhead{ \Teff } & \colhead{log$g$} & \colhead{[Fe/H]}  & \colhead{ $\xi_t$ }  & \colhead{Ref.}\\
\colhead{} & \colhead{} & \colhead{} & \colhead{ K} & \colhead{CGS} & \colhead{dex} & \colhead{ \kms }    & \colhead{} 
}
\decimalcolnumbers
\startdata
 \multicolumn8c{7000 $<$ \Teff $\leq$ 17500 K }  \\\hline
  HD~160762 &  $\iota$ Her &   B3 IV SPB  &   17500  &  3.8~  &  ~0.02    &   1.0   & ~1     \\         
  HD~17081  &  $\pi$ Cet   &   B7 IV E    &   12800  &  3.8~  &  ~0.0~    &   1.0   & ~2     \\         
  HD~22136  &              &   B8 V C     &   12700  &  4.2~  &  -0.28    &   1.1   & ~3    \\          
  HD~209459 &   21 Peg     &   B9.5 V C   &   10400  &  3.5~  &  ~0.0~    &   0.5   & ~2     \\         
  HD~48915  &   Sirius     &   A1 V+DA    &   ~9850  &  4.3~  &  ~0.4~    &   1.8   & ~4     \\         
  HD~72660  &              &   A0 V C     &   ~9700  &  4.1~  &  ~0.4~    &   1.8   & ~5    \\         
  HD~172167 &   Vega       &   A0 Va C    &   ~9550  &  3.95  &  -0.5~    &   2.0   & ~6    \\         
  HD~73666  &   40 Cnc     &   A1 V C     &   ~9380  &  3.78  &  ~0.16    &   1.9   & ~7     \\        
  HD~32115  &              &   A9 V C     &   ~7250  &  4.2~  &  ~0.00    &   2.3   & ~8     \\ \hline    
  \multicolumn8c{3900 $\leq$ \Teff $<$ 7000 K }  \\\hline                          
  HD~61421  &  Procyon     &   F5 IV-V    &   ~6580  &  4.00  &  -0.03    &   2.0   & ~9    \\                          
 HD~84937   &              &  F8 V        &   ~6350  &  4.09  &  -2.08    &   1.7   & 10    \\            
 HD~140283  &              &  F9 V        &   ~5780  &  ~3.7  &  -2.38    &   1.6   & 11    \\            
 HD~103095  &              &  K1 V        &   ~5130  &  4.66  &  -1.26    &   0.9   & 11    \\            
 HD~62509   &  Pollux      &  K0 IIIb     &   ~4860  &  2.90  &  ~0.13    &   1.5   & 12    \\                        
 HD~122563  &              &  G8 III      &   ~4600  &  1.6~  &  -2.56    &   2.0   & 10    \\            
 HD~124897  &  Arcturus    &  K1.5 III    &   ~4290  &  1.6~  &  -0.52    &   1.7   & 12    \\                      
 HD~29139   &  Aldebaran   &  K5+III      &   ~3930  &  1.11  &  -0.37    &   1.7   & 12   \\                      
\enddata                                                
\tablecomments{{\bf Note.} References: 
  (1) \citet{2012AA...539A.143N}; (2) \citet{2009AA...503..945F}; (3) \citet{2013AA...551A..30B}; 
 (4) \citet{1993AA...276..142H}; (5) \citet{2016MNRAS.461.1000S}; (6) \citet{2000AA...359.1085P}; (7) \citet{2007AA...476..911F}; (8) \citet{2011MNRAS.417..495F}; (9) \citet{2013ApJ...771...40B}.
 (10) \citet{mash_fe}; (11) \citet{2015ApJ...808..148S}; (12) \citet{2015AA...582A..49H} 
 Spectral types are extracted from the SIMBAD database. }
\end{deluxetable*}

\begin{deluxetable*}{lcccccc}
%\tablenum{2}
\tablecaption{Characteristics of observed spectra. \label{tab_obs}}
\tablewidth{0pt}
\tablehead{
\colhead{Star} & \colhead{Name} & \colhead{ V$^1$ } &   \colhead{Telescope/} & \colhead{Spectral range} & \colhead{$R$} & \colhead{$S/N$}   \\
\colhead{} & \colhead{} & \colhead{mag} & \colhead{Spectrograph } & \colhead{\AA\,} & \colhead{} & \colhead{} 
}
\decimalcolnumbers
\startdata
  HD~160762  &  $\iota$ Her & 3.8    &  1 & 3690-10\,480  & 65\,000   &  600  \\
  HD~17081   &  $\pi$ Cet   & 4.2    &  1 & 3690-10\,480  & 65\,000   &  600  \\
  HD~22136   &              & 6.9    &  1 & 3690-10\,480  & 65\,000   &  500  \\
  HD~209459  &   21 Peg     & 5.8    &  1 & 3690-10\,480  & 65\,000   &  600  \\
  HD~48915   &   Sirius     & $-$1.5 &  1 & 3690-10\,480  & 65\,000   &  500  \\
  HD~72660   &              & 5.8    &  1 & 3690-10\,480  & 65\,000   &  500  \\
  HD~172167  &   Vega       & 0.0    &  1 & 3690-10\,480  & 65\,000   &  500  \\
  HD~73666   &   40 Cnc     & 6.6    &  1 & 3690-10\,480  & 65\,000   &  660   \\
  HD~32115   &              & 6.3    &  1 & 3690-10\,480  & 65\,000   &  500  \\ 
  HD~61421   &  Procyon     & 0.4    &  2 & 3300-9900     & 80\,000   &  200  \\  %\hline
             &              &        &  3 & 12.219$\mu$m-12.229$\mu$m & 86\,000 & 200     \\
             &              &        &  3 & 12.320$\mu$m-12.323$\mu$m & 86\,000 & 200   \\
  HD~84937   &              & 8.3    &  2 & 3300-9900     & 80\,000   &  200  \\
  HD~140283  &              & 7.2    &  2 & 3300-9900     & 80\,000   &  200  \\
  HD~103095  &              & 6.4    &  4 & 3700-3900     & 60\,000   &  200  \\
  HD~62509   &  Pollux      & 1.1    &  1 & 3690-10\,480  & 65\,000   &  500  \\
             &              &        &  3 & 12.320$\mu$m-12.323$\mu$m &  100\,000  &  450   \\     
  HD~122563  &              & 6.2    &  2 & 3300-9900     & 80\,000   &  200   \\ 
  HD~124897  &  Arcturus    & $-$0.05&  1 & 3690-10\,480  & 65\,000   &  600  \\ 
             &              &        &  3 & 12.219$\mu$m-12.229$\mu$m &  100\,000  &  300     \\  
  HD~29139   &  Aldebaran   & 0.9    &  1 & 3690-10\,480  & 65\,000   &  600  \\
             &              &        &  3 & 12.219$\mu$m-12.229$\mu$m &  100\,000  &  300     \\   
\enddata                                                
\tablecomments{{\bf Notes:} $^1$V is a visual magnitude from the SIMBAD data base.
Telescope/spectrograph: 1 = CFHT/ESPaDOnS; 2 = VLT2/UVES; 3 = IRTF/TEXES; 4 = Shane/HES.}
\end{deluxetable*}

\subsection{Stellar sample, observations, and stellar parameters }   
 
 Our sample includes 17 bright stars with well-known atmospheric parameters (Table~\ref{tab_param}).

  All stars can be divided into three groups. The fisrt group includes 10 hot stars of F, A, B spectral types with effective temperatures from 6580~K to 17500~K. 
 
\begin{itemize}
  \item[]$\iota$~Her (HD~160762) is the hottest star in our sample. It is a single bright star, with $V$sin$i$ $\sim$ 6~\kms. Its atmospheric parameters were
 determined spectroscopically from all available H and He lines and multiple ionisation equilibria, and they were confirmed via the spectral energy distribution and the $Hipparcos$ distance \citep{2012AA...539A.143N}.
 
\item[]The star $\pi$~Cet (HD~17081) is an SB1 star with $V$sin$i$~$\sim$~20~\kms. This star was used by \citet{1993A&A...274..335S} as a normal comparison star in the abundance study of chemically peculiar stars.
 
\item[]The stars 21~Peg (HD~209459) and HD~22136 are 'normal' single stars with $V$sin$i$ $\sim$ 4~\kms\ and 15~\kms, respectively \citep{1981PASP...93..587S, 2009AA...503..945F, 2013AA...551A..30B}.
 For 21~Peg and $\pi$~Cet, their effective temperatures and surface
 gravities were spectroscopically determined from the Balmer lines \citep{2009AA...503..945F}. The fundamental parameters of HD~22136 were derived by \citet{2013AA...551A..30B} based on the Geneva and uvby$\beta$ photometry
 
 \item[]The Sirius binary system (HD 48915 = HIP 32349) is composed of a main sequence A1V star and a hot DA white dwarf.
 It is an astrometric visual binary system at a distance of only 2.64~pc. Their physical parameters are available with the 
 highest precision and were taken from \citet{1993AA...276..142H}. Sirius~A is classified as a hot metallic-line (Am) star.
 
\item[] Similar to Sirius HD~72660 was classified as hot Am star in the catalogue of \citet{2009A&A...498..961R}, but later it was reclassified as a 
 transition object between HgMn and hot Am stars \citep{2016MNRAS.456.3318G}.
 Its parameters 9700/4.10/0.45/1.8 were derived in \citet{2016MNRAS.461.1000S} by fitting the 4400 -- 5200 \AA\ and 6400 -- 6700 \AA\ spectral regions with
 SME (Spectroscopy Made Easy) program package \citep{1996A&AS..118..595V, 2017A&A...597A..16P}.

\item[] Effective temperature and surface gravity of Vega (HD~172167) were determined by \citet{2000AA...359.1085P} from the Balmer line wings and Mg\ione\ and Mg\ii\ lines.
 Vega is a rapidly rotating star seen pole-on. Rapid rotation causes a change of the surface shape from spherical to ellipsoidal one, therefore, the temperature and surface gravity vary from the pole to equator 
 \citep{2010ApJ...712..250H}. We ignore the non-spherical effects in our study and analyse Vega's flux spectrum using the average temperature and gravity. Vega was also classified as a mild $\lambda$~Bootis-type star \citep{1990ApJ...363..234V}.

\item[]HD~73666 is a Blue Straggler and a member of the Praesepe cluster.
 It is previously considered as an Ap (Si) star, but appears to have the abundances of a normal A-type star \citep{2007AA...476..911F}.
 This star is a primary component of SB1, as is the case for many other Blue Stragglers \citep{1996ASPC...90..337L}.
 The flux coming from the secondary star is negligible, as previously checked by \citet{1998A&A...338.1073B}, so the star
 was analysed ignoring the presence of the secondary. The physical parameters were taken from \citet{2007AA...476..911F}. 
 Both Fe\ione\ excitation (for the effective temperature) and Fe\ione/Fe\ii\ ionisation (for log~$g$) equilibria were used in the parameter determination. 

 \item[]HD~32115 is another single line spectroscopic binary. From radial velocity variations \citet{2006AJ....132.1490F} 
 derived an orbital period of 8.11128 days,  and
 concluded that the companion is either a late K-type or an early
 M-type dwarf. It allows us to consider that the spectral lines of the primary are
 not affected by the companion at spectral range of our study, so one can neglect a contribution of the secondary
  to the total flux. Atmospheric parameters of HD~32115 were determined by  \citet{2011MNRAS.417..495F} based on photometry, hydrogen line 
  profiles and wings of Mg\ione\ lines.   
  
  \item[]Procyon is a known spectroscopic binary sistem, consisting of F5 IV-V star (Procyon A) and a faint white dwarf (Procyon B).
 The proximity of Procyon to the Earth makes possible to measure directly its angular diameter. We adopted \Teff\ = 6582~K and log~$g$ = 4.00 found by \citet{2013ApJ...771...40B} using
 Hipparcos parallaxes and measured bolometric fluxes.   
 
 \end{itemize}
 
 The second group includes 4 metal-poor stars in the $-$2.56 $\leq$ [Fe/H] $\leq$ $-$1.26 metallicity range. Their characterisics were taken from \citet{2015MNRAS.453.1619A}.
 
 The third group is 3 K-giants from the sample of $Gaia$ FGK benchmark stars: Pollux (HD~62509), Arcturus (HD~124897), and Aldebaran (HD~29139). 
 Their V magnitudes range from $-$0.05 to 1.1. The stars are all nearby and their parameters should be fairly accurate. 
 We adopted the \Teff\ and log~$g$ from \citet{2015AA...582A..49H}, where the determinations were obtained in a systematic way from a compilation of angular diameter 
 measurements and bolometric fluxes and from a homogeneous mass determination based on stellar evolution models. The NLTE [Fe/H] values for these stars were taken from \citet{2014A&A...564A.133J}.
 
 For 12 stars in the sample, in the visible spectral range (3690 -- 10480\,\AA) we used the normalized spectra obtained with the Echelle SpectroPolarimetric Device
 for the Observation of Stars (ESPaDOnS) attached at the 3.6 m telescope of the
 Canada-France-Hawaii (CFHT) observatory. For 11 of them, spectra were extracted from the ESPaDOnS archive\footnote[6]{http://www.cfht.hawaii.edu/Instruments/Spectroscopy/Espadons/}, 
 while spectrum of HD~72660 was kindly provided by V. Khalack. 
 Spectral observations (3300 -- 9900\,\AA) of 5 stars were carried out with the UltraViolet and Visible Echelle Spectrograph (UVES) at the 8-m Very Large Telescope (VLT) of the European Southern Observatory (ESO).
 Spectra of Procyon, HD 84937, HD 122563 and HD 140283 were taken from the ESO Ultraviolet and Visual Echelle Spectrograph Paranal Observatory Project (UVESPOP) archive \citep{2003Msngr.114...10B}.  Spectrum of HD~103095 was obtained with Hamilton Echelle
 Spectrograph mounted on the Shane 3 m telescope of the Lick
 Observatory \citep[see, for details,][]{2015ApJ...808..148S}.
 All spectra were obtained with a high spectral resolving power, R, and high S/N ratio (more than 100). Characteristics of the observed spectra for individual stars are given in Table~\ref{tab_obs}.
 The infrared spectra of Procyon, Pollux, Arcturus, and Aldebaran were kindly provided by Nils Ryde \citep{2004ApJ...617..551R, 2008A&A...486..985S}.
 The observations were made with the Texas Echelon-cross-echelle Spectrograph attached at the 3 m NASA Infrared Telescope Facility (IRTF) \citep{2002PASP..114..153L}.
 Signal-to-noise ratios in the spectra vary but are generally high, reaching S/N $\sim$ 450 per pixel for Pollux and $\approx$300 for Arcturus and Aldebaran, in regions around the 12.22 $\mu$m line.

\subsection{Analysis of magnesium lines in stars with 6580 $<$ \Teff\ $\leq$ 17500 K} 

 In this section, we derive the magnesium abundances of ten selected stars with effective temperatures from 6580~K to 17500~K using lines of Mg\ione\ and Mg\ii\ in the visible and near IR spectral range. The Mg\ii\ 9218 and 9244~\AA\ lines are sitting at the ends of consecutive echelle orders, where a continuum fitting is problematic. Moreover, these lines 
 are lying in the wings of the Paschen P9 line at 9229.7~\AA. This makes a formation of Mg\ii\ 9218 and 9244~\AA\ to be very sensitive to correct modelling of the hydrogen Paschen line formation. As was
 shown by \citet{2018MNRAS.477.3343S}, the different treatment of hydrogen opacity in the Paschen-series region implemented in the codes for spectral synthesis leads to significant difference in the Paschen line absorption, thus introducing difficulties for careful analysis of spectral lines falling close to the Paschen line cores. We analyse the Mg\ii\ 9218 and 9244~\AA\ lines (Table~\ref{tab_hot}), but do not include them in calculations of the mean abundances.  
 
For Sirius and HD~72660, we use also the UV spectra kindly provided by J. Landstreet. The detailed description of spectra and their reduction is given in \citet{2011AA...528A.132L} and in \citet{2016MNRAS.456.3318G}. 
 We examined the Mg\ii\ 1737.62 and 2798.00~\AA\ and Mg\ione\ 1827.93 and 2852.13~\AA\ lines and found them to be nearly unblended and suitable for abundance analysis. 
Atomic parameters were taken from \citet{NIST_ASD} (oscillator strengths) and \citet{1995BABel.151..101D, 1996AAS..117..127D} (Stark damping data).

\begin{figure}
 \begin{center}
 \includegraphics[scale=0.45]{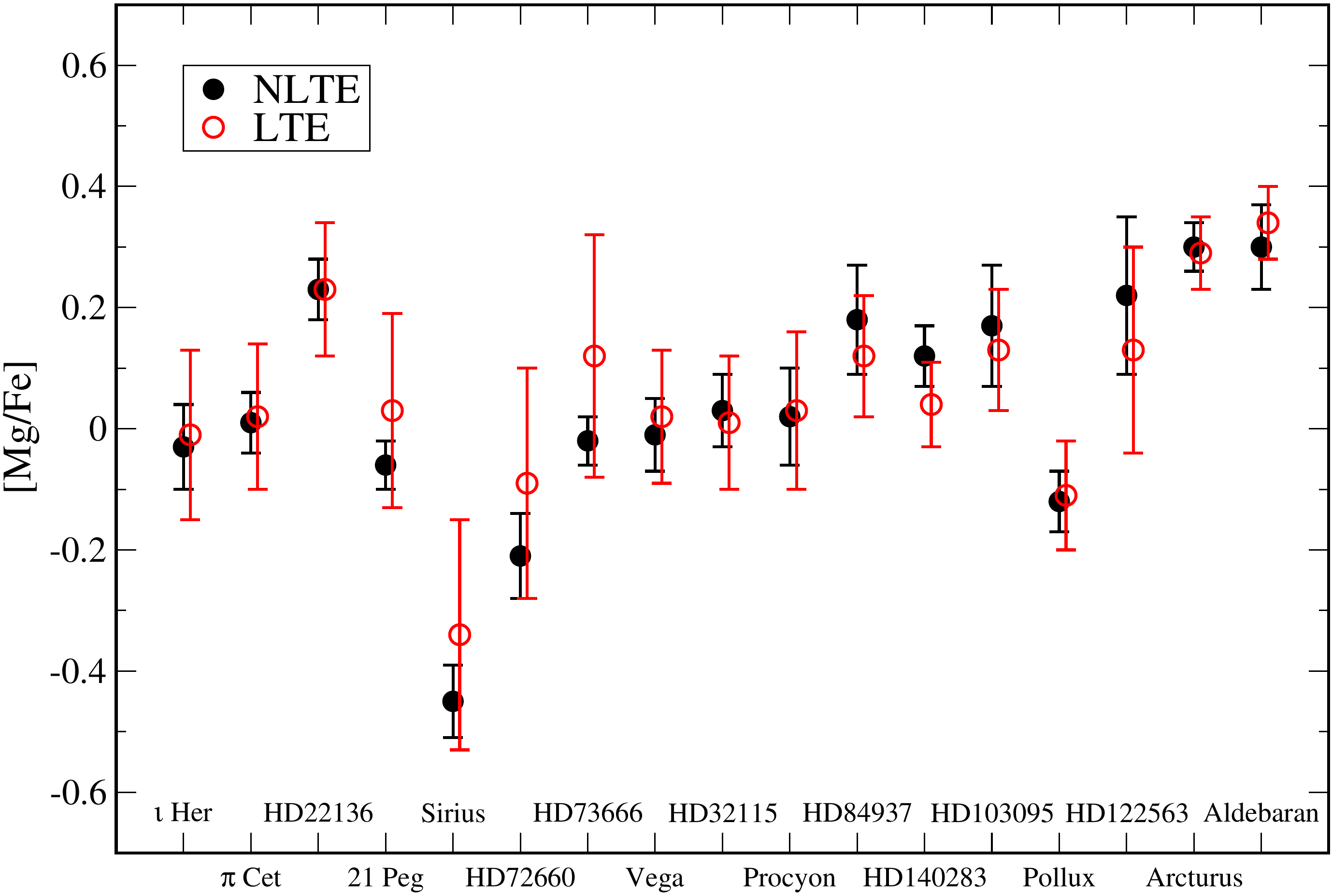}
 \caption{ [Mg/Fe] NLTE and LTE ratios for the sample stars. The average values were calculated with Mg\ione\ and Mg\ii\ lines, where available, except for HD~140283 and HD~84937, for which only lines of Mg\ione\ were employed. The error bars correspond to the dispersion in the single line measurements about the mean.}
 \label{Mg_Fe}
 \end{center}
 \end{figure}

 The abundance results are presented in Tables~\ref{tab_hot} and \ref{tabUV}. For each star, NLTE provides smaller line-to-line scatter that is well illustrated by Fig.~\ref{Mg_Fe}.
 For nine stars where lines of both ionization stages are available, the NLTE abundances derived from the Mg\ione\ and the Mg\ii\ lines agree within the error bars, while, in LTE, the abundance differences vary between +0.23~dex and $-0.21$~dex. 
 For most stars the obtained Mg NLTE abundance is close to the solar value. The exceptions are Vega, HD~72660, and HD~73666 whose classification indicates a peculiar origin.
  Below we comment briefly on some individual stars. 
  
{\bf $\iota$~Her.} This is the hottest star of our sample, therefore the magnesium abundance, log$\epsilon_{\rm Mg}$ = 7.55$\pm$0.07, was derived from the Mg\ii\ lines only.
   The NLTE corrections are negative for Mg\ii\ 4481, 7877, and 7896~\AA\ and positive for Mg\ii\ 4384, 4390, 4427, 4433, and 4739~\AA. Positive corrections do not exceed 
   0.06~dex for the first four lines.
 
{\bf $\pi$~Cet.} This star is 4700~K cooler than $\iota$~Her, and three weak lines of Mg\ione\ can be measured. 
    The NLTE effects for the Mg\ii\ lines are similar to those for $\iota$~Her, although they are slightly smaller.    

{\bf HD~22136.} This star was found to be a SPB-type variable \citep{2005AcA....55..375M}, and the asymmetric line profiles seen in its spectrum seem to support this classification. The asymmetry causes uncertainties in the line fitting procedure, depending on the line intensity. Nevertheless, the magnesium NLTE abundances based on 4 Mg\ione\ and 11 Mg\ii\ lines are consistent within 0.09~dex.
 The Mg NLTE abundance does not differ significantly from the solar value, however, it is worth noting that it is 0.2~dex higher than the overall stellar metallicity (see Table~\ref{tab_param}).
  
{\bf 21~Peg.}  In this star, the Mg\ione\ / Mg\ii\ ionization equilibrium is achieved in both NLTE and LTE, however, with much smaller line-to-line scatter in NLTE ($\sigma_{\rm Mg I}$ = 0.05~dex, $\sigma_{\rm Mg II}$ = 0.04~dex) than in LTE ($\sigma_{\rm Mg I}$ = 0.15~dex, $\sigma_{\rm Mg II}$ = 0.16~dex). This is illustrated well by Fig.~\ref{Mg_Fe}. Hereafter, the statistical abundance error is the dispersion in the single line measurements about the mean:
$\sigma = \Sigma (x - x_i )^2 /(N - 1)$. 
The NLTE abundance corrections are slightly positive ($\Delta_{\rm NLTE} \le$ 0.08~dex) for Mg\ione\ 4702 and 5528~\AA, close to zero for Mg\ione\ 4167~\AA\ and Mg\ii\ 3848, 3850, 4384, 4390, 4427, 4433, 4739~\AA, slightly negative ($\Delta_{\rm NLTE} \le$ 0.18~dex in absolute value) for Mg\ione\,b, 3829, 5167~\AA\ and Mg\ii\ 4481, 7896.04~\AA, however, pronounced NLTE effects are found for the Mg\ione\ 3832, 3838~\AA\ and Mg\ii\ 7877, 7896.37~\AA\ lines, with negative NLTE corrections up to $\Delta_{\rm NLTE} = -$0.48~dex. 
  
\begin{figure*}
\begin{center}
\includegraphics[scale=0.45]{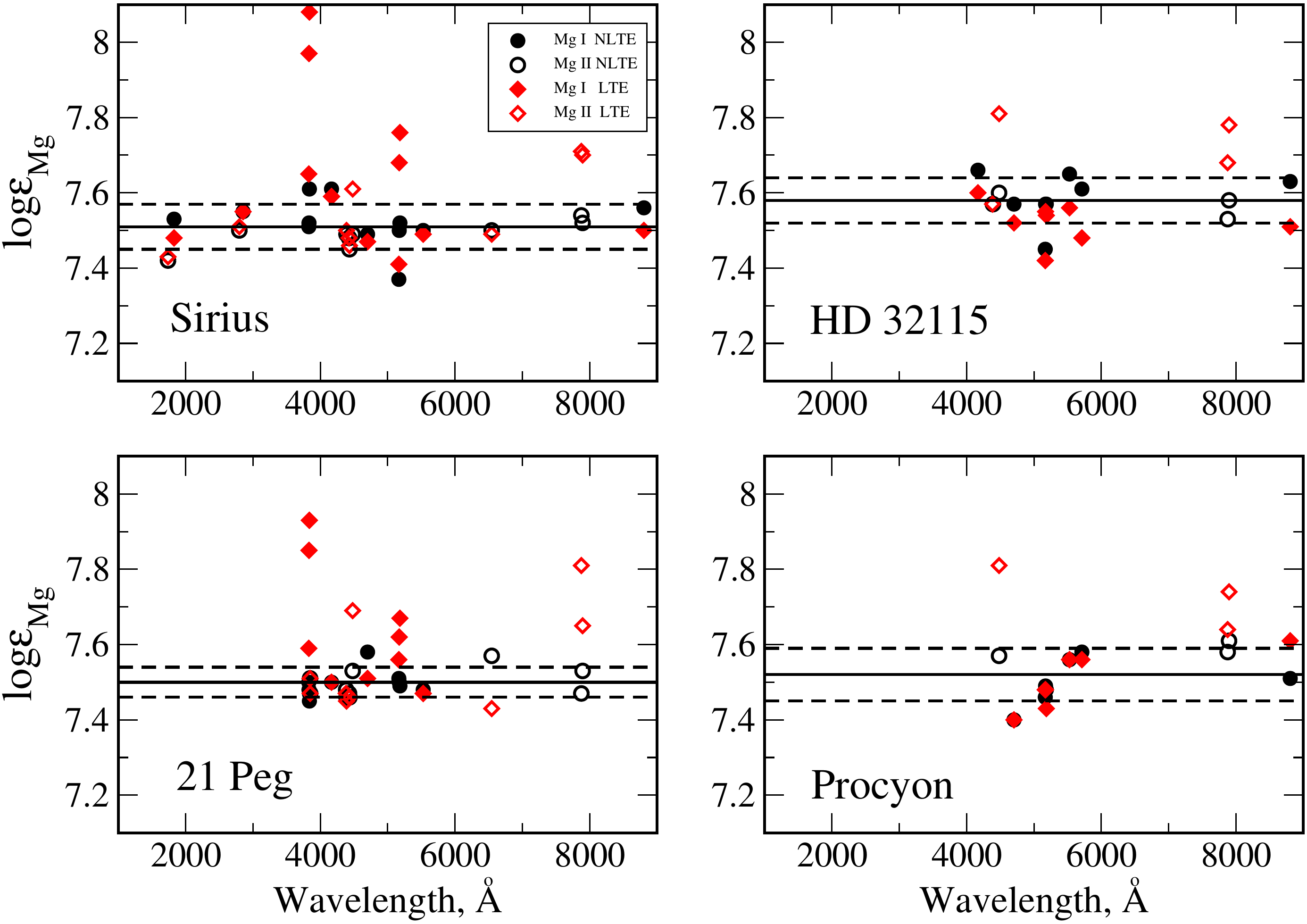}
\caption{Magnesium LTE (rhombi) and NLTE (circles) abundances of few program stars derived from lines of M\ione\ (filled symbols) and Mg\ii\ (open symbols) in wide spectral region. In each panel, the continuous line represents the NLTE abundance averaged over all M\ione\ and Mg\ii\ lines. Dashed lines represent the standard deviation. }
\label{Sirius}
\end{center}
\end{figure*}

{\bf Sirius.} The NLTE abundances from lines of the two ionization stages, Mg\ione\ and Mg\ii, are consistent, and they 
are close to the solar value, although the star is classified as a hot Am star with the metallicity of +0.4~dex (Table~\ref{tab_param}). 
 We find that abundances derived from the lines in the UV region (Table~\ref{tabUV}) agree with those derived from the visible and near-IR region (Table~\ref{tab_hot}), which means that the model atom developed in this paper provides consistent magnesium abundance determinations over a wide spectral range from UV to IR (Fig.~\ref{Sirius}). The NLTE abundance corrections for the investigated UV lines of Mg\ione\ and Mg\ii\ are smaller than 0.05~dex. 
  
{\bf HD~72660.} This star has the same metallicity as Sirius but, in contrast to Sirius, the Mg NLTE abundance, averaged over 11 Mg\ione\ and 10 Mg\ii\ lines, exceeds the solar value by 0.2~dex.  
Perhaps, this reflects a classification of HD~72660  as a transition object between HgMn and hot Am stars. The NLTE abundance derived from the Mg\ione\ UV lines, log$\epsilon_{\rm Mg}$ = 7.97$\pm$0.11 (Table~\ref{tabUV}), is 0.22~dex higher than that based on the Mg\ione\ visible and near-IR lines (Table~\ref{tab_hot}), although the discrepancy does not exceed 2$\sigma$. For Mg\ii, the NLTE abundances based on the UV (log$\epsilon_{\rm Mg}$ = 7.76$\pm$0.10) and visible (log$\epsilon_{\rm Mg}$ = 7.74$\pm$0.07) lines are consistent.

{\bf HD~73666.} 
For Mg\ione, the NLTE abundance corrections are either slightly positive ($\Delta_{\rm NLTE} \le$ 0.02~dex for 4167, 4702, 5528, and 5711~\AA) or negative, with $\Delta_{\rm NLTE}$ up to $-$0.55~dex. For most investigated lines of Mg\ii, the NLTE effects are minor. The exceptions are 4481, 7877, and 7896.04~\AA, with  $\Delta_{\rm NLTE}$ = $-0.12$, $-0.27$, and $-0.26$~dex, respectively. 
  
{\bf HD~32115.} NLTE leads to weakened lines of Mg\ione\ and, in contrast, strengthened lines of Mg\ii, such that the obtained NLTE abundances from the two ionization stages are consistent within 0.02~dex (Fig.~\ref{Sirius}). The LTE abundance based on the Mg\ii\ lines is 0.18~dex lower compared with that for Mg\ione. 

{\bf Vega.}  The NLTE effects for the Mg lines are very similar to those found for HD~73666.
The magnesium NLTE abundance is subsolar and
corresponds to the Vega's metallicity, such that [Mg/Fe] = $-$0.02.

{\bf Procyon.} The best fits to the Mg\ione\ 5183~\AA\ and Mg\ii\ 4481~\AA\ lines in Procyon are shown in Fig.\,\ref{pics_stars}. 
   Our NLTE calculations reproduce well the Mg\ione\ 12.22 and 12.32~$\mu$m lines observed in emission (Fig.\,\ref{pics_stars_emis} for 12.32~$\mu$m). 
   The best fits are achieved with log$\epsilon_{\rm Mg}$ = 7.50, in line with the abundances derived from the absorption lines of Mg\ione\ and Mg\ii.   
   The NLTE abundances from the Mg\ione\ (8 lines) and Mg\ii\ (3 lines) are fairly consistent (Fig.~\ref{Sirius}), while the LTE abundance difference amounts to Mg\ione\ -- Mg\ii\ = $-0.21$~dex.

 \begin{longrotatetable}
 \begin{deluxetable*}{cccccccccccccccccccccccccccc}
 \tabletypesize{\scriptsize}
 \tablecaption{NLTE abundances of the program stars. \label{tab_hot}}
\def\arraystretch{0.7}
\setlength{\tabcolsep}{2pt} 
\tabletypesize{\scriptsize }
\tablewidth{20pt}
\tablehead{
\colhead{$\lambda$ (\AA\,)} & \colhead{ \tiny NLTE} & \colhead{ \tiny LTE} & \colhead{$\Delta$} & \colhead{ \tiny NLTE} & \colhead{ \tiny LTE} & \colhead{$\Delta$} & \colhead{\tiny NLTE} & \colhead{ \tiny LTE} & \colhead{$\Delta$} & \colhead{ \tiny NLTE} & \colhead{ \tiny LTE} & \colhead{$\Delta$} & \colhead{ \tiny NLTE} & \colhead{ \tiny LTE} & \colhead{$\Delta$} & \colhead{ \tiny NLTE} & \colhead{ \tiny LTE} & \colhead{$\Delta$}& \colhead{\tiny NLTE} & \colhead{\tiny LTE} & \colhead{$\Delta$}& \colhead{\tiny NLTE} & \colhead{ \tiny LTE} & \colhead{$\Delta$} & \colhead{ \tiny NLTE} & \colhead{ \tiny LTE} & \colhead{$\Delta$} 
}
%\decimalcolnumbers
\startdata
%$\lambda$&jHer&                   & PiCet &   &           &HD22136 &  &           &21Peg&                  &   Sirius& &    &       & HD72660&               &HD73666 &  &              &Vega &  &                  & HD32115                 
%         &NLTE & LTE & Delta      &NLTE & LTE & Delta     &NLTE & LTE &Delta      &NLTE & LTE     &Delta   &   NLTE & LTE   & Delta &NLTE & LTE & Delta      &NLTE & LTE &Delta         &NLTE & LTE &Delta          &NLTE & LTE & Delta 
 \multicolumn{25}c{ }      \\
    &\multicolumn3c{$\iota$ Her}   &\multicolumn3c{$\pi$ Cet}&\multicolumn3c{HD22136}&\multicolumn3c{21Peg} &\multicolumn3c{Sirius}  &\multicolumn3c{HD72660} &\multicolumn3c{HD73666}   &\multicolumn3c{Vega}    & \multicolumn3c{HD32115} \\ \hline
 Mg\ione\ & \multicolumn{24}c{ }      \\
 3829     &\nodata&\nodata&\nodata &\nodata&\nodata&\nodata&\nodata&\nodata&\nodata&~7.50  &~7.59  &-0.09   &~7.51  &~7.65   &-0.14  & ~7.87 & ~8.05 &-0.18   &~7.67  &~7.93   & -0.26   &\nodata&\nodata&\nodata &\nodata&\nodata&\nodata    \\
 3832     &\nodata&\nodata&\nodata &\nodata&\nodata&\nodata&\nodata&\nodata&\nodata&~7.48  &~7.85  &-0.37   &~7.52  &~7.97   &-0.45  & ~7.82 & ~8.26 &-0.44   &~7.69  &~8.23   & -0.54   &\nodata&\nodata&\nodata &\nodata&\nodata&\nodata    \\
 3838     &\nodata&\nodata&\nodata &\nodata&\nodata&\nodata&\nodata&\nodata&\nodata&~7.45  &~7.93  &-0.48   &~7.61  &~8.08   &-0.47  & ~7.79 & ~8.26 &-0.47   &~7.74  &~8.29   & -0.55   &\nodata&\nodata&\nodata &\nodata&\nodata&\nodata    \\
 4167     &\nodata&\nodata&\nodata &\nodata&\nodata&\nodata&\nodata&\nodata&\nodata&~7.50  &~7.50  &~0.00   &~7.61  &~7.59   &~0.02  & ~7.71 & ~7.70 &~0.01   &~7.72  &~7.70   & ~0.02   &~7.10  &~7.09  &~0.01   & ~7.66 & ~7.60 &~0.06      \\
 %4571     &\nodata&\nodata&\nodata &\nodata&\nodata&\nodata&\nodata&\nodata&\nodata&\nodata&\nodata&\nodata &\nodata&\nodata &\nodata&\nodata&\nodata&\nodata &\nodata&\nodata &\nodata  &\nodata&\nodata&\nodata & ~7.25 & ~7.17 &~0.05      \\
 4702     &\nodata&\nodata&\nodata &\nodata&\nodata&\nodata&~7.56  &~7.55  &~0.01  &~7.58  &~7.51  &~0.07   &~7.49  &~7.47   &~0.02  & ~7.74 & ~7.75 &-0.01   &~7.73  &~7.71   & ~0.02   &~7.09  &~7.07  &~0.00   & ~7.57 & ~7.52 &~0.05      \\                                                                       
 5167     &\nodata&\nodata&\nodata &\nodata&\nodata&\nodata&\nodata&\nodata&\nodata&~7.51  &~7.56  & -0.05  &~7.37  &~7.41   &-0.04  & ~7.75 & ~7.87 &-0.12   &~7.58  &~7.76   &-0.18    &~7.03  &~7.14  &-0.04   & ~7.45 & ~7.42 &~0.03      \\
 5172     &\nodata&\nodata&\nodata &~7.56  &~7.61  &-0.05  &~7.60  &~7.64  &-0.04  &~7.50  &~7.62  & -0.12  &~7.50  &~7.68   &-0.18  & ~7.85 & ~8.08 &-0.23   &~7.69  &~8.05   &-0.36    &~7.07  &~7.29  &-0.16   & ~7.57 & ~7.55 &~0.02      \\
 5183     &\nodata&\nodata&\nodata &~7.61  &~7.67  &-0.06  &~7.58  &~7.62  &-0.04  &~7.49  &~7.67  & -0.18  &~7.52  &~7.76   &-0.24  & ~7.87 & ~8.17 &-0.30   &~7.69  &~8.11   &-0.32    &~7.00  &~7.30  &-0.23   & ~7.57 & ~7.54 &~0.03      \\
 5528     &\nodata&\nodata&\nodata &~7.61  &~7.47  &~0.16  &~7.55  &~7.46  &~0.09  &~7.60  &~7.52  &~0.08   &~7.50  &~7.49   &~0.01  & ~7.76 & ~7.75 &~0.01   &~7.74  &~7.73   &~0.01    &~7.09  &~7.07  &~0.00   & ~7.65 & ~7.57 &~0.08      \\
 5711     &\nodata&\nodata&\nodata &\nodata&\nodata&\nodata&\nodata&\nodata&\nodata&\nodata&\nodata&\nodata &\nodata&\nodata &\nodata& ~7.84 & ~7.83 &~0.01   &~7.76  &~7.74   &~0.02    &~7.08  &~7.07  &-0.01   & ~7.61 & ~7.48 &~0.13      \\
 8806     &\nodata&\nodata&\nodata &\nodata&\nodata&\nodata&\nodata&\nodata&\nodata&~7.84* &~7.50* &~0.34   &~7.56  &~7.50   &~0.06  & ~7.74 & ~7.86 &-0.12   &~7.78  &~7.87   &-0.09    &~7.15  &~7.09  &~0.07   & ~7.63 & ~7.61 &~0.02      \\  \hline                                                                                                                                                                                             
 Mean     &\nodata&\nodata&\nodata &~7.59  &~7.59  &\nodata&~7.57  &~7.57  &\nodata&~7.52  &~7.64  &\nodata &~7.52  &~7.71   &\nodata& ~7.80 & ~8.01 &\nodata &~7.71  &~7.97   &\nodata  &~7.08  &~7.12  &\nodata & ~7.59 & ~7.54 &\nodata    \\
 $\sigma$ &\nodata&\nodata&\nodata &~0.03  &~0.10  &\nodata&~0.02  &~0.08  &\nodata&~0.05  &~0.15  &\nodata &~0.07  &~0.22   &\nodata& ~0.06 & ~0.21 &\nodata &~0.05  &~0.22   &\nodata  &~0.05  &~0.07  &\nodata & ~0.07 & ~0.07 &\nodata    \\  \hline                                                                                                                                             
 Mg\ii\   & \multicolumn{24}c{ }      \\                                                                                                                                                    
 3848     &~7.63  &~7.59  &~0.04   &~7.64  &~7.61  &~0.03  &~7.44  &~7.44  &~0.00  &~7.51  &~7.51  &~0.00   &\nodata&\nodata &\nodata& ~7.71 & ~7.72 &-0.01   & ~7.67  & ~7.68  & -0.01  &\nodata&\nodata&\nodata &\nodata&\nodata&\nodata    \\
 3850     &\nodata&\nodata&\nodata &\nodata&\nodata&\nodata&~7.48  &~7.48  &~0.00  &~7.47  &~7.47  &~0.00   &\nodata&\nodata &\nodata& ~7.78 & ~7.78 &~0.00   & ~7.72  & ~7.73  & -0.01  &\nodata&\nodata&\nodata &\nodata&\nodata&\nodata    \\
 4384     &~7.48  &~7.43  &~0.05   &~7.58  &~7.54  &~0.04  &~7.46  &~7.42  &~0.04  &~7.48  &~7.47  &-0.01   &\nodata&\nodata &\nodata& ~7.72 & ~7.73 &-0.01   & ~7.71  & ~7.72  & -0.01  &\nodata&\nodata&\nodata &\nodata&\nodata&\nodata    \\
 4390     &~7.54  &~7.48  &~0.06   &~7.54  &~7.50  &~0.04  &~7.46  &~7.42  &~0.04  &~7.46  &~7.45  &-0.01   & ~7.49 &~7.50   &-0.01  & ~7.65 & ~7.65 &~0.00   & ~7.66  & ~7.66  & ~0.00  &~7.00  &~7.01   &-0.01  & ~7.57 & ~7.57 &~0.00      \\
 4427     &~7.57  &~7.51  &~0.06   &~7.58  &~7.55  &~0.03  &~7.50  &~7.47  &~0.03  &~7.47  &~7.46  &-0.01   & ~7.48 &~7.48   &~0.00  & ~7.65 & ~7.66 &-0.01   & ~7.64  & ~7.65  & -0.01  &~6.97  &~6.97   &~0.00  &\nodata&\nodata&\nodata    \\
 4433     &~7.55  &~7.49  &~0.06   &~7.58  &~7.54  &~0.05  &~7.48  &~7.44  &~0.04  &~7.46  &~7.46  &~0.00   & ~7.45 &~7.46   &-0.01  & ~7.64 & ~7.64 &~0.00   & ~7.64  & ~7.65  & -0.01  &~6.95  &~6.96   &-0.01  &\nodata&\nodata&\nodata    \\
 4481     &~7.41  &~7.77  &-0.36   &~7.54  &~7.80  &-0.26  &~7.52  &~7.74  &-0.22  &~7.53  &~7.69  &-0.16   & ~7.49 &~7.61   &-0.12  & ~7.75 & ~7.84 &-0.09   & ~7.74  & ~7.86  & -0.12  &~7.02  &~7.17   &-0.15  & ~7.60 & ~7.81 &-0.21      \\
 4739     &~7.64  &~7.53  &~0.11   &~7.46  &~7.46  &~0.00  &~7.52  &~7.47  &~0.05  &~7.45  &~7.45  &~0.00   &\nodata&\nodata &\nodata&\nodata&\nodata&\nodata & ~7.67  & ~7.66  & ~0.01  &\nodata&\nodata&\nodata &\nodata&\nodata&\nodata    \\
 6545     &\nodata&\nodata&\nodata &\nodata&\nodata&\nodata&\nodata&\nodata&\nodata&~7.57  &~7.43  &~0.14   & ~7.50 &~7.49   &~0.01  & ~7.83 & ~7.80 &~0.03   & ~7.66  & ~7.60  & ~0.06  &\nodata&\nodata&\nodata &\nodata&\nodata&\nodata    \\
 7877     &~7.52  &~7.75  &-0.23   &~7.58  &~7.80  &-0.22  &~7.48  &~7.66  &-0.18  &~7.47  &~7.81  &-0.34   & ~7.54 &~7.71   &-0.17  & ~7.76 & ~7.94 &-0.18   & ~7.67  & ~7.94  & -0.27  &\nodata&\nodata&\nodata & ~7.53 & ~7.68 &-0.15      \\              
 7896.04  &~7.53  &~7.77  &-0.24   &~7.52  &~7.71  &-0.19  &~7.44  &~7.66  &-0.22  &~7.53  &~7.65  &-0.12   & ~7.52 &~7.70   &-0.18  & ~7.75 & ~7.87 &-0.12   & ~7.69  & ~7.95  & -0.26  &\nodata&\nodata&\nodata & ~7.58 & ~7.78 &-0.20      \\
 7896.37  &\nodata&\nodata&\nodata &\nodata&\nodata&\nodata&\nodata&\nodata&\nodata&~7.48  &~7.90  &-0.42   &\nodata&\nodata &\nodata&\nodata&\nodata&\nodata &\nodata &\nodata &\nodata &\nodata&\nodata&\nodata &\nodata&\nodata&\nodata    \\
 9218$^*$ &\nodata&\nodata&\nodata &~7.88  &~8.32  &-0.44  &~7.78  &~8.10  &-0.32  &~7.59  &~8.42  &-0.83   &\nodata&\nodata &\nodata& ~7.94 & ~8.63 &-0.69   & ~7.79  & ~8.75  & -0.96  &\nodata&\nodata&\nodata & ~7.62 & ~8.36 &-0.74      \\
 9244$^*$ &\nodata&\nodata&\nodata &~7.88  &~8.20  &-0.32  &\nodata&\nodata&\nodata&~7.62  &~8.18  &-0.56   &\nodata&\nodata &\nodata& ~8.12 & ~8.66 &-0.54   & ~8.01  & ~8.72  & -0.71  &\nodata&\nodata&\nodata & ~7.64 & ~8.34 &-0.70      \\ \hline                                                                                                                                                 
 Mean     &~7.55  &~7.61  &\nodata &~7.56  &~7.63  &\nodata&~7.48  &~7.53  &\nodata&~7.49  &~7.59  &\nodata &~7.50  &~7.58   &\nodata&~7.74  &~7.78  &\nodata &~7.68   &~7.75  &\nodata  &~6.99  &~7.04  &\nodata & ~7.57 & ~7.72 &\nodata    \\
 $\sigma$ &~0.07  &~0.14  &\nodata &~0.05  &~0.12  &\nodata&~0.03  &~0.12  &\nodata&~0.04  &~0.16  &\nodata &~0.03  &~0.11   &\nodata&~0.07  &~0.10  &\nodata &~0.03   &~0.12  &\nodata  &~0.03  &~0.09  &\nodata & ~0.03 & ~0.11 &\nodata   \\\hline  
Mg\ione$-$Mg\ii\                                                                                                                                                                                                  
          &\nodata&\nodata&\nodata &~0.03  &-0.04  &\nodata&~0.09  &~0.04  &\nodata&~0.03  &~0.05  &\nodata &~0.02  &~0.13   &\nodata&~0.06  &~0.23  &\nodata &~0.03   &~0.22 &\nodata   &~0.09  &~0.08  &\nodata &~0.02  &-0.18  &\nodata   \\ \hline\hline 
           \multicolumn{28}c{ }      \\
          & \multicolumn3c{Procyon} & \multicolumn3c{HD122563} & \multicolumn3c{HD140283} &  \multicolumn3c{HD103095} & \multicolumn3c{HD84937} & \multicolumn3c{Arcturus} &\multicolumn3c{Pollux} & \multicolumn3c{Aldebaran} & \multicolumn3c{} \\\hline
%\decimalcolnumbers
%\startdata
% $\lambda$  & Procyon                  &HD122563               &HD140283&  &           &HD103095&               &HD84937&   &           &Arcturus&  &           &Pollux&                 &&Aldebaran& &&
%            &NLTE & LTE & Delta        & NLTE & LTE  &D        &NLTE & LTE & D         &NLTE & LTE              &NLTE & LTE &           &NLTE & LTE &           &NLTE & LTE &            &NLTE & LTE &&
 Mg\ione\     & \multicolumn{24}c{ }      \\
 4167        &\nodata&\nodata&\nodata   &\nodata&\nodata&\nodata&\nodata&\nodata&\nodata&\nodata&\nodata&\nodata&~5.68  &~5.66  &~0.02  &\nodata&\nodata&\nodata&\nodata&\nodata&\nodata &\nodata&\nodata&\nodata & & & \\
 4571        &\nodata&\nodata&\nodata &~5.16  &~5.00  &~0.16  &~5.29  &~5.12  &~0.17  &~6.36  &~6.35  &~0.01  &~5.48  &~5.42  &~0.06  &\nodata&\nodata&\nodata&~7.34  &~7.29  &~0.05   &\nodata&\nodata&\nodata & & & \\
 4702        &~7.40  &~7.40  &~0.00     &~5.19  &~5.17  &~0.02  &~5.35  &~5.31  &~0.04  &~6.32  &~6.33  &-0.01  &~5.68  &~5.65  &~0.03  &~7.31  &~7.33  &-0.02  &\nodata&\nodata&\nodata &\nodata&\nodata&\nodata & & &\\
 4730        &~7.68  &~7.67  &~0.01     &~5.28  &~5.28  &~0.00  &\nodata&\nodata&\nodata&~6.57  &~6.57  &~0.00  &~5.80  &~5.80  &~0.00  &\nodata&\nodata&\nodata&\nodata&\nodata&\nodata &\nodata&\nodata&\nodata & & &\\
 5167        &~7.46  &~7.48  &-0.02     &~5.09  &~4.92  &~0.17  &~5.26  &~5.22  &~0.04  &~6.43  &~6.42  &~0.01  &~5.61  &~5.55  &~0.06  &\nodata&\nodata&\nodata&\nodata&\nodata&\nodata &\nodata&\nodata&\nodata & & &\\
 5172        &~7.49  &~7.48  &~0.01     &~5.07  &~5.02  &~0.05  &~5.26  &~5.25  &~0.01  &~6.43  &~6.42  &~0.01  &~5.58  &~5.57  &~0.01  &~7.36  &~7.35  &~0.01  &\nodata&\nodata&\nodata &\nodata&\nodata&\nodata & & &\\
 5183        &~7.48  &~7.43  &~0.05     &~5.08  &~5.07  &~0.01  &~5.24  &~5.23  &~0.01  &~6.42  &~6.41  &~0.01  &~5.60  &~5.62  &-0.02  &~7.31  &~7.30  &~0.01  &~7.50  &~7.49  &~0.01   &\nodata&\nodata&\nodata & & &\\
 5528        &~7.56  &~7.56  &~0.00     &~5.26  &~5.25  &~0.01  &~5.35  &~5.33  &~0.02  &~6.45  &~6.45  &~0.00  &~5.70  &~5.68  &~0.02  &~7.37  &~7.42  &-0.03  &~7.55  &~7.56  &-0.01   &~7.47  &~7.54  &-0.07   & & & \\
 5711        &~7.58  &~7.56  &~0.02     &\nodata&\nodata&\nodata&\nodata&\nodata&\nodata&~6.61  &~6.61  &~0.00  &\nodata&\nodata&\nodata&~7.42  &~7.49  &-0.07  &~7.62  &~7.61  &~0.01   &~7.42  &~7.54  &-0.12   & & & \\
 8806        &~7.51  &~7.61  &-0.10     &~5.46  &~5.43  &~0.03  &~5.33  &~5.33  &~0.00  &~6.55  &~6.57  &-0.02  &~5.70  &~5.68  &~0.02  &~7.38  &~7.44  &-0.06  &~7.61  &~7.64  &-0.03   &~7.56  &~7.63  &-0.07   & & & \\ \hline
 Mean        &~7.53  &~7.53  &\nodata   &~5.22  &~5.17  &\nodata&~5.30  &~5.26  &\nodata&~6.47  &~6.47  &\nodata&~5.66  &~5.64  &\nodata&~7.34  &~7.37  &\nodata&~7.57  &~7.58  &\nodata &~7.49  &~7.57  &\nodata & & &  \\
$\sigma$     &~0.09  &~0.09  &\nodata   &~0.13  &~0.17  &\nodata&~0.05  &~0.07  &\nodata&~0.10  &~0.10  &\nodata&~0.09  &~0.10  &\nodata&~0.04  &~0.06  &\nodata&~0.06  &~0.07  &\nodata &~0.07  &~0.06  &\nodata & & &  \\ \hline
Mg\ii\       & \multicolumn{27}c{ }    \\
 4481        &~7.57  &~7.81  &-0.24     &\nodata&\nodata&\nodata&~5.54  &~5.58  &-0.04  &\nodata&\nodata&\nodata&~5.85  &~5.86  &-0.01  &\nodata&\nodata&\nodata&\nodata&\nodata&\nodata &\nodata&\nodata&\nodata & & & \\
 7877        &~7.58  &~7.64  &-0.06     &\nodata&\nodata&\nodata&\nodata&\nodata&\nodata&\nodata&\nodata&\nodata&\nodata&\nodata&\nodata&\nodata&\nodata&\nodata&\nodata&\nodata&\nodata &\nodata&\nodata&\nodata & & & \\
 7896.04     &~7.61  &~7.74  &-0.13     &\nodata&\nodata&\nodata&\nodata&\nodata&\nodata&\nodata&\nodata&\nodata&\nodata&\nodata&\nodata&\nodata&\nodata&\nodata&\nodata&\nodata&\nodata &\nodata&\nodata&\nodata & & & \\
 9244        &\nodata&\nodata&\nodata   &\nodata&\nodata&\nodata&\nodata&\nodata&\nodata&\nodata&\nodata&\nodata&\nodata&\nodata&\nodata&\nodata&\nodata&\nodata&~7.56  &~7.74  &-0.18   &\nodata&\nodata&\nodata & & & \\ \hline
 Mean        &~7.59  &~7.74  &\nodata   &\nodata&\nodata&\nodata&~5.54  &~5.58  &\nodata&\nodata&\nodata&\nodata&~5.85  &~5.86  &\nodata&\nodata&\nodata&\nodata&~7.56  &~7.74  &\nodata &\nodata&\nodata&\nodata & & & \\
$\sigma$     &~0.02  &~0.08  &\nodata   &\nodata&\nodata&\nodata&\nodata&\nodata&\nodata&\nodata&\nodata&\nodata&\nodata&\nodata&\nodata&\nodata&\nodata&\nodata&\nodata&\nodata&\nodata &\nodata&\nodata&\nodata & & & \\  \hline
Mg\ione\ $-$ Mg\ii\ &-0.06&-0.21&\nodata&\nodata&\nodata&\nodata&-0.24  &-0.32  &\nodata&\nodata&\nodata&\nodata&-0.19  & -0.22 &\nodata&\nodata&\nodata&\nodata&~0.01  &-0.16  &\nodata &\nodata&\nodata&\nodata & & & \\   \hline          
\enddata                                                                                                                                                                  
\tablecomments{{\bf Notes.} Here, $\Delta$ means $\Delta_{NLTE}$, Lines and abundances, which were not used in mean calculations are marked by (*).}
\end{deluxetable*}
%\tt\string\end\{longrotatetable}
 \end{longrotatetable}

 \begin{deluxetable*}{lccccccc}
\tablecaption{Lines of Mg\ione\ and Mg\ii\ in the UV spectra of HD~72660 and Sirius and the obtained NLTE abundances. \label{tabUV}}
%\tabletypesize{\scriptsize}
\tablehead{
\colhead{$\lambda$} & \colhead{Transition} & \colhead{\Eexc} & \colhead{log~$gf$} & \colhead{log $\Gamma_4$/$N_e$}  & \colhead{log $\Gamma_4$/$N_p$}  & HD~72660 & Sirius \\
\colhead{\AA\,} & \colhead{} & \colhead{eV} & \colhead{}&\multicolumn2c{rad s$^{-1}$cm$^3$ } & \multicolumn2c{log~$\epsilon_{\rm Mg}$} 
}
\colnumbers
\startdata
 Mg\ione\ &                                                    &      &           &          &           &      &        \\
 1827.93 & 3s$^{2}$ $^{1}$S$_{0}$ -- 5p $^{1}$P$_{1}^{\circ}$  &  0.00&  $-$1.667 & $-$4.47  & $-$4.97   &7.89  &7.53\\ % 7.43
 2852.13 & 3s$^{2}$ $^{1}$S$_{0}$ -- 3p $^{1}$P$_{1}^{\circ}$  &  0.00&    ~0.240 & $-$5.78  & $-$6.20   &8.05  &7.55 \\ %7.53
 Mg\ii\ &                                                      &      &           &          &           &      &  \\   
 1737.61  & 3p $^{2}$P$_{3/2}^{\circ}$ -- 4d $^{2}$D$_{3/2}$   &  4.43&$-$1.814   &$-$4.65   & $-$5.43   &      &    \\
 1737.63  & 3p $^{2}$P$_{3/2}^{\circ}$ -- 4d $^{2}$D$_{5/2}$   &  4.43&$-$1.859   &$-$4.65   & $-$5.43   &7.69  &7.42\\ %7.43
 2797.93  & 3p $^{2}$P$_{3/2}^{\circ}$ -- 3d $^{2}$D$_{3/2}$   &  4.43&$-$0.426   &$-$5.32   & $-$6.44   &      &    \\
 2798.00  & 3p $^{2}$P$_{3/2}^{\circ}$ -- 3d $^{2}$D$_{5/2}$   &  4.43&  ~0.528   &$-$5.32   & $-$6.44   &7.84  &7.50\\ %7.51
\enddata
\end{deluxetable*}

{\it Methodical notes and recommendations.}
  We can recommend the Mg\ii\ 3848, 3850, 4384, 4390, 4427, and 4433~\AA\ lines for Mg abundance determinations even at the LTE assumption in wide range of effective temperatures, from 7000 to 17500~K, due to small NLTE effects for these lines. 
 The Mg\ione\ 4167, 4571, 4702, 5528, 5167, 5172, and 5183~\AA\ lines can be safely used in the LTE analysis for stars with 7000~K $<$ \Teff\ $\leq$ 8000~K. For the hotter stars, with \Teff\ from 8000 to 9500~K, the NLTE effects are minor only for Mg\ione\ 4167, 4702, and 4528~\AA. 
 We caution against using the 
 Mg\ii\ 9218, 9244~\AA\ lines for Mg abundance determination due to their position in the wings of the Paschen P9 line and also pronounced NLTE effects, with negative NLTE corrections up to one order of magnitude. 

 \subsection{Analysis of magnesium lines in stars with 3900 $\leq$ \Teff\ $<$ 6350~K}

\begin{figure*}
  \begin{minipage}{170mm}
  %\begin{center}
  \parbox{0.3\linewidth}{\includegraphics[scale=0.19]{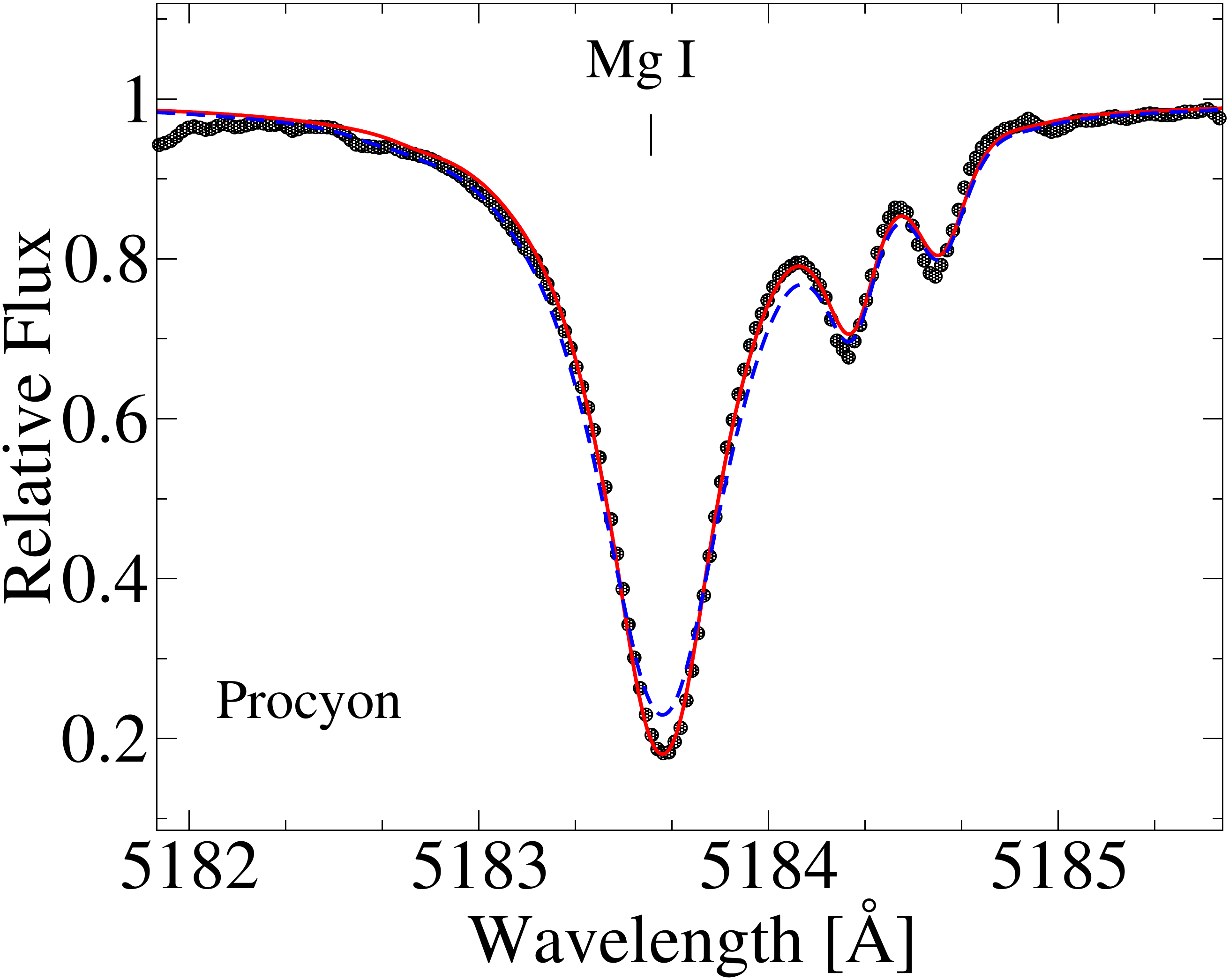}\\
  \centering}
  %\hspace{0.33\linewidth}
  \parbox{0.3\linewidth}{\includegraphics[scale=0.19]{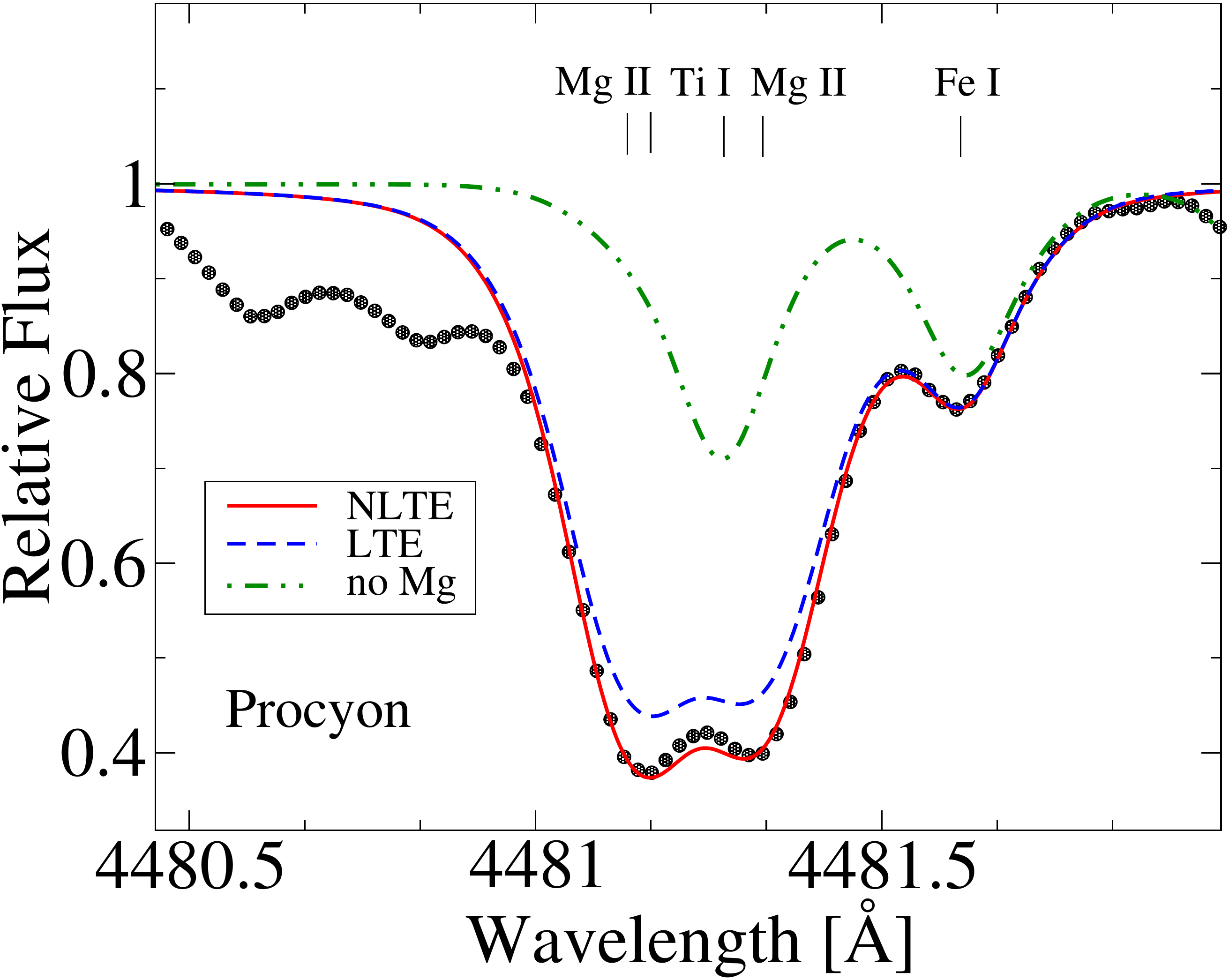}\\
  \centering}
  \parbox{0.3\linewidth}{\includegraphics[scale=0.19]{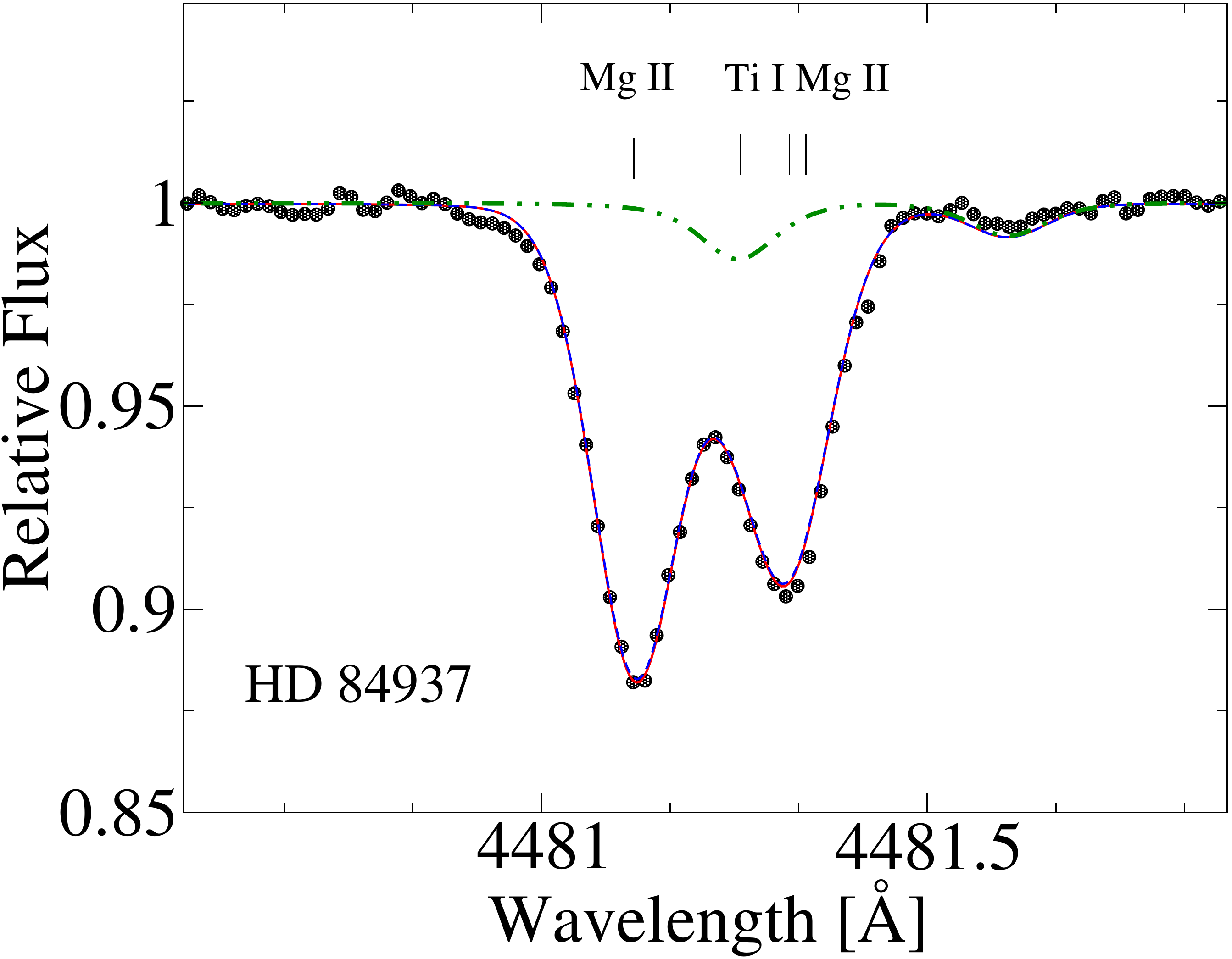}\\
  \centering}
  \hspace{1\linewidth}
  \hfill
  \\[0ex]
  \caption{Best NLTE fits (continuum curves) to the selected Mg\ione\ and Mg\ii\ lines in Procyon and HD~84937. The observed spectra are shown by
   black circles. For each line, the LTE profile (dashed curve) was computed with the magnesium abundance obtained from the NLTE analysis. The theoretical spectra without the Mg\ii\ 4481~\AA\ line are shown by the dashed-dot-dot curve.} 
  \label{pics_stars}
  \end{minipage}
  \end{figure*}

For seven stars with 3900~K $\leq$ \Teff\ $<$ 6350~K, the abundance results are presented in Table~\ref{tab_hot}. For each star NLTE provides smaller line-to-line scatter (Fig.~\ref{Mg_Fe}). For the three stars, abundances were determined from lines of the two ionization stages, Mg\ione\ and Mg\ii. The best fit to Mg\ii\ 4481~\AA\ in HD~84937 is shown in Fig.~\ref{pics_stars}.  The only line of Mg\ii, at 9244\,\AA, is observed in Pollux and it gives the NLTE abundance, which differs from that for Mg\ione, by only 0.01~dex. In LTE, we obtain Mg\ione\ -- Mg\ii\ = $-0.16$~dex.  

  However, the Mg\ione /Mg\ii\ ionization equilibrium is not achieved in HD~140283 and HD~84937, independent of either NLTE or LTE. In these two stars, we can measure the only line of Mg\ii, at 4481\,\AA, and it gives higher abundance than that from the Mg\ione\ lines, by 0.24 and 0.19~dex, respectively.
  Although Mg\ii\ 4481~\AA\ is contaminated by blends of Ti\ione\ 4481.259~\AA, Gd\ii\ 4481.054~\AA, and Ni\ione\ 4481.119~\AA, which contribute no more than
  15~\%\ of the total equivalent width, this can hardly be a reason of the obtained abundance discrepancy. The NLTE effects for the Mg\ione\ and Mg\ii\ lines in these stars are small and also cannot solve the problem.
  Unfortunately, we were unable to find the literature investigations of the Mg\ii\ 4481~\AA\ line in the atmospheres of cool stars.
  In order to understand an origin of such a discrepancy, it would be useful to perform full 3D NLTE calculations with our comprehensive model atom.

\begin{figure*}
  \begin{minipage}{170mm}
  \parbox{0.3\linewidth}{\includegraphics[scale=0.19]{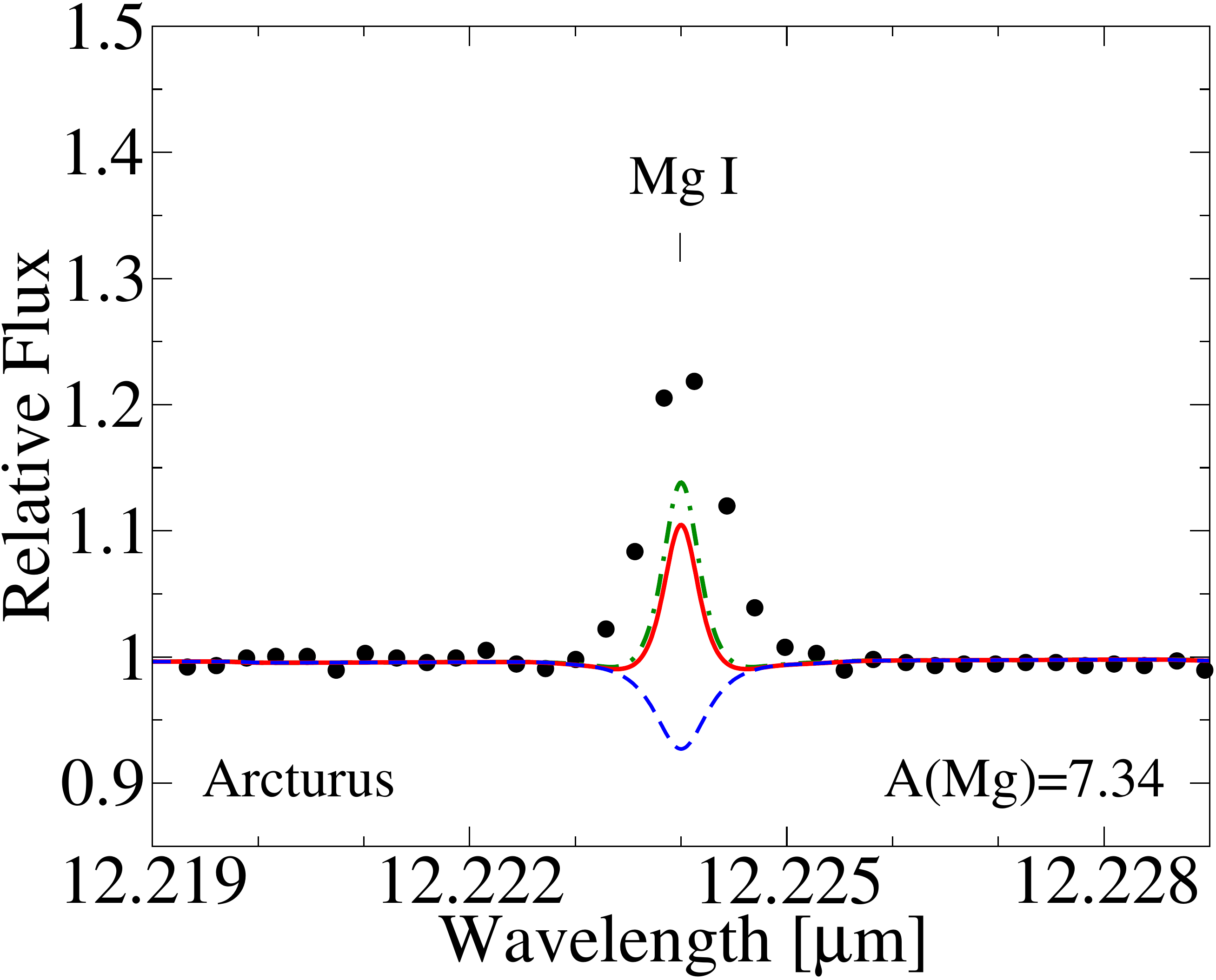}\\
  \centering}
  \parbox{0.3\linewidth}{\includegraphics[scale=0.19]{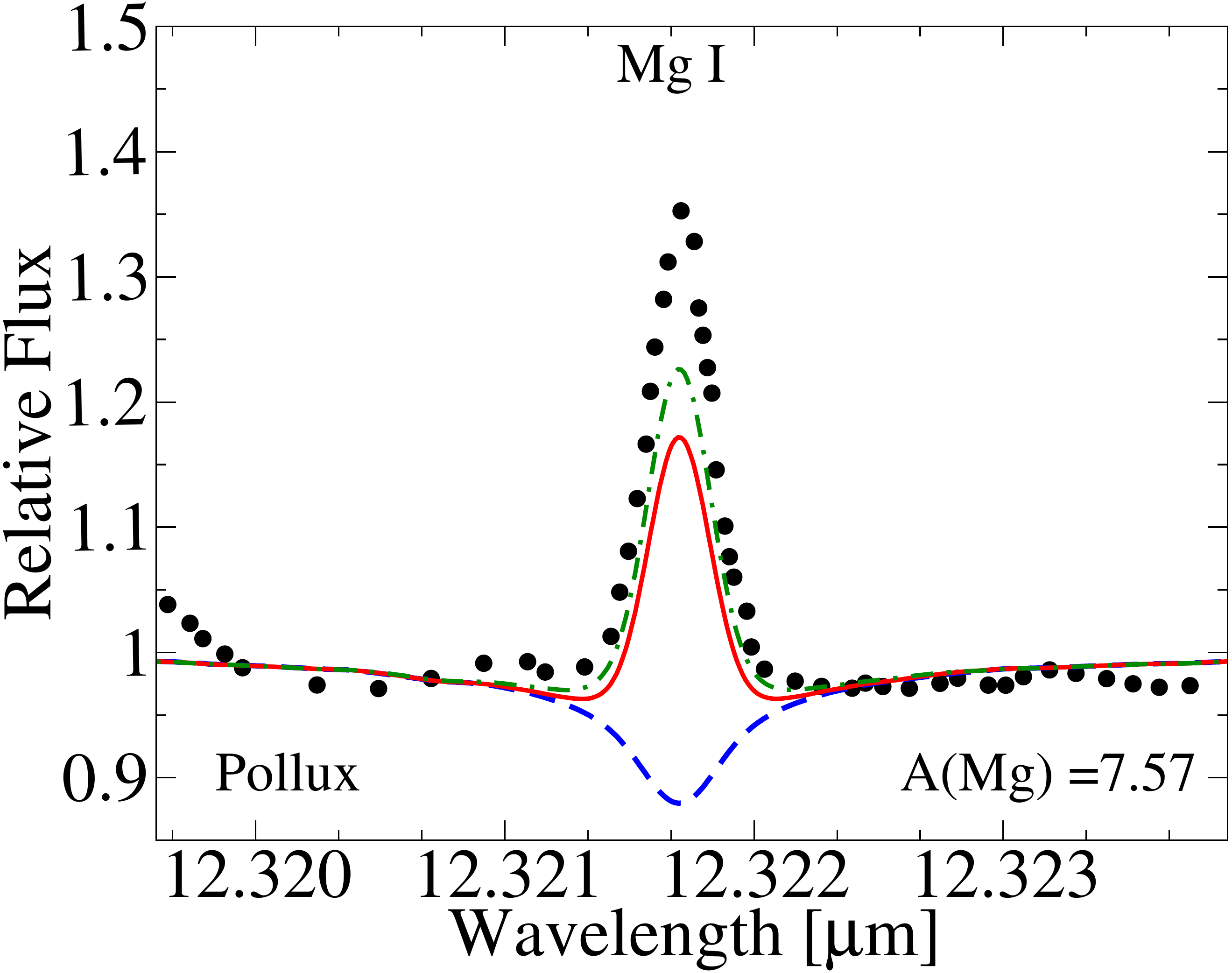}\\
  \centering}
  \parbox{0.3\linewidth}{\includegraphics[scale=0.19]{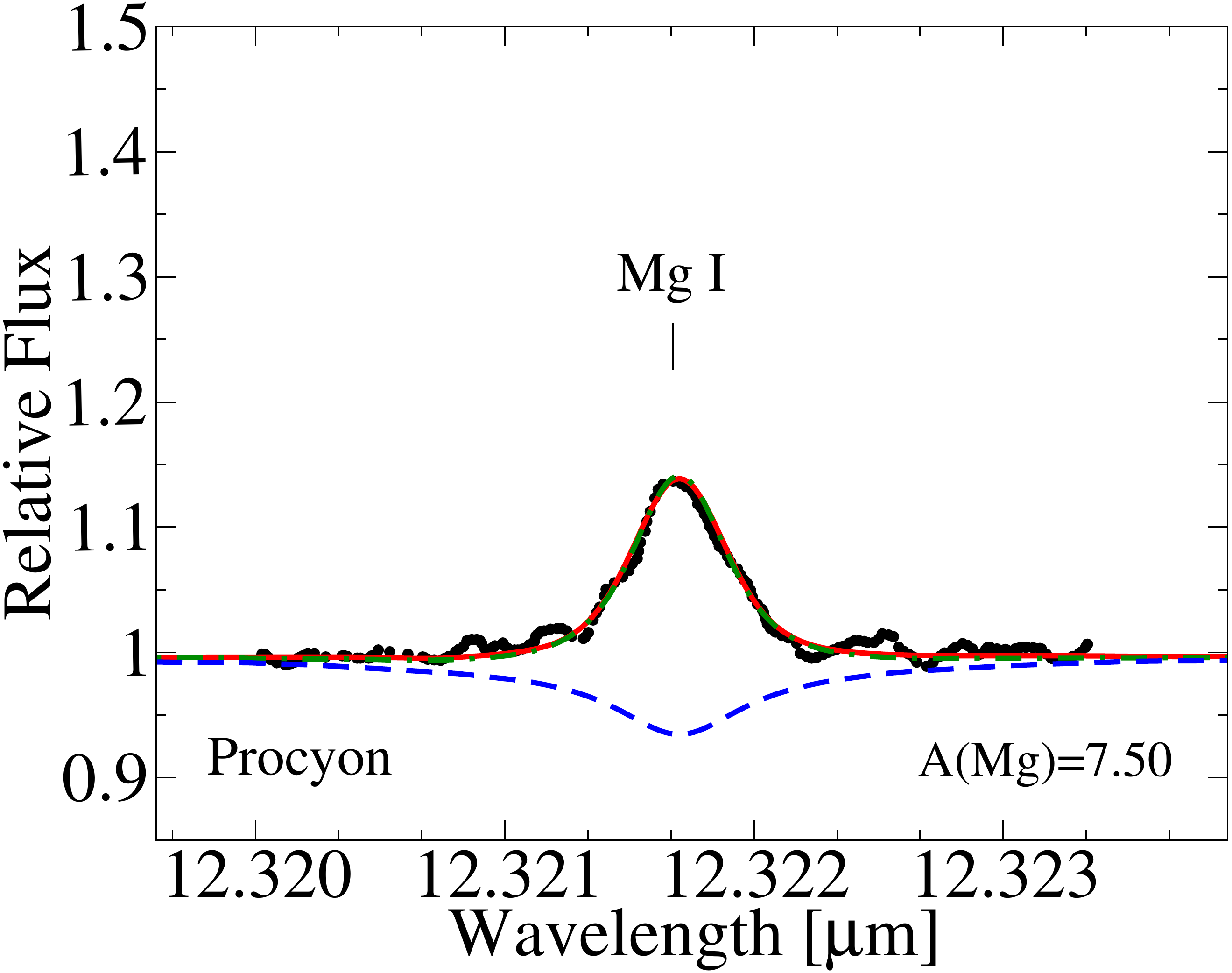}\\
  \centering}
  \hspace{1\linewidth}
  \hfill
  \\[0ex]
  \caption{Mg\ione\ emission lines in Arcturus, Pollux, and Procyon (black circles) compared with the synthetic spectra computed with log~$\epsilon_{\rm Mg}$ = 7.34, 7.57, and 7.50, respectively, at the LTE assumption (dashed curve) and in NLTE with the M15 (continuous curve) and IPM (dash-dotted curve) model atom.
}
  \label{pics_stars_emis}
  \end{minipage}
  \end{figure*}

As found by \citet{1996ASPC..109..723U} and \citet{2008A&A...486..985S}, the IR spectra of our three K giants reveal the emission in the Mg\ione\ 12.22 and 12.32~$\mu$m lines (Fig.~\ref{pics_stars_emis} shows Arcturus and Pollux), which is much stronger than in the corresponding solar and Procyon lines. Our NLTE calculations predict emission lines, however, they are too weak. We suggest that a formation of Mg\ione\ 12.22 and 12.32~$\mu$m in Aldebaran, Arcturus, and Pollux is affected by the chromospheric temperature rise, which is not accounted for by our classical radiative-equilibrium model atmospheres.

\section{Comparison with previous studies}\label{sec:comparison}

 In Table~\ref{tab2}, we compare the NLTE corrections for three lines of Mg\ione\ computed in this work, by \citet[][M13]{2013AA...550A..28M}, \citet[][O16]{2016A&A...586A.120O}, and \citet[][B17]{2017ApJ...847...15B}. The calculations were performed with the MARCS model atmospheres in this and O16 studies and with the MAFAGS-OS model atmospheres in M13 and B17 studies. We believe that using different types of model atmospheres in different papers does not influences on the abundance differences, that is the NLTE abundance corrections. All the four studies use very similar model atoms of Mg\ione\ with respect to the energy levels and the radiative data. The differences between the four studies are mainly due to different treatment of inelastic collisions. 
 
 Common recipes were applied by \citet{2013AA...550A..28M} and \citet{2017ApJ...847...15B}, namely, for electron-impact excitation: rate coefficients of \citet{1988ApJ...330.1008M}, where available, and for the remaining transitions formula of \citet{1962ApJ...136..906V} or $\Omega = 1$, if the transition is allowed or forbidden; for H\ione\ impact excitations and charge transfer processes: rate coefficients from
quantum mechanical calculations of \citet{2012A&A...541A..80B}. 

For electron-impact excitation, \citet{2016A&A...586A.120O} used the R-matrix calculations for transitions between the first nine levels of Mg\ione, the IPM \citep{1962PPS....79.1105S} for the remaining allowed transitions, and a temperature dependent collisional strengths for the  forbidden transitions, with separating the transitions involving electron exchange from those without electron exchange. For charge transfer processes, \citet{2016A&A...586A.120O} applied rate coefficients of \citet{2012A&A...541A..80B}, while, for the hydrogen-impact excitation, the latter data were complemented with the Kaulakys prescription for Rydberg levels as described by \citet{2015AA...579A..53O}. In this study, accurate data on electron-impact excitation are used for a larger number of transitions compared with that of \citet{2015AA...579A..53O}, that is, 369 transitions against 45 ones. From the other hand side, we use more simple approch compared with that of \citet{2015AA...579A..53O} to treat electron-impact excitation of those forbidden transitions, which are not available in \citet{2015A&A...577A.113M}, and we neglect the hydrogen-impact excitation of the Rydberg levels in Mg\ione.

 It can be seen from Table~\ref{tab2} that, for any of the three Mg\ione\ lines, there is no systematic discrepancy in $\Delta_{\rm NLTE}$ between this and the other three studies and between \citet{2016A&A...586A.120O} and the other three studies, which do not use H\ione\ impact excitations of the Rydberg states.

\begin{deluxetable*}{lccccccccccccccc}
%\tablenum{2}
\scriptsize
\tablecaption{Comparison of the NLTE abundance 
corrections for Mg\ione\ lines from this work with those from {\bf M13}: Mashonkina (2013), {\bf O16}: Osorio \& Barklem (2016) and {\bf B17}: Bergemann et al. (2017a). \label{tab2}}
\tablewidth{0pt}
\tablehead{
\colhead{\Teff} & \colhead{log$g$} & \colhead{[Fe/H]} & \colhead{$\xi_t$} &
\multicolumn4c{4571\,\AA\,} & \multicolumn4c{5528\,\AA\,} & \multicolumn4c{5711\,\AA\,} \\
\colhead{K} & \colhead{CGS} & \colhead{dex} & \colhead{\kms} & \colhead{This work} & \colhead{\bf{M13}} & \colhead{\bf{O16}} & \colhead{\bf{B17}} & \colhead{This work} & \colhead{\bf{M13}} & \colhead{\bf{O16}} & \colhead{\bf{B17}} &
\colhead{This work} & \colhead{\bf{M13}} &  \colhead{\bf{O16}} &\colhead{\bf{B17}}
}
\decimalcolnumbers
\startdata
 6000  & 4.0     &  $-$1.0 &  1      & ~0.11 & ~0.08 & ~0.03 & ~0.04  &  $-$0.03 &$-$0.03  &$-$0.03  &$-$0.03   &   ~0.02 &  ~0.04  &   ~0.03  &   ~0.03  \\
 6000  & 4.0     &  $-$2.0 &  1      & ~0.12 & ~0.08 & ~0.05 & ~0.07  &  ~0.01   &$-$0.02  &$-$0.01  &  ~0.04   &   ~0.02 &  ~0.03  &   ~0.03  &   ~0.05  \\
 6000  & 4.0     &  $-$3.0 &  1      & ~0.20 &       & ~0.12 & ~0.12  &  ~0.05   & ~0.07   & ~0.10   &  ~0.09   &   ~0.05 &         &   ~0.10  &   ~0.10  \\
 5000  & 2.0     &  $-$1.0 &  2      & ~0.16 & ~0.13 & ~0.20 & ~0.08  &  $-$0.04 &$-$0.16  &$-$0.07  &$-$0.06   &   ~0.03 &$-$0.05  &   ~0.02  & $-$0.03  \\
 5000  & 2.0     &  $-$2.0 &  2      & ~0.21 & ~0.16 & ~0.13 & ~0.13  &  $-$0.04 &$-$0.12  &$-$0.13  &$-$0.02   &   ~0.05 &  ~0.03  &   ~0.02  &   ~0.04  \\
 5000  & 2.0     &  $-$3.0 &  2      & ~0.34 & ~0.29 & ~0.23 & ~0.24  &   ~0.08  & ~0.10   & ~0.06   &  ~0.13   &   ~0.08 &  ~0.07  &   ~0.07  &   ~0.09  \\    
\enddata
\end{deluxetable*}
  
We also compare the obtained Mg NLTE abundances of our sample stars with the 1D NLTE results available in the literature.

{\bf Vega.} \citet{2001A&A...369.1009P} performed the NLTE calculations for Mg\ione/Mg\ii\ on the basis of the ATLAS9  
model atmosphere using 
collisional cross-sections from \citet{1991PhRvA..44.2874C} for the Mg\ione\ transitions and effective collision strengths from \citet{1995JPhB...28.4879S} for the Mg\ii\ transitions. 
   The departure coefficients computed by \citet{2001A&A...369.1009P} (Figure~5 in their paper) are very similar with ours (Fig.\,\ref{DC}, middle row). For an abundance comparison, the NLTE abundances derived by \citet{2001A&A...369.1009P} from individual lines were reduced to the $gf$ set adopted in this study (Table~\ref{tab3}). The corrected value, log$\epsilon_{\rm Mg}$ = 7.03$\pm$0.04, agrees well with our determination, log$\epsilon_{\rm Mg}$ = 7.05$\pm$0.06.

{\bf $\iota$~Her.} The Mg NLTE abundance was derived by \citet{2012AA...539A.143N} using four lines of Mg\ii. Their value, 
log$\epsilon_{\rm Mg}$ = 7.56$\pm$0.06 agrees perfectly with our determination,  log$\epsilon_{\rm Mg}$ = 7.55$\pm$0.07.

    \begin{deluxetable*}{lcccccccc|cccccc}
 %   \tablenum{7}
    \tablecaption{NLTE abundances from the Mg\ione\ and Mg\ii\ lines in Vega as determined in this work and by Przybilla et~al. (2001, {\bf P01} with the 9550 / 3.95 / $-$0.5 model atmosphere. \label{tab3}}
    \tablewidth{0pt}
    \tablehead{
    \colhead{} & \multicolumn8c{Mg\ione\,} & \multicolumn4c{Mg\ii\,} \\
    \colhead{$\lambda$, \AA\ } & \colhead{4167} & \colhead{4702} & \colhead{5167} & \colhead{5172} & \colhead{5183} & \colhead{5528} & \colhead{5711} & \colhead{8806} & \colhead{4390} &
    \colhead{4427} & \colhead{4433} & \colhead{4481} & \colhead{Mean} & \colhead{$\sigma$} 
    }
    \decimalcolnumbers
    \startdata
    This work         &  7.10   &  7.09   & 7.03   & 7.07 &  7.00   & 7.09   & 7.08   &  7.15   & 7.00 &  6.97   &  6.95 & 7.02 & 7.05 & 0.06    \\  
    {\bf P01}*        &  7.08   &  7.00   & \nodata& 7.05 &  7.08   & 7.03   & \nodata&  7.02   & 6.98 &  7.03   &  7.02 & 6.97 & 7.03 & 0.04 \\  
    \enddata
    \tablecomments{{\bf Notes.} *All abundances were corrected according to our oscillator strengths. }
\end{deluxetable*}
    
 {\bf Sirius} and {\bf HD~72660.} \citet{2011AA...528A.132L} and \citet{2016MNRAS.456.3318G} derived the LTE abundances of Sirius and HD~72660, respectively, using their UV spectra. They used the resonance lines of Mg\ii\ at 2795 and 2802~\AA\, as well as Mg\ii\ 1737.62~\AA\ for Sirius and Mg\ii\ 2798.00~\AA, Mg\ione\ 1827.93 and 2852.13~\AA\ for HD~72660. For Sirius few optical  Mg\ii\ lines were also added. As mentioned above, the NLTE effects for the Mg\ii\ and Mg\ione\ UV lines are minor, and we can compare our NLTE results with the LTE ones from the literature. While the abundance of HD~72660, log$\epsilon_{\rm Mg}$ = 7.80$\pm$0.22 \citep{2016MNRAS.456.3318G}, agrees within the error bars with our determinations (Table~\ref{tabUV}), for Sirius \citet{2011AA...528A.132L} obtained log$\epsilon_{\rm Mg}$ = 7.34$\pm$0.1, which is lower than our value, by 0.18~dex.  
 
\begin{figure*}
	\begin{center}
		%\begin{center}
		\includegraphics[scale=0.4]{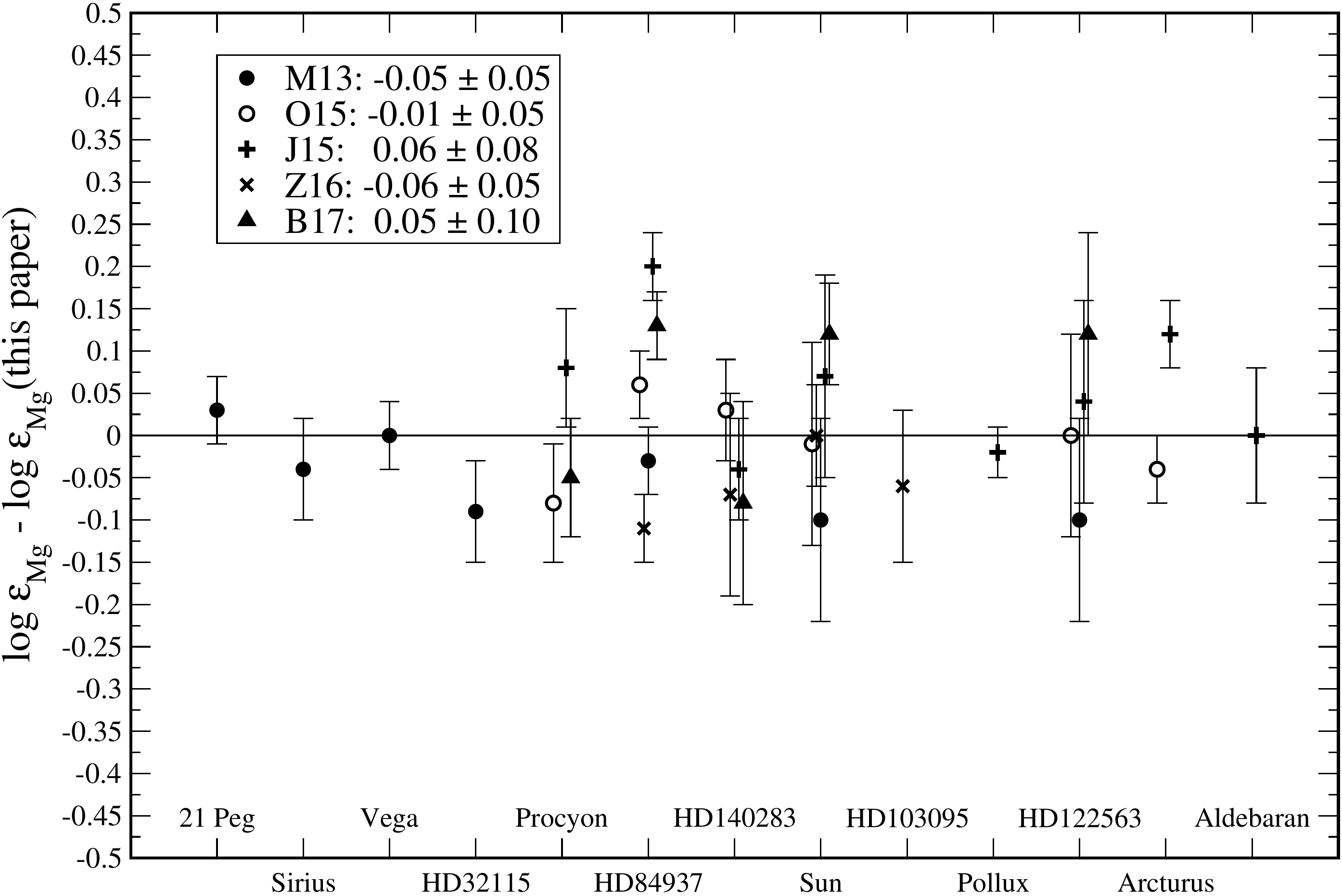}
		\caption{Abundance differences between the literature 1D NLTE analyses and this paper. The error bars correspond to that estimated in this paper. The average differences together with their standard deviations are indicated for different data sets in the left upper corner. } 
		\label{plot-comp}
	\end{center}
\end{figure*}

 Table~\ref{comp} and Fig.~\ref{plot-comp} show abundance comparisons for all other stars, for which the literature NLTE data are available. All the published Mg NLTE abundances are based on lines of Mg\ione, except those in \citet{2015AA...579A..53O}, where lines of Mg\ii\ were also used. Therefore for comparisons, we use our Mg NLTE abundances derived from lines of Mg\ione. We find the agreement between our results and those derived in the other NLTE studies rather satisfactory. 
 The largest abundance difference, of 0.08~dex, and the largest dispersion, of up to 0.12~dex, are obtained, when comparing our results with those of \citet{2017ApJ...847...16B} and \citet{2015AA...582A..81J}. In the latter case, this can be explained, in part, by using different sets of the Mg\ione\ lines.  

 \begin{deluxetable*}{lcccccccc}
%\tablenum{2}
%\scriptsize
\tablecaption{Comparison of the Mg NLTE abundances derived in this paper with the literature 1D NLTE data. Statistical errors are given in parentheses. \label{comp}}
\tablewidth{0pt}
\tablehead{
%\colhead{Star} & \colhead{Name} & \colhead{ \Teff } & \colhead{This paper} & \colhead{\citet{2013AA...550A..28M}} & \colhead{\citet{2015AA...579A..53O}} & \colhead{\citet{2015AA...582A..81J}}  &  %\colhead{\citet{2016ApJ...833..225Z}} &\colhead{\citet{2017ApJ...847...16B}}\\
\colhead{Star} & \colhead{Name} & \colhead{ \Teff } & \colhead{This paper} & \colhead{M13} & \colhead{O15} & \colhead{J15}  &  \colhead{Z16} &\colhead{B17}
}
\decimalcolnumbers
\startdata
  HD~209459 &   21 Peg     &  10400  & 7.52(05) & 7.57(07)&         &          &         &          \\         
  HD~48915  &   Sirius     &  ~9850  & 7.52(07) & 7.50(06)&         &          &         &          \\         
  HD~172167 &   Vega       &  ~9550  & 7.08(05) & 7.10(04)&         &          &         &          \\         
  HD~32115  &              &  ~7250  & 7.59(07) & 7.51(06)&         &          &         &          \\     
  HD~61421  &  Procyon     &  ~6580  & 7.53(09) &         & 7.45(04)& 7.61(08) &         & 7.48(12) \\                 
 HD~84937   &              &  ~6350  & 5.66(09) & 5.66(05)& 5.75(03)& 5.89(11) & 5.58(08)& 5.82(07) \\            
 HD~140283  &              &  ~5780  & 5.30(05) &         & 5.39(10)& 5.32(05) & 5.36(00)& 5.48(11) \\  
            &  Sun         &  ~5777  & 7.56(10) & 7.48(13)& 7.57(08)& 7.65(08) & 7.51(02)& 7.50(05) \\            
 HD~103095  &              &  ~5130  & 6.47(10) &         &         &          & 6.44(04)&          \\            
 HD~62509   &  Pollux      &  ~4860  & 7.57(06) &         &         & 7.56(07) &         &          \\                        
 HD~122563  &              &  ~4600  & 5.22(13) & 5.16(11)& 5.26(13)& 5.30(06) &         & 5.38(05) \\            
 HD~124897  &  Arcturus    &  ~4290  & 7.34(04) &         & 7.33(07)& 7.49(09) &         &          \\                      
 HD~29139   &  Aldebaran   &  ~3930  & 7.49(07) &         &         & 7.48(15) &         &          \\                      
\enddata                                                
\tablecomments{References: 
  M13 -- \citet{2013AA...550A..28M}; O15 -- \citet{2015AA...579A..53O}; J15 -- \citet{2015AA...582A..81J}; 
 Z16 -- \citet{2016ApJ...833..225Z}; B17 -- \citet{2017ApJ...847...16B}. 
For this paper, we give Mg abundances derived from the Mg\ione\ lines. }
\end{deluxetable*}

\section{Conclusions}\label{Sect:Conclusions}
  
  We constructed a comprehensive model atom for Mg\ione\ -- Mg\ii\ using the most up-to-date atomic data including that for inelastic collisions with neutral hydrogen atoms and electrons.
  Our model atom allows to analyze absoption and emission lines of Mg\ione\ and Mg\ii\ in wide range of stellar parameters.
Based on high S/N, high-resolution, and broad wavelength coverage spectra, the magnesium NLTE and LTE abundances were determined for the Sun and 17 stars
%The NLTE line formation for Mg\ione\ and Mg\ii\ was considered in classical 1D-LTE model atmospheres of the \textcolor{red}{17} stars
with reliable atmospheric parameters, which lie in the following ranges: 3900 $\le$ \Teff\ $\le$ 17500~K, 1.1 $\le$ log$g$ $\le$ 4.7, and $-$2.6 $\le$ [Fe/H] $\le$ +0.4.

 With the MARCS solar model atmosphere, we obtain the solar NLTE abundance, log~$\epsilon_{\rm Mg}$ = 7.54$\pm$0.11, from 10 lines of Mg\ione\ and log~$\epsilon_{\rm Mg}$ = 7.59$\pm$0.05 from 5 lines of Mg\ii. In LTE, the abundance difference between Mg\ione\ and Mg\ii\ amounts to $-0.20$~dex. 

  We find that, for each star, NLTE leads to smaller line-to-line scatter.
  For 10 stars with both Mg\ione\ and Mg\ii\ lines observed, NLTE provides consistent within 0.09~dex abundances from the two ionization stages, while the LTE abundance difference can be up to 0.23~dex in absolute value. However, we obtain an abundance discrepancy betweeen Mg\ione\ and Mg\ii\ in the two very metal-poor stars, HD~140283 and HD~84937. 
  The average abundance derived from the Mg\ione\ lines is lower than that from Mg\ii\ 4481~\AA, by 0.24 and 0.19~dex, respectively. Future investigations are required to explain such a discrepancy.

  We investigate the emission phenomena in the solar center disk intensity observations of the Mg\ione\ 7.3, 12.2, and 12.3~$\mu$m lines and in the spectra of 12.2 and 12.3 $\mu$m in Procyon, Arcturus, Pollux, and Aldebaran.
  Our standard NLTE modelling predicts the emission peaks in the Mg\ione\ IR lines, 
  however, the fits to the observed line profiles are not perfect except for Procyon, where Mg\ione\ 12.22 and 12.32~$\mu$m are reproduced very well.
  For solar Mg\ione\ 12.2 and 12.3~$\mu$m lines, their wings can be better fitted with a change in abundance of order 0.2 dex compared with the abundance derived from absorption lines of Mg\ione\ and Mg\ii. We find that the Mg\ione\ IR emission lines are strongly sensitive to collisional data variations. For example, we can fit solar Mg\ione\ 7.3, 12.2 and 12.3~$\mu$m lines by replacing our standard collisional recipe, which is based on using quantum mechanical electron-impact excitation rate coefficients of \citet{2015A&A...577A.113M}, with the theoretical approximations, that is the impact parameter method \citep{1962PPS....79.1105S} for the allowed transitions and $\Omega_{ij}$ = 1 for the forbidden ones.
  
  For K-giants, the predicted emission lines are too weak.
  We suggest that a formation of Mg\ione\ 12.22 and 12.32~$\mu$m in Aldebaran, Arcturus, and Pollux is affected by the chromospheric temperature rise, which is not accounted for by our classical radiative-equilibrium model atmospheres.  

Our NLTE calculations predict that lines of Mg\ione\ in the infrared spectral range may appear as emission lines depending on the atmospheric parameters. 
  The emission in Mg\ione\ 12.224 and 12.321~$\mu$m appears first at effective temperature of 4000~K (log~$g$ = 2.0), and it is strengthened towards higher \Teff. 
  The Mg\ione\ lines at 11.789~$\mu$m and 7.736~$\mu$m come into emission at \Teff\ $>$ 6000~K (log~$g$ = 2.0). 
 
For determinations of Mg abundance of stars with 7000 $\le$ \Teff\ $\le$ 17500~K, we can recommend the Mg\ii\ 3848, 3850, 4384, 4390, 4427, and 4433~\AA\ lines, even at the LTE assumption, due to small NLTE effects for these lines. For stars with 7000~K $<$ \Teff\ $\leq$ 8000~K, the Mg\ione\ 4167, 4571, 4702, 5528, 5167, 5172, and 5183~\AA\ lines can be safely used in the LTE analysis. 
  For the hotter stars, with \Teff\ from 8000 to 9500~K, the NLTE effects are minor only for Mg\ione\ 4167, 4702, and 4528~\AA. 
  We do not recommend to use the Mg\ii\ 9218 and 9244~\AA\ lines due to their position in the wings of the Paschen P9 line and also pronounced NLTE effects, with negative NLTE corrections up to one order of magnitude. 
  
In the models representing atmospheres of F-G-K-type stars of close-to-solar metallicity, the NLTE corrections for all Mg\ione\ lines, which we analysed, can be positive and negative depending on stellar parameters, but do not exceed 0.09~dex in absolute value. For the very metal-poor atmospheres, the NLTE corrections are close to zero for Mg\ione\ 4703, 4730, 8806, 4167, 5528, and 5711~\AA\, positive of up to 0.27~dex for Mg\ione\ 4571, 5167, 5172, and 5183~\AA, and slightly negative ($\le$ 0.04~dex in absolute value) for Mg\ii\ 4481~\AA.

   \software{DETAIL \citep{detail}, SynthV\_NLTE \citep{2016MNRAS.456.1221R}, binmag \citep{binmag3,2018ascl.soft05015K}, MARCS \citep{2008A&A...486..951G},  LLmodels code \citep{2004AA...428..993S}, SME \citep{1996A&AS..118..595V, 2017A&A...597A..16P}}.

\acknowledgments
  
   We thank the anonymous referee for valuable suggestions and comments. 
  We thank Prof. J. Landstreet for providing us with the UV spectra of HD~72660 and Sirius and N. Ryde for the infrared spectra of Procyon, Pollux, Arcturus, and Aldebaran. 
  This work was supported by the National Natural Science Foundation of China and Chinese Academy of Sciences joint fund on astronomy under grant No. U1331102 and by Sino-German Science Foundation under project No. GZ1183. This work is also partly supported by Young Scholars Program of Shandong University, Weihai.
  S.A. is grateful to China Postdoctoral international exchange program (ISS-SDU) for financial support.
  This research is based on observations obtained with MegaPrime/MegaCam, a joint project of CFHT and CEA/IRFU, at the Canada–France–Hawaii Telescope (CFHT) which is operated by the National Research Council (NRC) of Canada, the Institut National des Science de l Univers of the Centre National de la Recherche Scientifique (CNRS) of France, and the University of Hawaii.
  We made use of the NIST, SIMBAD, and VALD databases.

\bibliography{magnesium}

\begin{thebibliography}{}
\expandafter\ifx\csname natexlab\endcsname\relax\def\natexlab#1{#1}\fi
\providecommand{\url}[1]{\href{#1}{#1}}
\providecommand{\dodoi}[1]{doi:~\href{http://doi.org/#1}{\nolinkurl{#1}}}
\providecommand{\doeprint}[1]{\href{http://ascl.net/#1}{\nolinkurl{http://ascl.net/#1}}}
\providecommand{\doarXiv}[1]{\href{https://arxiv.org/abs/#1}{\nolinkurl{https://arxiv.org/abs/#1}}}

\bibitem[{{Abia} \& {Mashonkina}(2004)}]{2004MNRAS.350.1127A}
{Abia}, C., \& {Mashonkina}, L. 2004, \mnras, 350, 1127,
  \dodoi{10.1111/j.1365-2966.2004.07728.x}

\bibitem[{{Alexeeva} \& {Mashonkina}(2015)}]{2015MNRAS.453.1619A}
{Alexeeva}, S.~A., \& {Mashonkina}, L.~I. 2015, \mnras, 453, 1619,
  \dodoi{10.1093/mnras/stv1668}

\bibitem[{{Alexeeva} {et~al.}(2016){Alexeeva}, {Ryabchikova}, \&
  {Mashonkina}}]{2016MNRAS.462.1123A}
{Alexeeva}, S.~A., {Ryabchikova}, T.~A., \& {Mashonkina}, L.~I. 2016, \mnras,
  462, 1123, \dodoi{10.1093/mnras/stw1635}

\bibitem[{{Altrock} \& {Cannon}(1975)}]{1975SoPh...42..289A}
{Altrock}, R.~C., \& {Cannon}, C.~J. 1975, \solphys, 42, 289,
  \dodoi{10.1007/BF00149912}

\bibitem[{{Andrievsky} {et~al.}(2010){Andrievsky}, {Spite}, {Korotin}, {Spite},
  {Bonifacio}, {Cayrel}, {Fran{\c c}ois}, \& {Hill}}]{2010A&A...509A..88A}
{Andrievsky}, S.~M., {Spite}, M., {Korotin}, S.~A., {et~al.} 2010, \aap, 509,
  A88, \dodoi{10.1051/0004-6361/200913223}

\bibitem[{{Anstee} \& {O'Mara}(1995)}]{1995MNRAS.276..859A}
{Anstee}, S.~D., \& {O'Mara}, B.~J. 1995, \mnras, 276, 859,
  \dodoi{10.1093/mnras/276.3.859}

\bibitem[{{Aoki} {et~al.}(2006){Aoki}, {Frebel}, {Christlieb}, {Norris},
  {Beers}, {Minezaki}, {Barklem}, {Honda}, {Takada-Hidai}, {Asplund}, {Ryan},
  {Tsangarides}, {Eriksson}, {Steinhauer}, {Deliyannis}, {Nomoto}, {Fujimoto},
  {Ando}, {Yoshii}, \& {Kajino}}]{Aoki_he1327}
{Aoki}, W., {Frebel}, A., {Christlieb}, N., {et~al.} 2006, \apj, 639, 897,
  \dodoi{10.1086/497906}

\bibitem[{{Athay} \& {Canfield}(1969)}]{1969BAAS....1..272A}
{Athay}, R.~G., \& {Canfield}, R.~C. 1969, in \baas, Vol.~1, Bulletin of the
  American Astronomical Society, 272

\bibitem[{{Bagnulo} {et~al.}(2003){Bagnulo}, {Jehin}, {Ledoux}, {Cabanac},
  {Melo}, {Gilmozzi}, \& {ESO Paranal Science Operations
  Team}}]{2003Msngr.114...10B}
{Bagnulo}, S., {Jehin}, E., {Ledoux}, C., {et~al.} 2003, The Messenger, 114, 10

\bibitem[{{Bailey} \& {Landstreet}(2013)}]{2013AA...551A..30B}
{Bailey}, J.~D., \& {Landstreet}, J.~D. 2013, \aap, 551, A30,
  \dodoi{10.1051/0004-6361/201220671}

\bibitem[{{Barklem} {et~al.}(2012){Barklem}, {Belyaev}, {Spielfiedel},
  {Guitou}, \& {Feautrier}}]{2012A&A...541A..80B}
{Barklem}, P.~S., {Belyaev}, A.~K., {Spielfiedel}, A., {Guitou}, M., \&
  {Feautrier}, N. 2012, \aap, 541, A80, \dodoi{10.1051/0004-6361/201219081}

\bibitem[{{Barklem} \& {O'Mara}(1997)}]{1997MNRAS.290..102B}
{Barklem}, P.~S., \& {O'Mara}, B.~J. 1997, \mnras, 290, 102,
  \dodoi{10.1093/mnras/290.1.102}

\bibitem[{{Barklem} {et~al.}(2000){Barklem}, {Piskunov}, \& {O'Mara}}]{BPM}
{Barklem}, P.~S., {Piskunov}, N., \& {O'Mara}, B.~J. 2000, Astron. and
  Astrophys. Suppl. Ser., 142, 467, \dodoi{10.1051/aas:2000167}

\bibitem[{{Bergemann} {et~al.}(2017{\natexlab{a}}){Bergemann}, {Collet},
  {Amarsi}, {Kovalev}, {Ruchti}, \& {Magic}}]{2017ApJ...847...15B}
{Bergemann}, M., {Collet}, R., {Amarsi}, A.~M., {et~al.} 2017{\natexlab{a}},
  \apj, 847, 15, \dodoi{10.3847/1538-4357/aa88cb}

\bibitem[{{Bergemann} {et~al.}(2017{\natexlab{b}}){Bergemann}, {Collet},
  {Sch{\"o}nrich}, {Andrae}, {Kovalev}, {Ruchti}, {Hansen}, \&
  {Magic}}]{2017ApJ...847...16B}
{Bergemann}, M., {Collet}, R., {Sch{\"o}nrich}, R., {et~al.}
  2017{\natexlab{b}}, \apj, 847, 16, \dodoi{10.3847/1538-4357/aa88b5}

\bibitem[{{Boyajian} {et~al.}(2013){Boyajian}, {von Braun}, {van Belle},
  {Farrington}, {Schaefer}, {Jones}, {White}, {McAlister}, {ten Brummelaar},
  {Ridgway}, {Gies}, {Sturmann}, {Sturmann}, {Turner}, {Goldfinger}, \&
  {Vargas}}]{2013ApJ...771...40B}
{Boyajian}, T.~S., {von Braun}, K., {van Belle}, G., {et~al.} 2013, \apj, 771,
  40, \dodoi{10.1088/0004-637X/771/1/40}

\bibitem[{{Brault} \& {Noyes}(1983)}]{1983ApJ...269L..61B}
{Brault}, J., \& {Noyes}, R. 1983, \apjl, 269, L61, \dodoi{10.1086/184056}

\bibitem[{{Burkhart} \& {Coupry}(1998)}]{1998A&A...338.1073B}
{Burkhart}, C., \& {Coupry}, M.~F. 1998, \aap, 338, 1073

\bibitem[{{Butler} \& {Giddings}(1985)}]{detail}
{Butler}, K., \& {Giddings}, J. 1985, Newsletter on the analysis of
  astronomical spectra, No. 9, University of London

\bibitem[{{Carlsson} {et~al.}(1992){Carlsson}, {Rutten}, \&
  {Shchukina}}]{1992A&A...253..567C}
{Carlsson}, M., {Rutten}, R.~J., \& {Shchukina}, N.~G. 1992, \aap, 253, 567

\bibitem[{{Chang} \& {Noyes}(1983)}]{1983ApJ...275L..11C}
{Chang}, E.~S., \& {Noyes}, R.~W. 1983, \apjl, 275, L11, \dodoi{10.1086/184161}

\bibitem[{{Chang} \& {Schoenfeld}(1991)}]{1991ApJ...383..450C}
{Chang}, E.~S., \& {Schoenfeld}, W.~G. 1991, \apj, 383, 450,
  \dodoi{10.1086/170802}

\bibitem[{{Clark} {et~al.}(1991){Clark}, {Csanak}, \&
  {Abdallah}}]{1991PhRvA..44.2874C}
{Clark}, R.~E.~H., {Csanak}, G., \& {Abdallah}, Jr., J. 1991, \pra, 44, 2874,
  \dodoi{10.1103/PhysRevA.44.2874}

\bibitem[{{Cunto} {et~al.}(1993){Cunto}, {Mendoza}, {Ochsenbein}, \&
  {Zeippen}}]{1993BICDS..42...39C}
{Cunto}, W., {Mendoza}, C., {Ochsenbein}, F., \& {Zeippen}, C.~J. 1993,
  Bulletin d'Information du Centre de Donnees Stellaires, 42, 39

\bibitem[{{Deeming}(1960)}]{1960MNRAS.121...52D}
{Deeming}, T.~J. 1960, \mnras, 121, 52, \dodoi{10.1093/mnras/121.1.52}

\bibitem[{{Dimitrijevi{\'c}} \&
  {Sahal-Br{\'e}chot}(1995)}]{1995BABel.151..101D}
{Dimitrijevi{\'c}}, M.~S., \& {Sahal-Br{\'e}chot}, S. 1995, Bulletin
  Astronomique de Belgrade, 151, 101

\bibitem[{{Dimitrijevic} \& {Sahal-Brechot}(1996)}]{1996AAS..117..127D}
{Dimitrijevic}, M.~S., \& {Sahal-Brechot}, S. 1996, \aaps, 117, 127

\bibitem[{{Edvardsson}(1988)}]{1988A&A...190..148E}
{Edvardsson}, B. 1988, \aap, 190, 148

\bibitem[{{Fekel} {et~al.}(2006){Fekel}, {Williamson}, {Buggs}, {Onuoha}, \&
  {Smith}}]{2006AJ....132.1490F}
{Fekel}, F.~C., {Williamson}, M., {Buggs}, C., {Onuoha}, G., \& {Smith}, B.
  2006, \aj, 132, 1490, \dodoi{10.1086/507023}

\bibitem[{{Fossati} {et~al.}(2007){Fossati}, {Bagnulo}, {Monier}, {Khan},
  {Kochukhov}, {Landstreet}, {Wade}, \& {Weiss}}]{2007AA...476..911F}
{Fossati}, L., {Bagnulo}, S., {Monier}, R., {et~al.} 2007, \aap, 476, 911,
  \dodoi{10.1051/0004-6361:20078320}

\bibitem[{{Fossati} {et~al.}(2009){Fossati}, {Ryabchikova}, {Bagnulo},
  {Alecian}, {Grunhut}, {Kochukhov}, \& {Wade}}]{2009AA...503..945F}
{Fossati}, L., {Ryabchikova}, T., {Bagnulo}, S., {et~al.} 2009, \aap, 503, 945,
  \dodoi{10.1051/0004-6361/200811561}

\bibitem[{{Fossati} {et~al.}(2011){Fossati}, {Ryabchikova}, {Shulyak},
  {Haswell}, {Elmasli}, {Pandey}, {Barnes}, \& {Zwintz}}]{2011MNRAS.417..495F}
{Fossati}, L., {Ryabchikova}, T., {Shulyak}, D.~V., {et~al.} 2011, \mnras, 417,
  495, \dodoi{10.1111/j.1365-2966.2011.19289.x}

\bibitem[{{Fuhrmann}(1998)}]{Fuhrmann1998}
{Fuhrmann}, K. 1998, \aap, 338, 161

\bibitem[{{Fuhrmann} {et~al.}(1997){Fuhrmann}, {Pfeiffer}, {Frank}, {Reetz}, \&
  {Gehren}}]{1997A&A...323..909F}
{Fuhrmann}, K., {Pfeiffer}, M., {Frank}, C., {Reetz}, J., \& {Gehren}, T. 1997,
  \aap, 323, 909

\bibitem[{{Gigas}(1988)}]{1988A&A...192..264G}
{Gigas}, D. 1988, \aap, 192, 264

\bibitem[{{Golriz} \& {Landstreet}(2016)}]{2016MNRAS.456.3318G}
{Golriz}, S.~S., \& {Landstreet}, J.~D. 2016, \mnras, 456, 3318,
  \dodoi{10.1093/mnras/stv2658}

\bibitem[{{Gratton} {et~al.}(1999){Gratton}, {Carretta}, {Eriksson}, \&
  {Gustafsson}}]{Gratton1999}
{Gratton}, R.~G., {Carretta}, E., {Eriksson}, K., \& {Gustafsson}, B. 1999,
  \aap, 350, 955

\bibitem[{{Green} {et~al.}(1957){Green}, {Rush}, \&
  {Chandler}}]{1957ApJS....3...37G}
{Green}, L.~C., {Rush}, P.~P., \& {Chandler}, C.~D. 1957, \apjs, 3, 37,
  \dodoi{10.1086/190031}

\bibitem[{{Gustafsson} {et~al.}(2008){Gustafsson}, {Edvardsson}, {Eriksson},
  {J{\o}rgensen}, {Nordlund}, \& {Plez}}]{2008A&A...486..951G}
{Gustafsson}, B., {Edvardsson}, B., {Eriksson}, K., {et~al.} 2008, \aap, 486,
  951, \dodoi{10.1051/0004-6361:200809724}

\bibitem[{{Heiter} {et~al.}(2015){Heiter}, {Jofr{\'e}}, {Gustafsson}, {Korn},
  {Soubiran}, \& {Th{\'e}venin}}]{2015AA...582A..49H}
{Heiter}, U., {Jofr{\'e}}, P., {Gustafsson}, B., {et~al.} 2015, \aap, 582, A49,
  \dodoi{10.1051/0004-6361/201526319}

\bibitem[{{Hibbert} {et~al.}(1993){Hibbert}, {Biemont}, {Godefroid}, \&
  {Vaeck}}]{1993AAS...99..179H}
{Hibbert}, A., {Biemont}, E., {Godefroid}, M., \& {Vaeck}, N. 1993, \aaps, 99,
  179

\bibitem[{{Hill} {et~al.}(2010){Hill}, {Gulliver}, \&
  {Adelman}}]{2010ApJ...712..250H}
{Hill}, G., {Gulliver}, A.~F., \& {Adelman}, S.~J. 2010, \apj, 712, 250,
  \dodoi{10.1088/0004-637X/712/1/250}

\bibitem[{{Hill} \& {Landstreet}(1993)}]{1993AA...276..142H}
{Hill}, G.~M., \& {Landstreet}, J.~D. 1993, \aap, 276, 142

\bibitem[{{Idiart} \& {Thevenin}(2000)}]{2000ApJ...541..207I}
{Idiart}, T., \& {Thevenin}, F. 2000, \apj, 541, 207, \dodoi{10.1086/309416}

\bibitem[{{Jofr{\'e}} {et~al.}(2014){Jofr{\'e}}, {Heiter}, {Soubiran},
  {Blanco-Cuaresma}, {Worley}, {Pancino}, {Cantat-Gaudin}, {Magrini},
  {Bergemann}, {Gonz{\'a}lez Hern{\'a}ndez}, {Hill}, {Lardo}, {de Laverny},
  {Lind}, {Masseron}, {Montes}, {Mucciarelli}, {Nordlander}, {Recio Blanco},
  {Sobeck}, {Sordo}, {Sousa}, {Tabernero}, {Vallenari}, \& {Van
  Eck}}]{2014A&A...564A.133J}
{Jofr{\'e}}, P., {Heiter}, U., {Soubiran}, C., {et~al.} 2014, \aap, 564, A133,
  \dodoi{10.1051/0004-6361/201322440}

\bibitem[{{Jofr{\'e}} {et~al.}(2015){Jofr{\'e}}, {Heiter}, {Soubiran},
  {Blanco-Cuaresma}, {Masseron}, {Nordlander}, {Chemin}, {Worley}, {Van Eck},
  {Hourihane}, {Gilmore}, {Adibekyan}, {Bergemann}, {Cantat-Gaudin},
  {Delgado-Mena}, {Gonz{\'a}lez Hern{\'a}ndez}, {Guiglion}, {Lardo}, {de
  Laverny}, {Lind}, {Magrini}, {Mikolaitis}, {Montes}, {Pancino},
  {Recio-Blanco}, {Sordo}, {Sousa}, {Tabernero}, \&
  {Vallenari}}]{2015AA...582A..81J}
---. 2015, \aap, 582, A81, \dodoi{10.1051/0004-6361/201526604}

\bibitem[{{Keller} {et~al.}(2014){Keller}, {Bessell}, {Frebel}, {Casey},
  {Asplund}, {Jacobson}, {Lind}, {Norris}, {Yong}, {Heger}, {Magic}, {da
  Costa}, {Schmidt}, \& {Tisserand}}]{2014Natur.506..463K}
{Keller}, S.~C., {Bessell}, M.~S., {Frebel}, A., {et~al.} 2014, \nat, 506, 463,
  \dodoi{10.1038/nature12990}

\bibitem[{{Kochukhov}(2010)}]{binmag3}
{Kochukhov}, O. 2010

\bibitem[{{Kochukhov}(2018)}]{2018ascl.soft05015K}
---. 2018, {BinMag: Widget for comparing stellar observed with theoretical
  spectra}, Astrophysics Source Code Library.
\newblock \doeprint{1805.015}

\bibitem[{{Kramida} {et~al.}(2018){Kramida}, {Ralchenko, Yu.}, {Reader}, \&
  {NIST ASD Team}}]{NIST_ASD}
{Kramida}, A., {Ralchenko, Yu.}, {Reader}, J., \& {NIST ASD Team}. 2018, {NIST
  Atomic Spectra Database (ver. 5.5.2), [Online]. Available:
  {\tt{https://physics.nist.gov/asd}} [2018, January 25]. National Institute of
  Standards and Technology, Gaithersburg, MD.}

\bibitem[{{Kurucz} {et~al.}(1984){Kurucz}, {Furenlid}, {Brault}, \&
  {Testerman}}]{1984sfat.book.....K}
{Kurucz}, R.~L., {Furenlid}, I., {Brault}, J., \& {Testerman}, L. 1984, {Solar
  flux atlas from 296 to 1300 nm}

\bibitem[{{Kurucz} \& {Peytremann}(1975)}]{KP}
{Kurucz}, R.~L., \& {Peytremann}, E. 1975, SAO Special Report, 362, 1

\bibitem[{{Lacy} {et~al.}(2002){Lacy}, {Richter}, {Greathouse}, {Jaffe}, \&
  {Zhu}}]{2002PASP..114..153L}
{Lacy}, J.~H., {Richter}, M.~J., {Greathouse}, T.~K., {Jaffe}, D.~T., \& {Zhu},
  Q. 2002, \pasp, 114, 153, \dodoi{10.1086/338730}

\bibitem[{{Landstreet}(2011)}]{2011AA...528A.132L}
{Landstreet}, J.~D. 2011, \aap, 528, A132, \dodoi{10.1051/0004-6361/201016259}

\bibitem[{{Leonard}(1996)}]{1996ASPC...90..337L}
{Leonard}, P.~J.~T. 1996, in Astronomical Society of the Pacific Conference
  Series, Vol.~90, The Origins, Evolution, and Destinies of Binary Stars in
  Clusters, ed. E.~F. {Milone} \& J.-C. {Mermilliod}, 337

\bibitem[{{Lodders} {et~al.}(2009){Lodders}, {Palme}, \&
  {Gail}}]{2009LanB...4B...44L}
{Lodders}, K., {Palme}, H., \& {Gail}, H.-P. 2009, Landolt B{\"o}rnstein,
  \dodoi{10.1007/978-3-540-88055-4_34}

\bibitem[{{Luo} {et~al.}(1989){Luo}, {Pradhan}, {Saraph}, {Storey}, \&
  {Yu}}]{1989JPhB...22..389L}
{Luo}, D., {Pradhan}, A.~K., {Saraph}, H.~E., {Storey}, P.~J., \& {Yu}, Y.
  1989, Journal of Physics B Atomic Molecular Physics, 22, 389,
  \dodoi{10.1088/0953-4075/22/3/006}

\bibitem[{{Mashonkina}(2013)}]{2013AA...550A..28M}
{Mashonkina}, L. 2013, \aap, 550, A28, \dodoi{10.1051/0004-6361/201220761}

\bibitem[{{Mashonkina} {et~al.}(2011){Mashonkina}, {Gehren}, {Shi}, {Korn}, \&
  {Grupp}}]{mash_fe}
{Mashonkina}, L., {Gehren}, T., {Shi}, J.-R., {Korn}, A.~J., \& {Grupp}, F.
  2011, \aap, 528, A87, \dodoi{10.1051/0004-6361/201015336}

\bibitem[{{Mashonkina} {et~al.}(2017){Mashonkina}, {Jablonka}, {Sitnova},
  {Pakhomov}, \& {North}}]{2017A&A...608A..89M}
{Mashonkina}, L., {Jablonka}, P., {Sitnova}, T., {Pakhomov}, Y., \& {North}, P.
  2017, \aap, 608, A89, \dodoi{10.1051/0004-6361/201731582}

\bibitem[{{Mashonkina} {et~al.}(1996){Mashonkina}, {Shimanskaya}, \&
  {Sakhibullin}}]{1996ARep...40..187M}
{Mashonkina}, L.~I., {Shimanskaya}, N.~N., \& {Sakhibullin}, N.~A. 1996,
  Astronomy Reports, 40, 187

\bibitem[{{Matteucci} \& {Brocato}(1990)}]{1990ApJ...365..539M}
{Matteucci}, F., \& {Brocato}, E. 1990, \apj, 365, 539, \dodoi{10.1086/169508}

\bibitem[{{Mauas} {et~al.}(1988){Mauas}, {Avrett}, \&
  {Loeser}}]{1988ApJ...330.1008M}
{Mauas}, P.~J., {Avrett}, E.~H., \& {Loeser}, R. 1988, \apj, 330, 1008,
  \dodoi{10.1086/166530}

\bibitem[{{Merle} {et~al.}(2011){Merle}, {Thevenin}, {Pichon}, \&
  {Bigot}}]{2011MNRAS.418..863M}
{Merle}, T., {Thevenin}, F., {Pichon}, B., \& {Bigot}, L. 2011, \mnras, 418,
  863, \dodoi{10.1111/j.1365-2966.2011.19540.x}

\bibitem[{{Merle} {et~al.}(2015){Merle}, {Th{\'e}venin}, \&
  {Zatsarinny}}]{2015A&A...577A.113M}
{Merle}, T., {Th{\'e}venin}, F., \& {Zatsarinny}, O. 2015, \aap, 577, A113,
  \dodoi{10.1051/0004-6361/201525914}

\bibitem[{{Mihalas}(1972)}]{1972ApJ...177..115M}
{Mihalas}, D. 1972, \apj, 177, 115, \dodoi{10.1086/151690}

\bibitem[{{Mishenina} {et~al.}(2004){Mishenina}, {Soubiran}, {Kovtyukh}, \&
  {Korotin}}]{2004A&A...418..551M}
{Mishenina}, T.~V., {Soubiran}, C., {Kovtyukh}, V.~V., \& {Korotin}, S.~A.
  2004, \aap, 418, 551, \dodoi{10.1051/0004-6361:20034454}

\bibitem[{{Molenda-Zakowicz} \& {Polubek}(2005)}]{2005AcA....55..375M}
{Molenda-Zakowicz}, J., \& {Polubek}, G. 2005, \actaa, 55, 375

\bibitem[{{Murcray} {et~al.}(1981){Murcray}, {Goldman}, {Murcray}, {Bradford},
  {Murcray}, {Coffey}, \& {Mankin}}]{1981ApJ...247L..97M}
{Murcray}, F.~J., {Goldman}, A., {Murcray}, F.~H., {et~al.} 1981, \apjl, 247,
  L97, \dodoi{10.1086/183598}

\bibitem[{{Nieva} \& {Przybilla}(2012)}]{2012AA...539A.143N}
{Nieva}, M.-F., \& {Przybilla}, N. 2012, \aap, 539, A143,
  \dodoi{10.1051/0004-6361/201118158}

\bibitem[{{Nordlander} {et~al.}(2017){Nordlander}, {Amarsi}, {Lind}, {Asplund},
  {Barklem}, {Casey}, {Collet}, \& {Leenaarts}}]{2017A&A...597A...6N}
{Nordlander}, T., {Amarsi}, A.~M., {Lind}, K., {et~al.} 2017, \aap, 597, A6,
  \dodoi{10.1051/0004-6361/201629202}

\bibitem[{{Osorio} \& {Barklem}(2016)}]{2016A&A...586A.120O}
{Osorio}, Y., \& {Barklem}, P.~S. 2016, \aap, 586, A120,
  \dodoi{10.1051/0004-6361/201526958}

\bibitem[{{Osorio} {et~al.}(2015){Osorio}, {Barklem}, {Lind}, {Belyaev},
  {Spielfiedel}, {Guitou}, \& {Feautrier}}]{2015AA...579A..53O}
{Osorio}, Y., {Barklem}, P.~S., {Lind}, K., {et~al.} 2015, \aap, 579, A53,
  \dodoi{10.1051/0004-6361/201525846}

\bibitem[{{Pehlivan Rhodin} {et~al.}(2017){Pehlivan Rhodin}, {Hartman},
  {Nilsson}, \& {Jonsson}}]{2017AA...598A.102P}
{Pehlivan Rhodin}, A., {Hartman}, H., {Nilsson}, H., \& {Jonsson}, P. 2017,
  Astron. and Astrophys., 598, A102, \dodoi{10.1051/0004-6361/201629849}

\bibitem[{{Piskunov} \& {Valenti}(2017)}]{2017A&A...597A..16P}
{Piskunov}, N., \& {Valenti}, J.~A. 2017, \aap, 597, A16,
  \dodoi{10.1051/0004-6361/201629124}

\bibitem[{{Przybilla} {et~al.}(2001){Przybilla}, {Butler}, {Becker}, \&
  {Kudritzki}}]{2001A&A...369.1009P}
{Przybilla}, N., {Butler}, K., {Becker}, S.~R., \& {Kudritzki}, R.~P. 2001,
  \aap, 369, 1009, \dodoi{10.1051/0004-6361:20010164}

\bibitem[{{Przybilla} {et~al.}(2000){Przybilla}, {Butler}, {Becker},
  {Kudritzki}, \& {Venn}}]{2000AA...359.1085P}
{Przybilla}, N., {Butler}, K., {Becker}, S.~R., {Kudritzki}, R.~P., \& {Venn},
  K.~A. 2000, \aap, 359, 1085

\bibitem[{{Przybilla} {et~al.}(2011){Przybilla}, {Nieva}, \&
  {Butler}}]{2011JPhCS.328a2015P}
{Przybilla}, N., {Nieva}, M.-F., \& {Butler}, K. 2011, Journal of Physics
  Conference Series, 328, 012015, \dodoi{10.1088/1742-6596/328/1/012015}

\bibitem[{{Renson} \& {Manfroid}(2009)}]{2009A&A...498..961R}
{Renson}, P., \& {Manfroid}, J. 2009, \aap, 498, 961,
  \dodoi{10.1051/0004-6361/200810788}

\bibitem[{{Ryabchikova} {et~al.}(2016){Ryabchikova}, {Piskunov}, {Pakhomov},
  {Tsymbal}, {Titarenko}, {Sitnova}, {Alexeeva}, {Fossati}, \&
  {Mashonkina}}]{2016MNRAS.456.1221R}
{Ryabchikova}, T., {Piskunov}, N., {Pakhomov}, Y., {et~al.} 2016, \mnras, 456,
  1221, \dodoi{10.1093/mnras/stv2725}

\bibitem[{{Rybicki} \& {Hummer}(1991)}]{rh91}
{Rybicki}, G.~B., \& {Hummer}, D.~G. 1991, \aap, 245, 171

\bibitem[{{Ryde} {et~al.}(2004){Ryde}, {Korn}, {Richter}, \&
  {Ryde}}]{2004ApJ...617..551R}
{Ryde}, N., {Korn}, A.~J., {Richter}, M.~J., \& {Ryde}, F. 2004, \apj, 617,
  551, \dodoi{10.1086/425265}

\bibitem[{{Ryde} \& {Richter}(2004)}]{2004ApJ...611L..41R}
{Ryde}, N., \& {Richter}, M.~J. 2004, \apjl, 611, L41, \dodoi{10.1086/423618}

\bibitem[{{Sadakane}(1981)}]{1981PASP...93..587S}
{Sadakane}, K. 1981, \pasp, 93, 587, \dodoi{10.1086/130892}

\bibitem[{{Sasso} {et~al.}(2017){Sasso}, {Andretta}, {Terranegra}, \&
  {Gomez}}]{2017A&A...604A..50S}
{Sasso}, C., {Andretta}, V., {Terranegra}, L., \& {Gomez}, M.~T. 2017, \aap,
  604, A50, \dodoi{10.1051/0004-6361/201730676}

\bibitem[{{Seaton}(1962{\natexlab{a}})}]{1962PPS....79.1105S}
{Seaton}, M.~J. 1962{\natexlab{a}}, Proceedings of the Physical Society, 79,
  1105, \dodoi{10.1088/0370-1328/79/6/304}

\bibitem[{{Seaton}(1962{\natexlab{b}})}]{Seaton1962}
---. 1962{\natexlab{b}}, Atomic and Molecular Processes(New York: Academic
  Press) (New York: Academic Press)

\bibitem[{{Shulyak} {et~al.}(2004){Shulyak}, {Tsymbal}, {Ryabchikova},
  {St{\"u}tz}, \& {Weiss}}]{2004AA...428..993S}
{Shulyak}, D., {Tsymbal}, V., {Ryabchikova}, T., {St{\"u}tz}, C., \& {Weiss},
  W.~W. 2004, \aap, 428, 993, \dodoi{10.1051/0004-6361:20034169}

\bibitem[{{Sigut} \& {Pradhan}(1995)}]{1995JPhB...28.4879S}
{Sigut}, T.~A.~A., \& {Pradhan}, A.~K. 1995, Journal of Physics B Atomic
  Molecular Physics, 28, 4879, \dodoi{10.1088/0953-4075/28/22/018}

\bibitem[{{Sitnova} {et~al.}(2015){Sitnova}, {Zhao}, {Mashonkina}, {Chen},
  {Liu}, {Pakhomov}, {Tan}, {Bolte}, {Alexeeva}, {Grupp}, {Shi}, \&
  {Zhang}}]{2015ApJ...808..148S}
{Sitnova}, T., {Zhao}, G., {Mashonkina}, L., {et~al.} 2015, \apj, 808, 148,
  \dodoi{10.1088/0004-637X/808/2/148}

\bibitem[{{Sitnova} {et~al.}(2016){Sitnova}, {Mashonkina}, \&
  {Ryabchikova}}]{2016MNRAS.461.1000S}
{Sitnova}, T.~M., {Mashonkina}, L.~I., \& {Ryabchikova}, T.~A. 2016, \mnras,
  461, 1000, \dodoi{10.1093/mnras/stw1202}

\bibitem[{{Sitnova} {et~al.}(2018){Sitnova}, {Mashonkina}, \&
  {Ryabchikova}}]{2018MNRAS.477.3343S}
---. 2018, \mnras, 477, 3343, \dodoi{10.1093/mnras/sty810}

\bibitem[{{Smith} \& {Dworetsky}(1993)}]{1993A&A...274..335S}
{Smith}, K.~C., \& {Dworetsky}, M.~M. 1993, \aap, 274, 335

\bibitem[{{Snijders} \& {Lamers}(1975)}]{1975A&A....41..245S}
{Snijders}, M.~A.~J., \& {Lamers}, H.~J.~G.~L.~M. 1975, \aap, 41, 245

\bibitem[{{Sundqvist} {et~al.}(2008){Sundqvist}, {Ryde}, {Harper}, {Kruger}, \&
  {Richter}}]{2008A&A...486..985S}
{Sundqvist}, J.~O., {Ryde}, N., {Harper}, G.~M., {Kruger}, A., \& {Richter},
  M.~J. 2008, \aap, 486, 985, \dodoi{10.1051/0004-6361:200809778}

\bibitem[{{Tafelmeyer} {et~al.}(2010){Tafelmeyer}, {Jablonka}, {Hill},
  {Shetrone}, {Tolstoy}, {Irwin}, {Battaglia}, {Helmi}, {Starkenburg}, {Venn},
  {Abel}, {Francois}, {Kaufer}, {North}, {Primas}, \&
  {Szeifert}}]{2010A&A...524A..58T}
{Tafelmeyer}, M., {Jablonka}, P., {Hill}, V., {et~al.} 2010, \aap, 524, A58,
  \dodoi{10.1051/0004-6361/201014733}

\bibitem[{{Uitenbroek} \& {Noyes}(1996)}]{1996ASPC..109..723U}
{Uitenbroek}, H., \& {Noyes}, R.~W. 1996, in Astronomical Society of the
  Pacific Conference Series, Vol. 109, Cool Stars, Stellar Systems, and the
  Sun, ed. R.~{Pallavicini} \& A.~K. {Dupree}, 723

\bibitem[{{Valenti} \& {Piskunov}(1996)}]{1996A&AS..118..595V}
{Valenti}, J.~A., \& {Piskunov}, N. 1996, \aaps, 118, 595

\bibitem[{{van Regemorter}(1962)}]{1962ApJ...136..906V}
{van Regemorter}, H. 1962, \apj, 136, 906, \dodoi{10.1086/147445}

\bibitem[{{Venn}(1995)}]{1995ApJS...99..659V}
{Venn}, K.~A. 1995, \apjs, 99, 659, \dodoi{10.1086/192201}

\bibitem[{{Venn} \& {Lambert}(1990)}]{1990ApJ...363..234V}
{Venn}, K.~A., \& {Lambert}, D.~L. 1990, \apj, 363, 234, \dodoi{10.1086/169334}

\bibitem[{{Woosley} \& {Weaver}(1995)}]{1995ApJS..101..181W}
{Woosley}, S.~E., \& {Weaver}, T.~A. 1995, \apjs, 101, 181,
  \dodoi{10.1086/192237}

\bibitem[{{Zhao} {et~al.}(1998){Zhao}, {Butler}, \&
  {Gehren}}]{1998A&A...333..219Z}
{Zhao}, G., {Butler}, K., \& {Gehren}, T. 1998, \aap, 333, 219

\bibitem[{{Zhao} {et~al.}(2016){Zhao}, {Mashonkina}, {Yan}, {Alexeeva},
  {Kobayashi}, {Pakhomov}, {Shi}, {Sitnova}, {Tan}, {Zhang}, {Zhang}, {Zhou},
  {Bolte}, {Chen}, {Li}, {Liu}, \& {Zhai}}]{2016ApJ...833..225Z}
{Zhao}, G., {Mashonkina}, L., {Yan}, H.~L., {et~al.} 2016, \apj, 833, 225,
  \dodoi{10.3847/1538-4357/833/2/225}

\end{thebibliography}
\bibliographystyle{aasjournal}

\end{document}